\documentclass[iop,numberedappendix,twocolappendix]{emulateapj}

\usepackage{chngcntr}

\usepackage{placeins}

\usepackage{float} 
\usepackage{url}
\usepackage{natbib}

\usepackage{graphicx}
\usepackage{color}
\usepackage{rotating}
\usepackage{lineno}

\usepackage[breaklinks, colorlinks, citecolor=blue]{hyperref} 

\def\,{\thinspace}
\def\lsim{\mathrel{\raise .4ex\hbox{\rlap{$<$}\lower 1.2ex\hbox{$\sim$}}}}
\def\gsim{\mathrel{\raise .4ex\hbox{\rlap{$>$}\lower 1.2ex\hbox{$\sim$}}}}

\def\simprop{\mathrel{\raise .4ex\hbox{\rlap{$\propto$}\lower 1.2ex\hbox{$\sim$}}}}
\def\deg{\ifmmode^\circ\else$^\circ$\fi}
\def\pdeg{\ifmmode $\setbox0=\hbox{$^{\circ}$}\rlap{\hskip.11\wd0 .}$^{\circ}
          \else \setbox0=\hbox{$^{\circ}$}\rlap{\hskip.11\wd0 .}$^{\circ}$\fi}
\def\arcs{\ifmmode {^{\scriptstyle\prime\prime}}
          \else $^{\scriptstyle\prime\prime}$\fi}
\def\arcm{\ifmmode {^{\scriptstyle\prime}}
          \else $^{\scriptstyle\prime}$\fi}
\newdimen\sa  \newdimen\sb
\def\parcs{\sa=.07em \sb=.03em
     \ifmmode \hbox{\rlap{.}}^{\scriptstyle\prime\kern -\sb\prime}\hbox{\kern -\sa}
     \else \rlap{.}$^{\scriptstyle\prime\kern -\sb\prime}$\kern -\sa\fi}
\def\parcm{\sa=.08em \sb=.03em
     \ifmmode \hbox{\rlap{.}\kern\sa}^{\scriptstyle\prime}\hbox{\kern-\sb}
     \else \rlap{.}\kern\sa$^{\scriptstyle\prime}$\kern-\sb\fi}
\def\GHz{\ifmmode $\,GHz$\else \,GHz\fi}
\def\MJysr{\ifmmode \,$MJy\,sr\mo$\else \,MJy\,sr\mo\fi}
\def\microns{\ifmmode \,\mu$m$\else \,$\mu$m\fi}

\def\kms{\ifmmode $\,km\,s$^{-1}\else \,km\,s$^{-1}$\fi}

\newcommand{\perbeam}{beam$^{-1}$}

\providecommand{\sorthelp}[1]{}

\newcommand{\dhiglsarchive}{www.cita.utoronto.ca/DHIGLS}

\newcommand{\uv}{$u$\,-\,$v$}
\newcommand{\colnum}{19}
\newcommand{\colunits}{$10^{\colnum}$~cm$^{-2}$}

\newcommand{\noisecenter}{3.0}
\newcommand{\factoredge}{9.5} 
\newcommand{\weightedge}{0.011}
\newcommand{\noiseedge}{28.5}

\newcommand{\noisesingledeltacorrected}{3.2} %

\newcommand{\expon}{\gamma}

\newcommand{\fwhmp}{108} 

\newcommand{\psfk}{\tilde{\phi}}

\newcommand{\kmin}{$k_{\rm min}$}
\newcommand{\kmax}{$k_{\rm max}$}
\newcommand{\kminvalst}{0.018\,arcmin$^{-1}$}
\newcommand{\kmaxvalst}{0.5\,arcmin$^{-1}$}
\newcommand{\bminst}{13}
\newcommand{\bmaxst}{361}

\newcommand{\bmincross}{13}
\newcommand{\bmaxcross}{35}
\newcommand{\kmincross}{0.018}
\newcommand{\kmaxcross}{0.0485}

\newcommand{\ccal}{1.12}
\newcommand{\estccal}{0.01} 
\newcommand{\esyccal}{0.03} 

\newcommand{\degree}{$^{\circ}$}
\newcommand{\h}{\hbox{\kern 0.20em $^{\rm h}$}}
\newcommand{\m}{\hbox{\kern 0.20em $^{\rm m}$}}
\newcommand{\s}{\hbox{\kern 0.20em $^{\rm s}$}}

\newcommand{\HI}{\ion{H}{1}}
\newcommand{\hi}{\ion{H}{1}}

\newcommand{\cmm}{${\rm cm}^{-2}$}

\newcommand{\nh}{$N_{ {\mathrm{H}} \, \mathrm{I}} $}
\newcommand{\nhm}{N_{ {\mathrm{H}} \, \mathrm{I}} }
\newcommand{\vnh}{n_{ {\mathrm{H}} \, \mathrm{I}} }
\newcommand{\wh}{$W_{{\mathrm{H}} \, \mathrm{I}} $}
\newcommand{\vlsr}{v_{\rm LSR}}
\newcommand{\ghigls}{GHIGLS}
\newcommand{\dhigls}{DHIGLS}
\newcommand{\draost}{DRAO~ST}
\newcommand{\dgm}{\draost\,+\,\ghigls}

\newcommand{\DFi}{DF-i} 
\newcommand{\ENi}{EN-i} 
\newcommand{\DRi}{DR-i} 
\newcommand{\POi}{PO-i} 
\newcommand{\UMi}{UM-i} 
\newcommand{\URSA}{URSA} 
\newcommand{\DF}{DF} 
\newcommand{\EN}{EN} 
\newcommand{\DR}{DR} 
\newcommand{\PO}{PO} 
\newcommand{\UM}{UM} 
\newcommand{\MC}{MC} 
\newcommand{\MG}{MG} 
\newcommand{\DFG}{DF} 
\newcommand{\ENG}{EN} 
\newcommand{\DRG}{DR} 
\newcommand{\POG}{PO} 
\newcommand{\UMG}{UM} 
\newcommand{\MCG}{MC} 
\newcommand{\MGG}{MG} 

\newcommand{\Trb}{T_{\rm b}}
\newcommand{\Trc}{T_{\rm c}}
\newcommand{\Trp}{T_{\rm p}}
\newcommand{\Trn}{T_{\rm n}}
\newcommand{\Trs}{T_{\rm s}}

\newcommand{\vrl}{v}

\newcommand{\Planck}{\textit{Planck}}
\newcommand{\Herschel}{\textit{Herschel}}
\newcommand{\Spitzer}{\textit{Spitzer}}
\newcommand{\IRAS}{\textit{IRAS}}

\newcommand{\ISO}{\textit{ISO}}

\newcommand{\ncpl}{NCPL} 

\slugcomment{Submitted to The Astrophysical Journal}

\shorttitle{\dhigls: Deep \hi\ Surveys with the DRAO ST}
\shortauthors{Blagrave et al.}

\begin{document}

\title{\dhigls: DRAO \HI\  Intermediate Galactic Latitude Survey}

\author{K. Blagrave\altaffilmark{1},
          P. G. Martin\altaffilmark{1},
          G. Joncas\altaffilmark{2},
          R. Kothes\altaffilmark{3},
          J. M. Stil\altaffilmark{4},
          M. A. Miville-Desch{\^e}nes\altaffilmark{1,5},\\
          Felix J.~Lockman\altaffilmark{6},
	   A.R. Taylor\altaffilmark{7,8,9}
          }

\altaffiltext{1}{Canadian Institute for Theoretical Astrophysics,
University of Toronto, 60 St. George Street, Toronto, ON M5S~3H8,
Canada; \email{blagrave@cita.utoronto.ca}}

\altaffiltext{2}{Universit\'{e} Laval, Qu\'{e}bec, PQ, Canada}

\altaffiltext{3}{Dominion Radio Astrophysical Observatory, Herzberg
Institute of Astrophysics, National Research Council of Canada, Box
248, Penticton, BC, Canada V2A 6J9}

\altaffiltext{4}{Department of Physics and Astronomy, University of
Calgary, 2500 University Drive NW, Calgary, AB, Canada T2N 1N4}

\altaffiltext{5}{Institut d'Astrophysique Spatiale, CNRS (UMR8617)
Universit\'{e} Paris-Sud 11, B\^{a}timent 121, Orsay, France}

\altaffiltext{6}{National Radio Astronomy Observatory, Green Bank, WV
  24944, USA}

\altaffiltext{7}{University of Calgary, Calgary, AB, Canada}

\altaffiltext{8}{University of Cape Town, Cape Town, South Africa}

\altaffiltext{9}{University of the Western Cape, Cape Town, South Africa}
	

\begin{abstract}
Observations of Galactic \HI\ gas for seven targeted regions at
intermediate Galactic latitude are presented at $1\arcmin$ angular
resolution using data from the DRAO Synthesis Telescope (ST) and the
Green Bank Telescope (GBT).
The \dhigls\ data are the most extensive arcminute resolution
measurements of the diffuse atomic interstellar medium beyond those in
the Galactic plane.
The acquisition, reduction, calibration, and mosaicking of the
\draost\ data and the cross calibration and incorporation of the
short-spacing information from the GBT are described.
The high quality of the resulting \dhigls\ products enables a variety of 
new studies in directions of low Galactic column density.
We analyze the angular power spectra of maps of the integrated \HI\
emission (column density) from the data cubes for several distinct
velocity ranges.  Fitting power spectrum models based on a power law,
but including the effects of the synthesized beam and noise at high
spatial frequencies, we find exponents ranging from $-2.5$ to $-3.0$.
Power spectra of maps of the centroid velocity for these components
give similar results.  These exponents are interpreted as being
representative of the 3D density and 3D velocity fields of the atomic
gas, respectively.
We find evidence for dramatic changes in the \HI\ structures in
channel maps over even small changes in velocity.  This narrow line
emission has counterparts in absorption spectra against bright
background radio sources, quantifying that the gas is cold and dense
and can be identified as the cold neutral medium phase.
Fully reduced \dhigls\ \hi\ data cubes and other data products are
available at \url{\dhiglsarchive}.
\end{abstract}


\maketitle


\section{Introduction}
\label{intro}

Atomic hydrogen is the most ubiquitous neutral component in the
interstellar medium (ISM).  Observations of its spin-flip transition
at 21\,cm have been used to map the distribution of \hi\ on the sky,
where it is found to exist not only in the Galactic plane but also at
all Galactic latitudes and with a variety of velocities
\citep[e.g.,][]{kalb05}.  Its distribution traces the neutral atomic
ISM in both the warm phase (warm neutral medium, or WNM) and the
colder denser phase (cold neutral medium, or CNM)
and reflects the influence of the gravitational potential of the Milky
Way galaxy and thermodynamic and dynamical effects.
Although the spatial power spectrum of \HI\ emission at intermediate
and high Galactic latitudes is quite steep indicating that most
structure in \HI\ occurs on the largest scales
\citep[e.g.,][]{mamd2003}, there is important structure at these
latitudes on angular scales $\la3\arcmin$ as evidenced in \Herschel\
observations \citep[e.g.,][]{mamd2010}.  These scales are not sampled
by existing 
single-dish
21\,cm surveys: the southern hemisphere-based \hi\ Galactic All-Sky
Survey \citep[GASS,][]{kalb15}; and in the north the Effelsberg-Bonn
\hi\ Survey \citep[EBHIS,][]{winkel2016} and the deeper, targeted
Green Bank Telescope (GBT) survey \citep[\ghigls,][]{mart15}.

To address this structure, arcminute resolution observations with an
interferometer are required.  Surveys in the Galactic plane include
the Canadian Galactic Plane Survey \citep[CGPS,][]{taylor2003}, the
Southern Galactic Plane Survey \citep[SGPS,][]{mccl2005}, and the VLA
Galactic Plane Survey \citep[VGPS,][]{stil2006}.
\hi\ has been studied at even higher resolution (20\arcsec) in the
THOR survey \citep{beuther16}, which supplements the VGPS with data
from the VLA C-array configuration.
Surveys at higher latitude are less common but some have been carried
out at the Dominion Radio Astrophysical Observatory (DRAO\footnote{
\url{www.nrc-cnrc.gc.ca/eng/solutions/facilities/drao.html}}) with the
Synthesis Telescope
(\draost)
\citep[e.g.,][]{mart94,mamd2003} and with the VLA closer to the plane
\citep[e.g.,][]{pido2015}.  As described in this paper we built on the
legacy of the CGPS by using the \draost\ to map \hi\ in seven regions
at intermediate Galactic latitude during the period 2004 to 2012. This
program is called \dhigls\ (DRAO \HI\ Intermediate Galactic Latitude
Survey).  The intermediate-latitude regions that we selected for the
\draost\ observations sample relatively faint \hi\ emission, with both
low and intermediate values of the neutral hydrogen column density
(\nh).  They also sample a variety of gas components distinguishable
by their velocity (``velocity components,'' hereafter VCs), spanning
low, intermediate, and high velocity gas (LVC, IVC, and HVC,
respectively).

The kinematic information in \hi\ spectra provides essential
diagnostics of various physical properties of the gas.  The power
spectrum of a map of \hi\ emission can provide insight into the
turbulent cascade in the ISM and the statistical description of the
density and velocity fields.  A direct correlation between the
observed power law exponent and the energy and/or density structure,
however, is not necessarily straightforward.  From theory
\citep{lazarian2000} and from fractional Brownian motion simulations
\citep{mamd2003b}, the exponent of the power spectrum is expected to
vary with the line-of-sight thickness of the medium, becoming steeper
as this thickness exceeds the measured transverse spatial scales.

\citet{mamd2003} provide an extensive discussion of empirical results.
Observations of \HI\ in the Galactic plane
\citep{green1993,dickey2001,green2007} result in power spectra with
power-law exponents ranging from $-2.2$ to $-3.0$ and values in the
inner Galaxy span $-3$ to $-4$.
\citet{henn12} presented a graphical summary of spectral indices as a
function of scale for various components of the ISM (see their
Figure~10).  Power laws associated with \HI\ components appear to vary
from shallow ($-2.75$) for \HI\ in absorption \citep{desh2000} to
steep ($-3.6 \pm 0.2$) for LVC \HI\ in emission in the Ursa Major
Galactic cirrus studied by \citet{mamd2003} using two pointings of the
\draost.  The difference has been attributed to the \hi\ lines in
question tracing material at different temperatures: the colder ISM
seen in \HI\ absorption has a shallower power law than the relatively
warm ISM that contributes to the \HI\ emission.

High resolution \HI\ spectra can also give insight to the properties
of interstellar dust.  Morphological spatial detail in maps of \hi\
varies as a function of velocity, and so any dust closely correlated
with a velocity component of the gas leaves a related morphological
imprint in the dust emission map.  This approach has been used to show
that there is dust of significant emissivity associated with IVC gas
\citep{mart94}.
The selection of fields targeted by \ghigls\ enabled exploration of
the different kinematics and spatial distributions of \hi\ gas and the
dust evolution at diverse stages of Galactic evolution.
\ghigls\ data were used in combination with \Planck\ data on thermal
dust emission to find the emissivity, opacity, and temperature of dust
associated with both LVC and IVC gas \citep{planck2011-7.12}.
At higher \nh, excess dust emission above the correlation with \hi\
suggests the presence of molecular hydrogen.  With this motivation,
\citet{barr2010} used \draost\ spectra to investigate the atomic to
molecular hydrogen transition in the diffuse ISM.  Short-spacing
information was supplied by the 26-m DRAO single dish.
These \draost\ syntheses are a subset of those used for the \DF\ and
\UM\ regions (Table~\ref{beamsize}).
However, the \dhigls\ products presented here improve and expand on
those data by increasing the size and depth of the mosaics and
integrating short-spacing data from the 100-m GBT.

In the first part of the paper, which deals with the processing and
quality assurance of the data,
Section~\ref{thefields} describes the \dhigls\ regions.
In Section~\ref{drao_observations} the spectral line observations with
the \draost\ are introduced.
Mosaicking of these data is described in Section~\ref{makingmosaics}.
The process of combining the mosaicked \draost\ interferometric data
with single dish data from \ghigls\ fields, which produces the
\dhigls\ products released with this paper, is discussed in
Section~\ref{gbt_observations}.

The second part of the paper illustrates some of the science enabled
by the data.
Properties of the gas in distinct ranges of velocity (velocity
components, or VCs) are introduced in Section~\ref{pvcs}, in
particular maps of integrated \hi\ emission (column density) and
centroid velocities of the VCs.
Several complementary power spectrum analyses of the \hi\ gas are
presented in Section~\ref{interpretation}.  
An exploration of some features of the observed CNM gas follows in
Section~\ref{showoff}.
Section~\ref{conclusions} summarizes the \dhigls\ project and the main
results of our first investigations with these new data.
A description of the data available in the \dhigls\ online archive is
provided there.

There are several supporting Appendices.
Appendix~\ref{appdfmps} details our implementation of analysis using
the angular power spectrum.
Complementary ranges of data in the \uv\ domain from interferometers
and single dishes are assessed in Appendix~\ref{appen:short}.
Appendix~\ref{crosscalib} describes the cross calibration of the
\draost\ and \ghigls\ data.
The dependence of optical depth corrections on the angular resolution
of the observations is explored in Appendix~\ref{hiopac}.
Appendix~\ref{discdetailsursa} revisits the results of some previous
measurements of power spectra in the ISM.
Supplementary details of the data and analysis for many \dhigls\
regions are presented in Appendix~\ref{additionalfields}.

\begin{figure}
\centering
\hspace*{-0.25cm} 
\includegraphics[angle=0, width=1.07\linewidth]{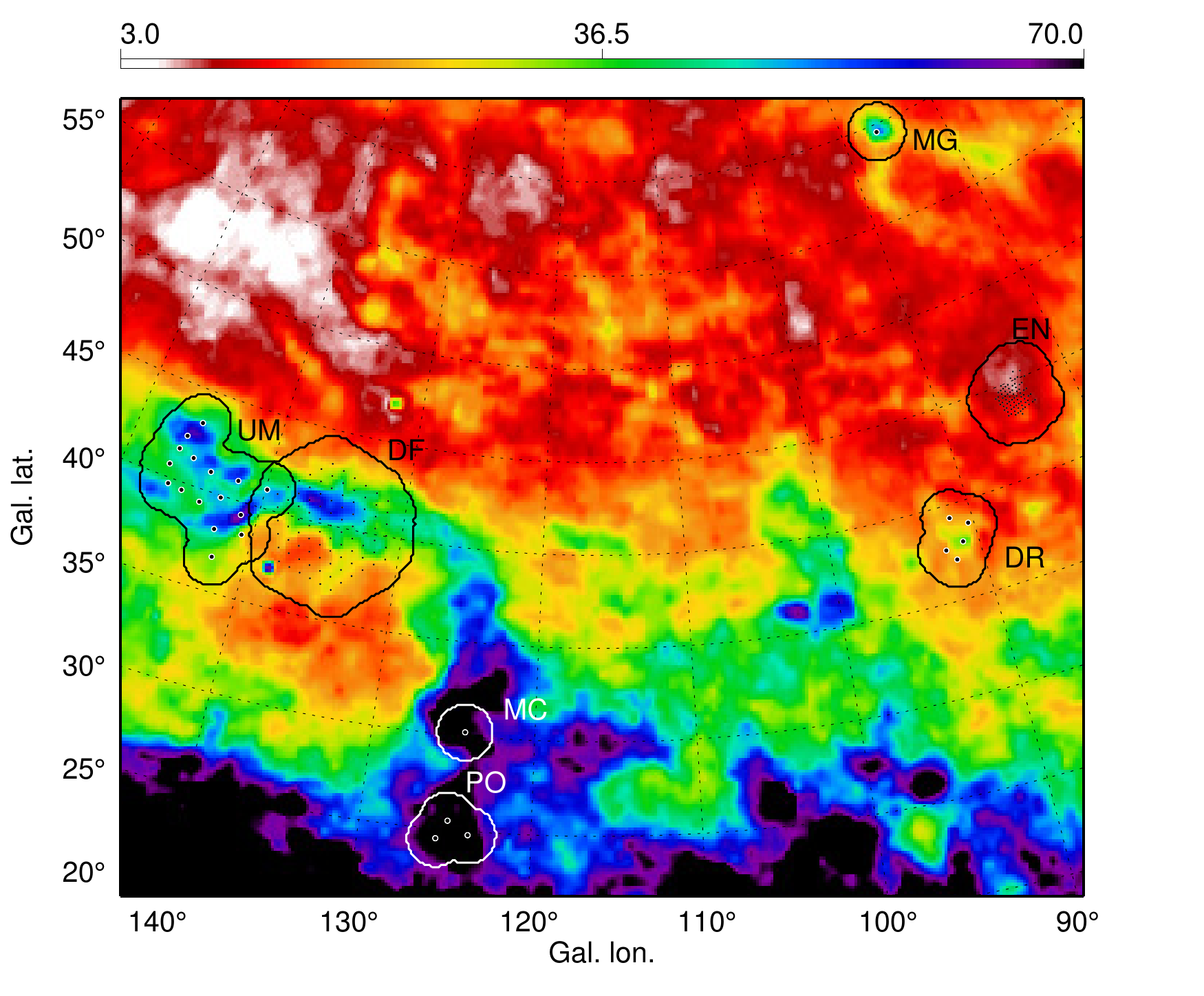}
\caption{
Locations of five \draost\ mosaics and two single-pointing syntheses
overlaid on a map of integrated \HI\ emission from LAB in units of
\colunits.  More details regarding the denser pointing centers are
given in Section~\ref{mosaicdef} and Appendix~\ref{umsummary}.
}
\label{field_locations}
\end{figure}

\begin{table*}
\caption{\dhigls\ Regions Observed with the \draost\ and GBT}
\label{beamsize}
\centering
\begin{tabular}{ccccccccc}
\hline \hline
\dhigls & \ghigls & 
Location\tablenotemark{a}
& Size\tablenotemark{a} & \draost\ & $v_{c}$\tablenotemark{b} & $\Delta v$\tablenotemark{b} &  Major\tablenotemark{c} & Minor\tablenotemark{c} \\
Region & Field & ($\alpha, \delta$) & (deg$^2$) & Syntheses  & (\kms) & (\kms)  & (arcsec) & (arcsec) \\
\hline
DF & SPIDER & $10^{\rm h}30^{\rm m}, 73\deg48\arcmin$ & 57.4 & DF00--DF90\tablenotemark{d} & $-60$ & $0.824$ & $55.8\pm0.7$ & $53.7\pm0.3$  \\
 \\
\EN & N1 & $16^{\rm h}14^{\rm m}, 54\deg49\arcmin$ & 14.6 & EN01--EN76  & $-60$ & $0.824$ & $65.7\pm0.8$ & $53.7\pm0.4$ \\
\\
MG & G86\tablenotemark{e} & $14^{\rm h}41^{\rm m}, 49\deg13\arcmin$ & 7.1 & MG & $-20$ & $0.412$ & $71.8$ & $53.4$ \\
\\
\DR & DRACO &  $16^{\rm h}48^{\rm m}, 61\deg38\arcmin$ & 12.5 & MD& $-70$ & $0.824$  &$61.3\pm0.5$ & $53.7\pm0.1$ \\
  & & & & PT--PW & $0$ & $0.824$  & & \\
\\
\PO & POL\tablenotemark{e}  & $02^{\rm h}57^{\rm m}, 87\deg07\arcmin$ & 10.7 & LU  LW  ME  & $-25$ & $0.412$ & $53.9\pm0.2$ & $53.3\pm0.2$\\
\\
MC & POL\tablenotemark{e} & $11^{\rm h}27^{\rm m}, 86\deg42\arcmin$ & 7.4 & MC  & $-35$ & $0.412$ & $54.7$ & $52.9$\\
\\
\UM & UMA\tablenotemark{f} & $09^{\rm h}41^{\rm m}, 68\deg33\arcmin$ & 40.7 & FP2\tablenotemark{g} NP & $-60$ & $0.824$  & $57.4\pm0.7$ & $53.6\pm0.3$ \\ 
& &  & & MB\tablenotemark{h} MH\tablenotemark{h} & $-20$ & $0.824$ & & \\
& &  & & NN\tablenotemark{i} NW NX PK--PS & $0$ & $0.824$  & & \\
\hline
\end{tabular}
\tablenotetext{1}{
Central pixel of the final \dhigls\ product
and area within solid line contour in Figure 1.}
\tablenotetext{2}{Configuration of the 256-channel spectrometer: central channel velocity, 
$v_{c}$, relative to the Local Standard of Rest (LSR), and channel spacing, $\Delta v$.
}
\tablenotetext{3}{Mean and dispersion of FWHM of synthesized beams for
groups of syntheses that have been assembled into a mosaic.}
\tablenotetext{4}{Syntheses DF66/76/81/87 excluded
(Figure~\ref{field_locations} and Section~\ref{register}).}
\tablenotetext{5}{Used \ghigls\ raw data reprocessed to a ``FINE" data
cube with channel spacing 0.32~\kms\ rather than 0.80~\kms\ 
(Section~\ref{gbtobs}).
}
\tablenotetext{6}{\ghigls\ \ncpl\ combined field used for slightly larger
spatial coverage 
(Section~\ref{gbtobs}).
}
\tablenotetext{7}{Same pointing as ``Field 2" in \citet{mamd2003}.}
\tablenotetext{8}{Used the original MB and MH syntheses with channel
spacing 0.412~\kms\ and their repeats MB2 and MH2 with channel spacing
0.824~\kms\
(Section~\ref{drao_observations}).
}
\tablenotetext{9}{Same pointing as ``Field 1" in \citet{mamd2003},
originally the EL field in \citet{jonc92} with early 4-antenna ST.}
\end{table*}

\section{The \dhigls\ Surveys}
\label{thefields}

Figure~\ref{field_locations} shows the locations of the \dhigls\
regions surveyed, overlaid on the map of \HI\ emission integrated over
velocity from the Leiden/Argentine/Bonn (LAB) survey \citep{kalb05}.
We note that by design all regions lie within \ghigls\ fields as
indicated in column~2 of Table~\ref{beamsize}.  The \ghigls\ \hi\ data
from the 100-m Robert C. Byrd Green Bank Telescope
\citep[GBT,][]{Prestage2009} at the National Radio Astronomy
Observatory (NRAO\footnote{ The National Radio Astronomy Observatory
is a facility of the National Science Foundation operated under
cooperative agreement by Associated Universities, Inc.}) over larger
areas are used to fill in the low spatial frequency data missing in
the interferometric observations (Section~\ref{gbt_observations}).
In some later figures and in the \dhigls\ data products these lower
resolution single-dish data are also used to provide important spatial
context for the \dhigls\ observations outside the region studied with
the \draost.

\draost\ data for our intermediate Galactic latitude mosaics were
obtained between 2002 and 2012.  The two deepest of these mosaics were
originally conceived as a complement to anticipated \Planck\ and
\Herschel\ observations.  We refer to the one toward the \ghigls\
field SPIDER as the ``DRAO Deep Field" and denote both the region
covered by mosaicked ST data and the final \dhigls\ product as \DF.
The second of these is toward the \ghigls\ field N1; we denote the
mosaicked region and product as \EN.  Refer to Table~\ref{beamsize}
for a summary of the additional details of these and further \dhigls\
mosaicked regions \DR, \PO, and \UM\ and individual syntheses \MG\ and
\MC.

\begin{figure*}
\centering
\includegraphics[clip=true,trim=10 20 10 10 ,angle=0,width=0.33\linewidth]{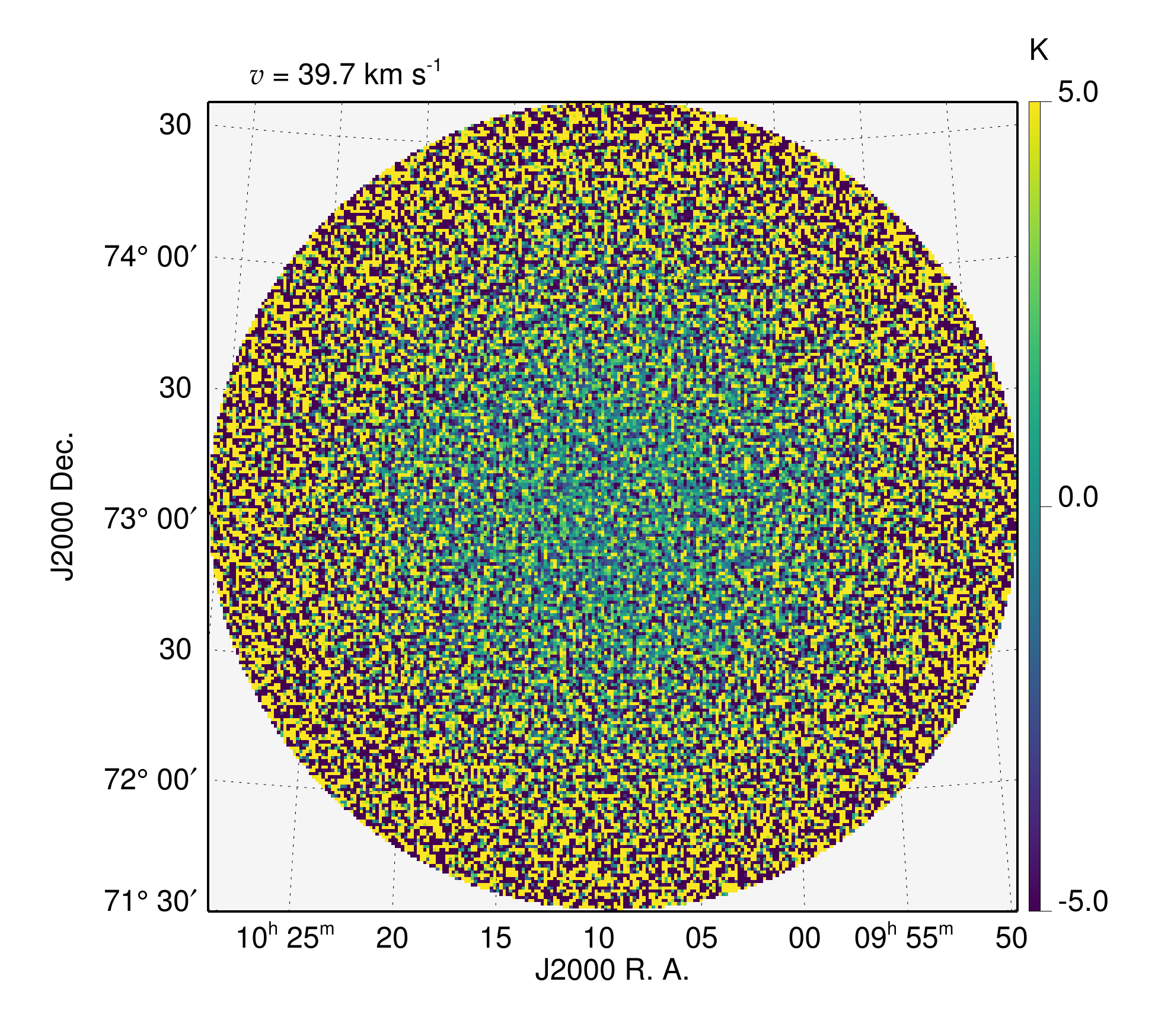}
\includegraphics[clip=true,trim = 10 20 10 10, angle=0,width=0.33\linewidth]{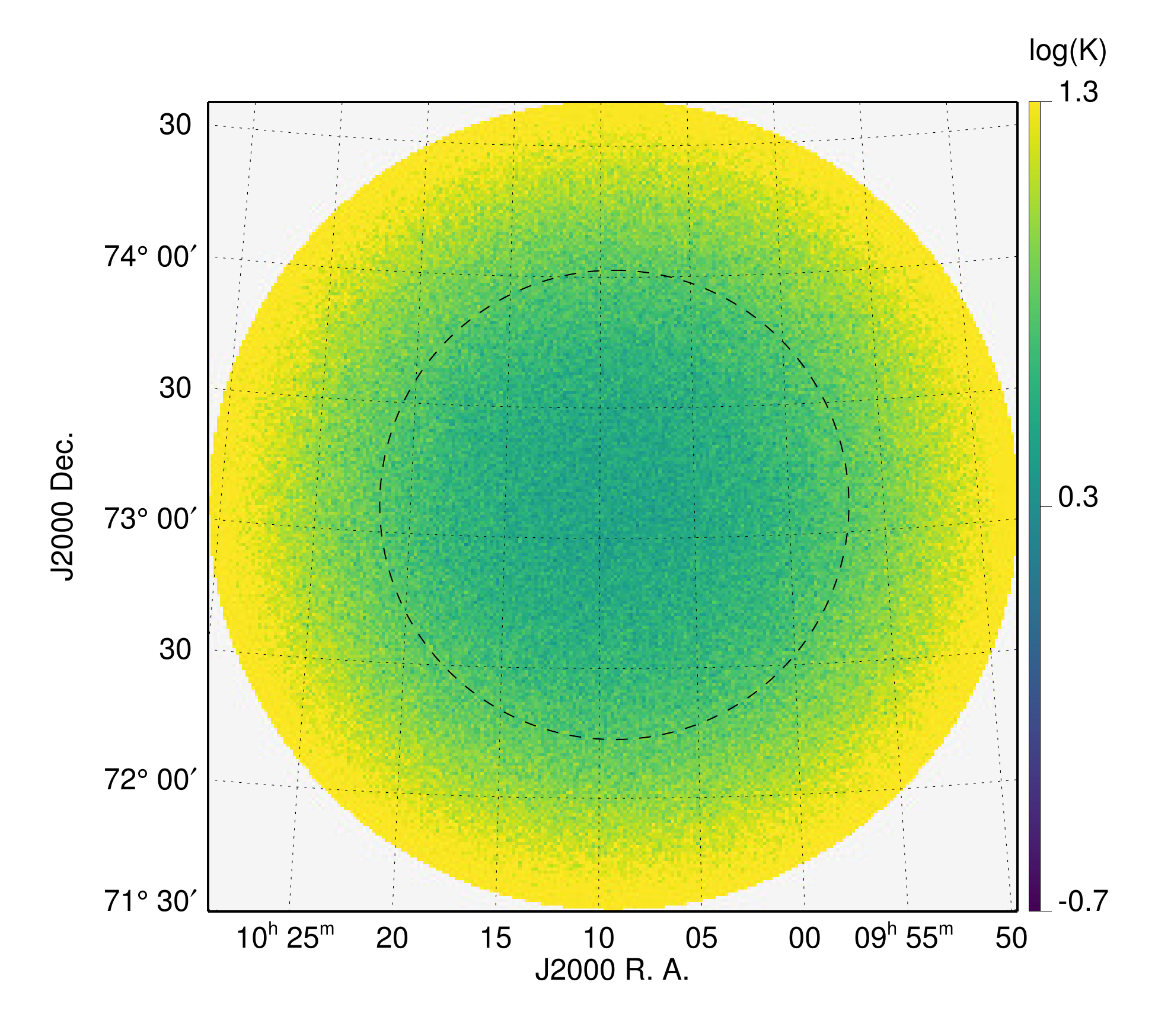}
\includegraphics[clip=true,trim = 10 20 10 10, angle=0,width=0.33\linewidth]{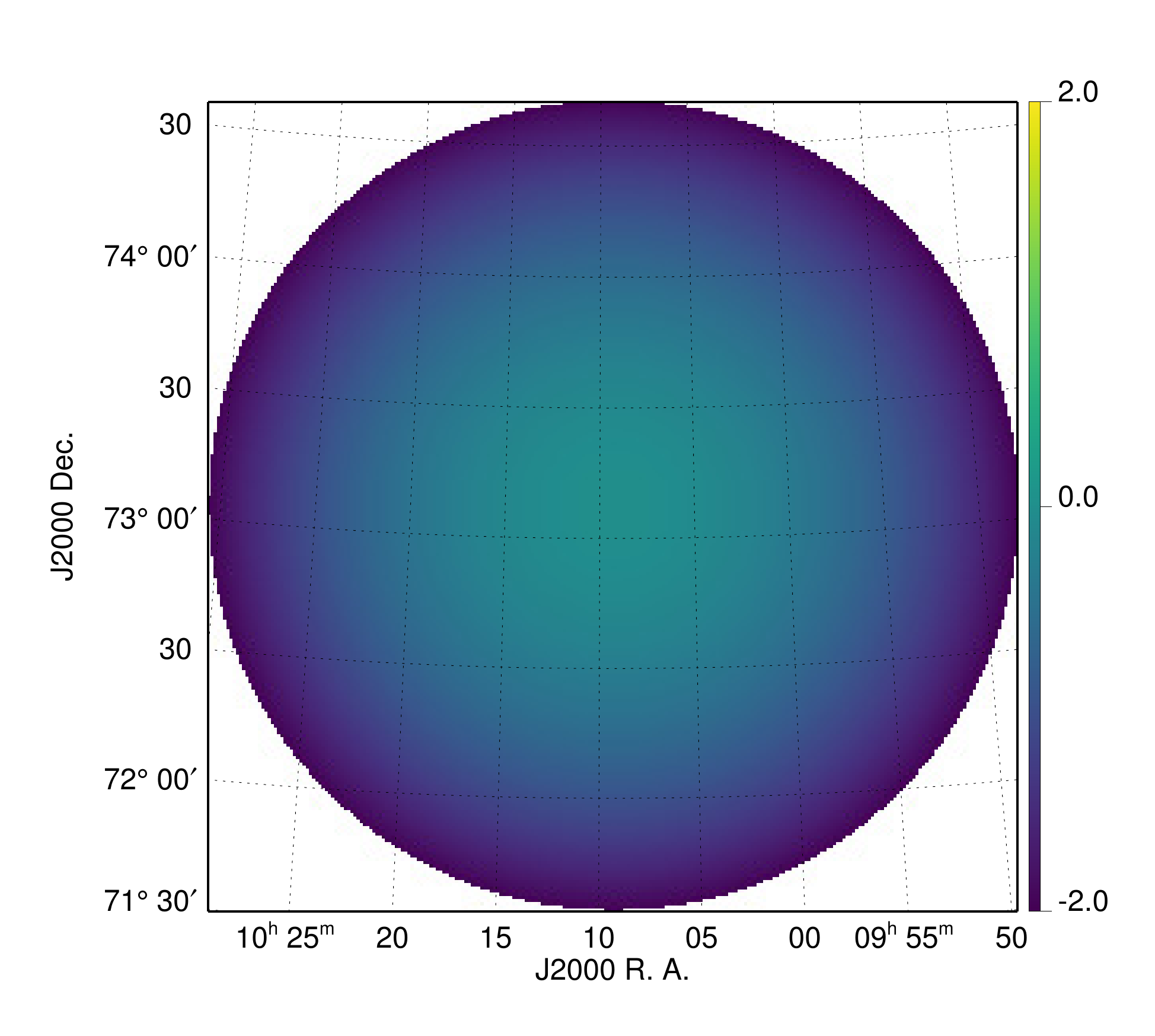}
\caption{
Left: single emission-free channel map from the representative single
synthesis DF20, part of the \DF\ mosaic. Correction for the primary
beam amplifies the noise away from the center.
Middle: noise map as determined 
empirically
from the scatter over a number of emission-free channels of DF20, with
logarithmic colorbar.  Noise at center is \noisecenter~K.  Dashed
circle of radius 54\arcmin\ locates the FWHM of the primary beam of a
9~m antenna in the interferometer; there the noise is a factor of 2
larger than at the center because of the primary beam correction.  At
the outer boundary of radius 93\arcmin\ the noise has increased by a
factor of \factoredge\ compared to the center, to \noiseedge~K.
Right: corresponding 
model 
weight map, in logarithmic form, normalized to unity at the centre.
Weight is 1/4 at the FWHM circle and \weightedge\ at the outer
truncation radius.
}
\label{noiseDF20}
\end{figure*}

\subsection{Description of \dhigls\ Regions and Science Goals}
\label{science}

With the \draost\ one cannot
simply carry out an all-sky survey to map Galactic
\hi\ at arcminute resolution.
Scarce
resources (i.e., the number of syntheses) must be focused
strategically.  This section presents briefly why we selected certain
fields to study.  The details of the ``resource allocation'' to the
\dhigls\ regions and within them (i.e., how the syntheses are placed),
which relate closely to the science goals, are deferred to
Section~\ref{mosaicdef} and are summarized in column 5 of
Table~\ref{beamsize}.  Included in the science considerations is the
availability of ancillary data (e.g., infrared imaging), which
broadens the range of scientific investigations that are enabled by
the \dhigls\ data.   
To complement studies in the Galactic plane, the \dhigls\ regions were
selected to be at intermediate latitude and to have low to
intermediate \nh\ within a variety of VCs. 

SPIDER\footnote{\url{www.cita.utoronto.ca/GHIGLS/page1.php?v=GHIGLS_SPIDER_mini}.
See the movies on
\url{www.cita.utoronto.ca/GHIGLS/page1.php?v=GHIGLS_NCPL_mini} 
for the larger-scale context of the structure in this field
as a function of velocity.
 }
is a 10\deg\ \ghigls\ field at the top of the arch of the North
Celestial Pole Loop (\ncpl), a giant gas structure north of the
Galactic Plane with a cylindrical morphology \citep{meyer91}.  The
related \dhigls\ mosaicked region \DF\ within this field focuses on
the highly structured diffuse LVC emission.  The IVC emission is much
fainter and the HVC negligible.
Targeted mapping of the stunning dust emission at even higher
resolution has been carried out with \Herschel.

The European Large-Area \ISO\ Survey (ELAIS) N1 field is an
``extragalactic window" targeted by the \Spitzer\ SWIRE survey
\citep{lons03} and most recently the HerMES legacy program on
\Herschel\ \citep{oliver2012}.  Its low column density can be
appreciated in Figure~\ref{field_locations}.  The \ghigls\ field N1
spans 5\deg.\footnote{
\url{www.cita.utoronto.ca/GHIGLS/page1.php?v=GHIGLS_N1_mini}.}
The \dhigls\ mosaicked region \EN\ focuses on the central lowest \nh\
region.  It has little LVC emission, negligible IVC emission, but
striking HVC emission.  At least at the GBT resolution the HVC gas
morphology is not immediately obviously traced by dust and
quantitatively the HVC has a low dust emissivity
\citep{planck2011-7.12}.  The higher resolution \dhigls\ observations
enable a deeper search for HVC-correlated dust emission.

In addition to the Galactic ISM science, another goal for the two
deepest \dhigls\ regions (\DF\ and \EN) is to provide higher
resolution \hi\ data to complement the characterization of foregrounds
at the 5\arcmin\ \Planck\ resolution.

DRACO \citep{herb93} was selected because of its prominent and
distinctive IVC gas, which has a clear dust signature with \IRAS\ and
\Planck\ \citep{planck2011-7.12}.  Targeted \Herschel\ dust emission
observations were carried out as well \citep{mamd2016}.  The dynamics
of the interaction of the IVC gas with the Galactic thick disk is
fascinating and the transition from atomic to molecular gas in dense
regions near the ``interaction surface" can also be studied.  Of
course, the new observations here cover only the \hi\ component.
Differences in dust properties between LVC and IVC gas could provide
evidence of dust evolution.

G86 is a prominent degree-sized high-latitude cloud found to have a
clear signature of dust correlated with the IVC gas using early DRAO
observations \citep{mart94}.  A more sensitive single-pointing
\draost\ synthesis MG has been carried out, along with targeted
\Herschel\ imaging.

The \dhigls\ region \UM\ within the \ghigls\ field UMA probes clouds
with LVC gas spanning the transition from the diffuse \hi\ phase to
the molecular phase.
A small two-pointing subregion of \UM, which we refer to as \URSA, has
been studied previously using early \draost\ data (\citealp{mamd2003};
revisited in Appendix~\ref{discdetailsursa}) and imaged with
\Herschel.

The small \dhigls\ region \PO\ is the brightest part of the \ghigls\
field POL and has molecular emission.  Part of this region has been
imaged with \Herschel\ \citep{mamd2010}.  We studied one other
single-field synthesis, MC, within POL.

\section{Spectral-line Mapping with the DRAO Synthesis Telescope}
\label{drao_observations}

For a complete discussion of the \draost\ and data reduction, refer to
\citet{landecker2000} and \citet{taylor2003}.  Here we summarize some
essential details for the \hi\ data.  The \draost\ is an array of
seven approximately 8.6~m antennas along an east-west baseline.  A
full synthesis contains data from baselines (antenna separations) in
multiples of $L=4.286$~m from $3L$ ($12.86$~m) to $144L$ ($617.18$~m)
and takes 144 hours of observing time ($12$~h $\times$ 12 antenna
configurations).  We note that while there are 21 baselines from 7
antennas, for \hi\ data the S21 correlator forms only those baseline
products needed for complete sampling of the Fourier plane, the 12
combinations involving a fixed and a movable antenna
\citep{landecker2000}; other combinations produce baselines that are
redundant.  For a field at Declination $\delta$ this sampling in the
Fourier domain results in a synthesized beam of $49\arcs \times
49\arcs \csc(\delta)$ at 1420 MHz \citep{landecker2000}.

For diffuse \hi\ emission such as studied here, with structure on all
scales, the presence of sidelobes in the dirty synthesized beam can
both affect the fidelity of the image and reduce its sensitivity.  To
suppress the sidelobes a Gaussian taper is applied in the Fourier
domain, attenuating the visibilities for the maximum baseline by
20\,\%.  Away from the main lobe the response of the synthesized beam
falls below 0.5\,\% everywhere in the field of view used (radius
93\arcmin) and is typically more than an order of magnitude less (see
Figure 2 in \citealp{taylor2003}).
This tapering results in a final beam of about $58\arcs \times 58\arcs
\csc(\delta)$.  To ensure proper sampling, maps from single syntheses
are on grids with a pixel separation of 21\farcs875 while final
\dhigls\ mosaicked products are on grids with a pixel separation of
18\arcs.

In this paper, velocities are relative to the Local Standard of Rest
($\vlsr$).  The spectrometer has 256-channels and the coverage was
centered on $\vlsr = v_{c}$ with $v_{c}$ varying for each mosaic as in
Table~\ref{beamsize}.
Channels were spaced every 0.824~\kms\ and the velocity resolution was
1.32~\kms.  As noted in Table~\ref{beamsize}, a few syntheses were
observed with a finer channel spacing of 0.41~\kms\, (or a central
channel velocity of 0~\kms) leading to reduced spectral coverage
(toward negative velocities).

\subsection{Noise and Weight Map for a Single Synthesis}
\label{singlenoise} 

Understanding the noise properties of any telescope is necessary to
inform an observing strategy.
Three maps related to the noise in a single synthesis are illustrated
in Figure~\ref{noiseDF20}.  The differences in each of these
representations are discussed below.

Inspection of an emission-free channel map from this single \dhigls\
synthesis indicates that the typical rms noise for 0.824~\kms\
channels is 18~mJy~beam$^{-1}$ (or $\noisesingledeltacorrected \sin
\delta$~K).
In fact, noise estimates for all syntheses compare well to the value
measured
for the original phase of the CGPS carried out in the previous decade
\citep{taylor2003}, $3.2 \sin\delta$~K.

For astrophysical applications of a single synthesis, we need to
correct each channel map by the primary beam of a 9~m antenna (the
effect of its attenuation), which amplifies the noise away from the
center.  We note that the primary beam is not a Gaussian, but is well
described by a $\cos^6$ function of radial offset \citep{taylor2003}
with a FWHM of \fwhmp\arcmin.  The corrected emission-free channel
maps are therefore noticeably noisier around the periphery, as shown
in the left panel of Figure~\ref{noiseDF20}.

The noise can also be quantified by finding the rms in the spectrum
over many emission-free channels pixel by pixel; after correction for
the primary beam this gives us an 
empirical
``noise map" as in the middle panel.
In making a mosaic, the contribution from a single synthesis is
weighted according to the inverse square of the noise.  The 
model for the
weight map is therefore the square of the primary beam, here
normalized to unity at the pointing center (there is no dependence of
the central value on $\delta$ as in the noise map).  This map is shown
in the right panel.

Each of the maps is truncated at a radius of 93\arcmin, where the
noise has risen by a factor 9.5 compared to the value at the pointing
center and the weight has fallen to 0.011
(see Figure~\ref{noiseDF20}).  As adopted for the CGPS,
data beyond this radius are not used in assembling the mosaics
(Section~\ref{mosaicmake}).

\begin{figure}
\centering
\includegraphics[angle=0,width=0.9\linewidth]{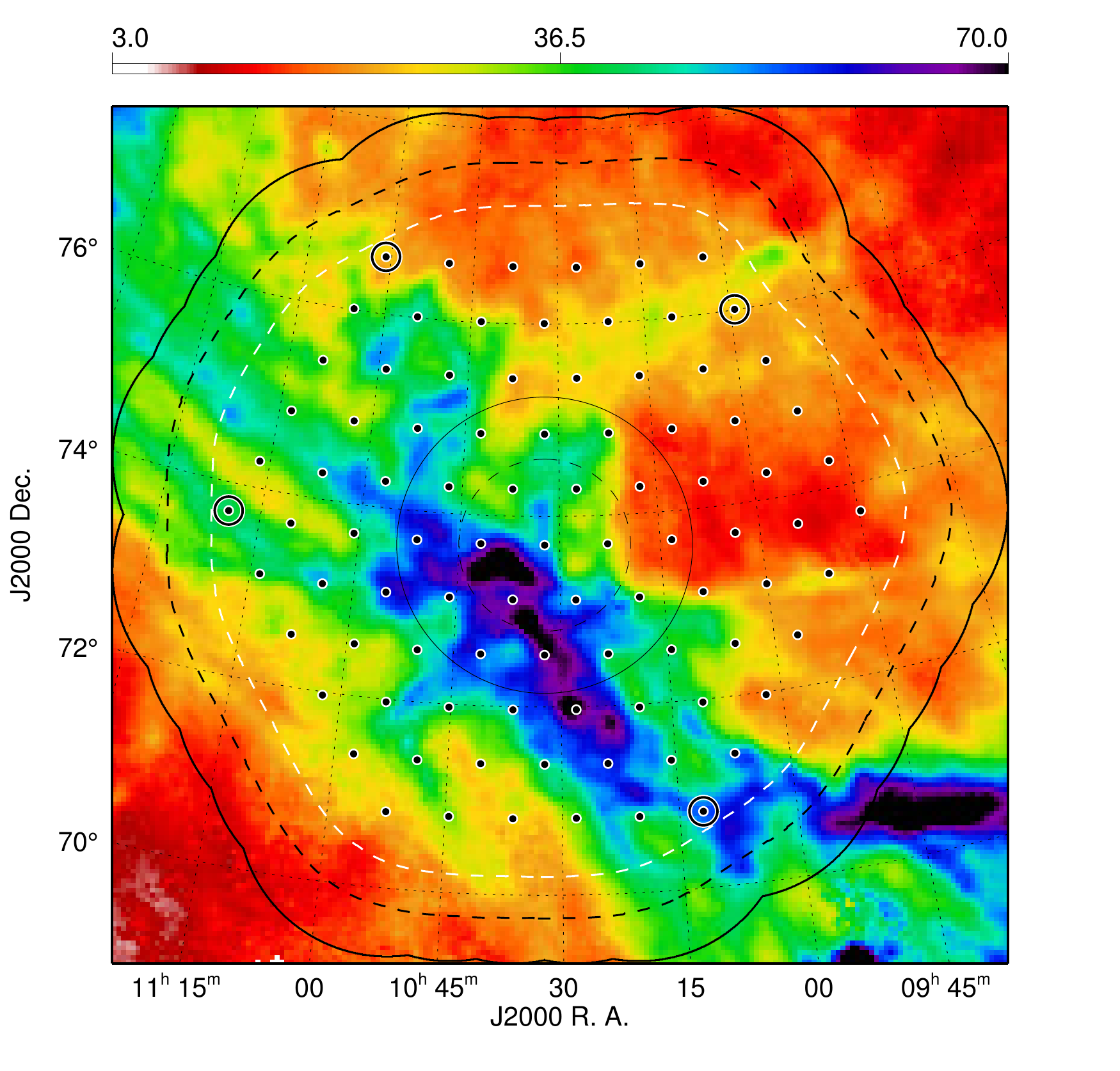}
\caption{
Synthesis pointing centers chosen for the mosaic of the \DF\ region,
each of the 91 marked by a bullet; 4 marked with a circle not included
because of calibration uncertainty.  Background is integrated emission
from \ghigls\ \ncpl\ combined field in units of \colunits.
For the highlighted pointing DF00 ($10^{\rm h}32^{\rm m}17^{\rm s},\,
73^\circ42\arcmin0\arcsec$) primary beam FWHM and mosaic truncation
radius (Section~\ref{mosaicmake}) are shown with dashed- and
solid-line circles, respectively (corresponding circles seen in single
synthesis field DF20 in Figure~\ref{noiseDF20}).  Dashed- and
solid-line contours near edge of mosaic trace constant noise levels
as discussed in Section~\ref{mosaicnoise}.
}
\label{coverageDF}
\end{figure}

\begin{figure}
\centering
\includegraphics[angle=0,width=0.9\linewidth]{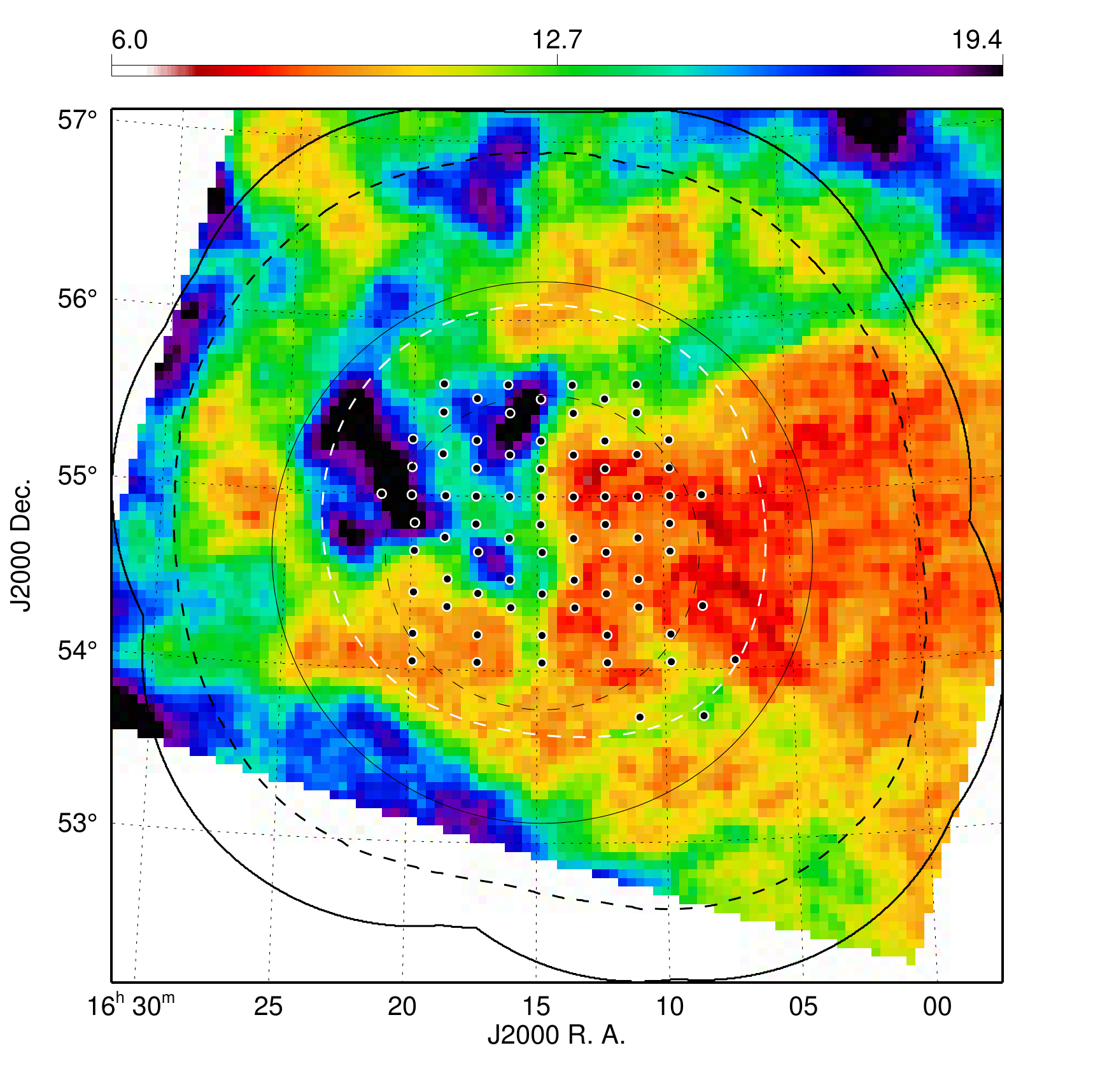}
\caption{
Same as Figure~\ref{coverageDF}, but for the 76 pointings making up
the mosaic of the \EN\ region, with highlighted pointing EN18
$(16^{\rm h}14^{\rm m}48^{\rm s},\, 54^\circ40\arcmin57\arcsec)$.
Note the difference in angular scale and the colorbar between these
two figures.
}
\label{coverageEN}
\end{figure}

\subsection{Planning Mosaics}
\label{mosaicdef} 

In a single synthesis the rms noise level for a single channel varies
across the field (Figure~\ref{noiseDF20}).  The noise level can be
made more uniform and/or greatly reduced by assembling a number of
single syntheses as a mosaic.  For example, the CGPS mosaic pattern
was that of a close-packed hexagonal grid with pointing centres
separated by $112\arcmin$, a bit larger than the FWHM, a compromise
between uniform sensitivity and spatial coverage at 1420~MHz.
With this in mind, for the much fainter \EN\ region we chose a pattern
of single pointings to emphasize sensitivity over spatial coverage
with spacings of about $12\arcmin$, while for the brighter \DF\ we
opted for more extended coverage with spacings of about $40\arcmin$.
Figures~\ref{coverageDF} and \ref{coverageEN} show the pointing
centers for \DF\ and \EN, 
numbering 91 and 76 respectively (see Table~\ref{beamsize}).  
These figures are in ICRS coordinates ($\alpha, \delta$) and the NCP
projection \citep{cala02} native to the \draost\ data.  The central
position of each synthesis field is marked by a bullet.  For scale, we
show the FWHM of the primary as a dashed circle of diameter
$108\arcmin$.  Also shown as the solid circle is the $93\arcmin$
truncation radius ($186\arcmin$ diameter) used in the production of
the final mosaic.

The pointing centers for the other mosaics of \dhigls\ regions are
given in Figure~\ref{field_locations} and Figure~\ref{coverageUM}.

\begin{figure}
\centering
\includegraphics[angle=90, width=1.0\linewidth]{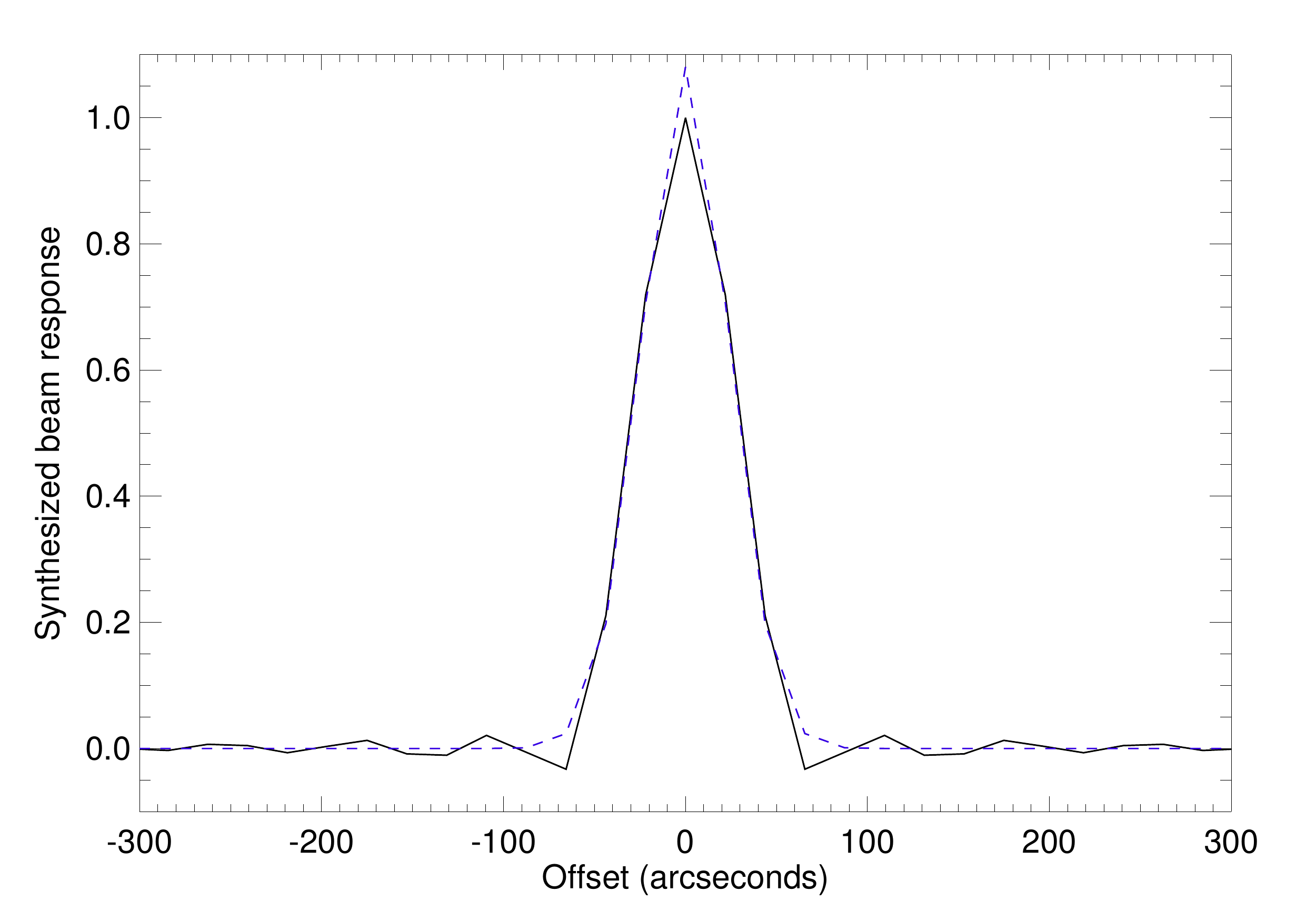}
\caption{
Cross section in Declination of the synthesized beam.  Overlaid red
line indicates that the Gaussian model overpredicts the peak
brightness of a point source.
}
\label{beamfit1D}
\end{figure}

\subsection{Characterizing the Beam}
\label{beamdef} 

A synthesized beam for each of the fields was produced corresponding
to the visibility coverage in the Fourier domain.  This beam was
normalized to unity at its center.  Because much of the data
processing involves approximating the beam with a Gaussian, the
synthesized beam was fitted in the image domain with a two-dimensional
(2D) Gaussian to determine the effective beam size. The effective beam
size was consistently smaller than the half-power beam-width (HPBW) of
$58\arcs$ and consistently peaked at 1.08 with a relative beam
integral of 1.08 (see Figure~\ref{beamfit1D}).  The mean Gaussian FWHM
of the synthesized beam and its dispersion for the mosaic is given in
Table~\ref{beamsize} for each region.  

\subsection{Continuum Removal}
\label{contremoval}

Two continuum maps are produced from an average of the channels at
each end of the spectrum that are free of \hi\ emission.  
The channel selection is performed independently for each synthesis
with the median number of channels used for a continuum map being 30,
separately for each end.
The continuum contribution at each channel in the spectral cube is
then determined by linear interpolation between these two continuum
maps.  Empirically there is only a minimal frequency dependence of the
continuum over the frequency range of the data cube and so in the few
exceptional cases where there are emission-free channels at only one
end of the spectrum the resulting single continuum map is subtracted
directly from the spectral cube.

In a number of fields there are very bright point sources that produce
correspondingly bright artifacts in the form of grating rings.
Subtraction of the continuum will in principle remove the continuum
source and any artifacts around it.
However, within the spectral cube the continuum emission could be
absorbed by diffuse \hi\ in some channels, resulting in a weaker point
source and a concomitantly weaker set of artifacts in those channels.
Following subtraction of the unattenuated continuum, a scaled negative
imprint of the source and artifacts would remain in those absorbing
channels.  For weaker point sources, the artifacts are hidden in the
noise but to mitigate such potential residual contamination for bright
sources it is useful to create a clean cube prior to producing the
continuum maps, by deconvolving those bright point sources with a
model beam.
This cleaning process follows routines developed for the \draost\ by
\citet{willis99} and is performed on each channel in a given cube.  It
is performed only on those cubes having continuum point sources with
flux density $F > 1$~Jy\,\perbeam\ and this cleaning proceeds down to
the level of 300~mJy\,\perbeam\ as was done in the CGPS
\citep{taylor2003}.

\subsection{Position Registration and Flux Density Scale}
\label{register}

An averaged continuum map was produced for each synthesis from the two
continuum maps.  Each point source in this map was fitted with a 2D
Gaussian \citep[with the {\tt fluxfit} routine included in the DRAO
Export Software Package {\tt madr},][]{higgs1997} to determine its sky
coordinates ($\alpha, \delta$) and flux density $F_{\rm DRAO}$,
limiting the width of each fit to be no less than 95\% of the
effective beam parameters determined in Section~\ref{beamdef}.

The fitted values were compared with those in the NVSS\footnote{
\url{www.cv.nrao.edu/nvss/}.}
database \citep{condon1998} to determine for each field a weighted
average offset in sky coordinates ($\Delta \alpha$, $\Delta \delta$)
and a weighted average flux density ratio, $F_{\rm DRAO}/F_{\rm
NVSS}$.  \citet{taylor2003} discuss this in detail.  We find values of
the flux density ratio $F_{\rm DRAO}/F_{\rm NVSS}$ in the range 1.2 to
1.5; the typical uncertainty for each determination is 0.03.

Because the synthesized beam is not quite Gaussian, $F_{\rm DRAO}$
measured by Gaussian fitting will be overestimated by a factor of
about 1.08 (Section~\ref{beamdef}).  We note that in the conversion
from surface brightness to brightness temperature (see
Section~\ref{makingmosaics}) this factor of 1.08 returns in the
amplitude of the Gaussian-fitted beam and so it cancels out.

\begin{figure}
\centering
\includegraphics[angle=90,width=1.06\linewidth]{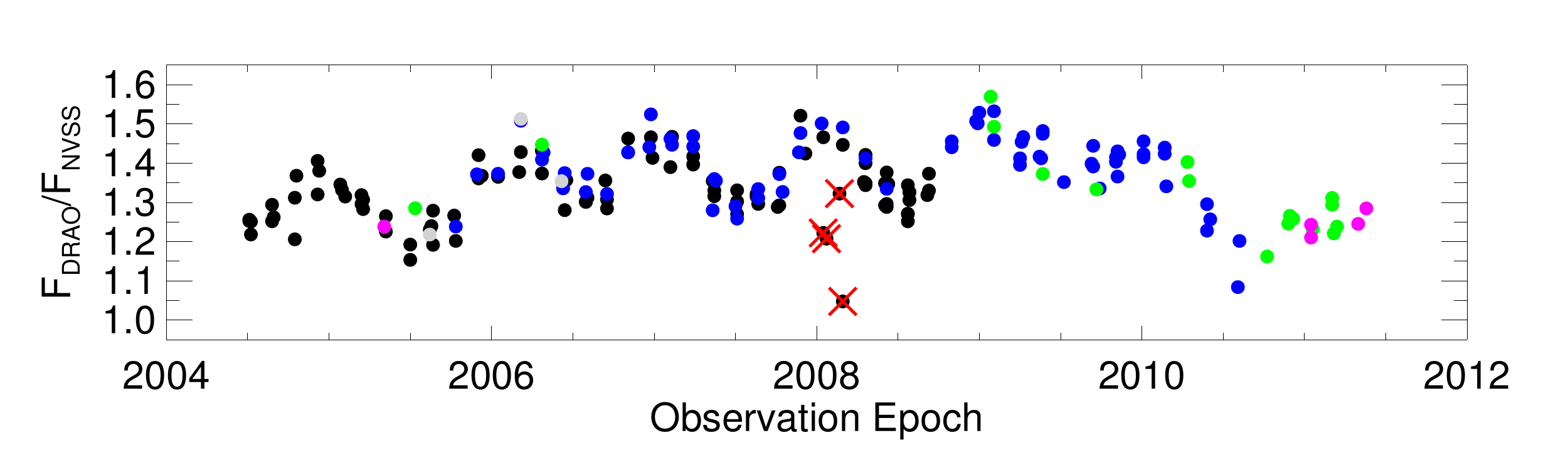}
\caption{
Temporal variability of the $F_{\rm DRAO}/F_{\rm NVSS}$ scale factor
(typical uncertainty 0.03).
Syntheses for regions \DF\ (black), \EN\ (blue), \UM\ (green), \DR\
(magenta), and other pointings (grey) show a consistent trend.  Four
syntheses in \DF\ that have strong asymmetric ring artifacts in the
point-source continuum maps also show the strongest deviations from
the trend (marked with a red cross), supporting the view that the
scale factors for these are unreliable.  Off the plot is a lone field
FP2 at epoch 2012.8 with scale factor 1.53.
}
\label{corrdate}
\end{figure}

A plot of $F_{\rm DRAO}/F_{\rm NVSS}$ for all \draost\ syntheses as a
function of observation date reveals intriguing systematic variations
(Figure~\ref{corrdate}).  This is present in the data used in a single
mosaic (e.g., regions \DF\ and \EN\ colored separately) and for the
interleaved observations, ruling out a dependence on Declination.
There might be a seasonal trend, though it is not exactly the same
year to year.  
We note that changes in 
the equivalent VLA to NVSS ratio cited as part of the VGPS
\citep{stil2006},
though also not fully understood, 
were shown to be related to the band-averaged scalar amplitude of the
continuum-subtracted visibilities, which is in turn connected to the
noise.

The empirical pattern of the temporal variation is useful for
confirming suspected anomalous $F_{\rm DRAO}/F_{\rm NVSS}$ scale
factors.  A few fields have strongly asymmetric ring artifacts in the
point-source continuum images and were suspected of having inaccurate
ratios, and we found that these fields (DF66, DF76, DF81, and DF87),
observed in late 2007/early 2008, have the largest deviations from the
trend seen in Figure~\ref{corrdate} (each marked with a red cross).
Therefore, these fields at the edge of the \DF\ region were excluded.

\section{Mosaics Made Using Interferometric Data from the \draost}
\label{makingmosaics}

In this section we focus on mosaics assembled from the \draost\ data
at many adjacent pointing centers.  To distinguish this intermediate
product that uses only data from the interferometer from the final
\dhigls\ product that also incorporates short-spacing information from
the GBT data (Section~\ref{gbt_observations}) we append ``-i'' to the
name of the region mosaicked, for example \DFi.

\subsection{Assembling the Mosaic}
\label{mosaicmake}

As described in Section~\ref{mosaicdef}, our \draost\ observations
were designed with the intent of combining several syntheses with
adjacent pointing centers into a mosaic.
For each synthesis the \draost\ data cube (in Jy~beam$^{-1}$) was
converted to brightness temperature, $\Trb$, using the beam parameters
determined in Section~\ref{beamdef} and the scale factor determined
from the comparison with NVSS (Section~\ref{register}).
The channels of each synthesized cube were then interpolated to a
common velocity grid using a spline interpolation.  This interpolation
has very little effect on the spectra, but is required to ensure a
consistent velocity grid among all fields.

The prepared data from the individual syntheses were then assembled
into mosaics using the routine {\tt SUPERTILE} in {\tt madr}.  As
described by \citet{dewdney2002} and \citet{taylor2003}, {\tt
  SUPERTILE} reads in the position registration results for each of
the fields (Section~\ref{register}), adjusts the position of each
field accordingly, divides by the \draost\ primary beam (which
amplifies the noise away from the beam center), truncates each data
set at a radius of $93\arcmin$ relative to its pointing center, and
combines the data, keeping track of the beam shape and weights
throughout the process.

\begin{figure}
\centering
\includegraphics[clip=true,trim=10 20 10 10 ,angle=0,width=0.8\linewidth]{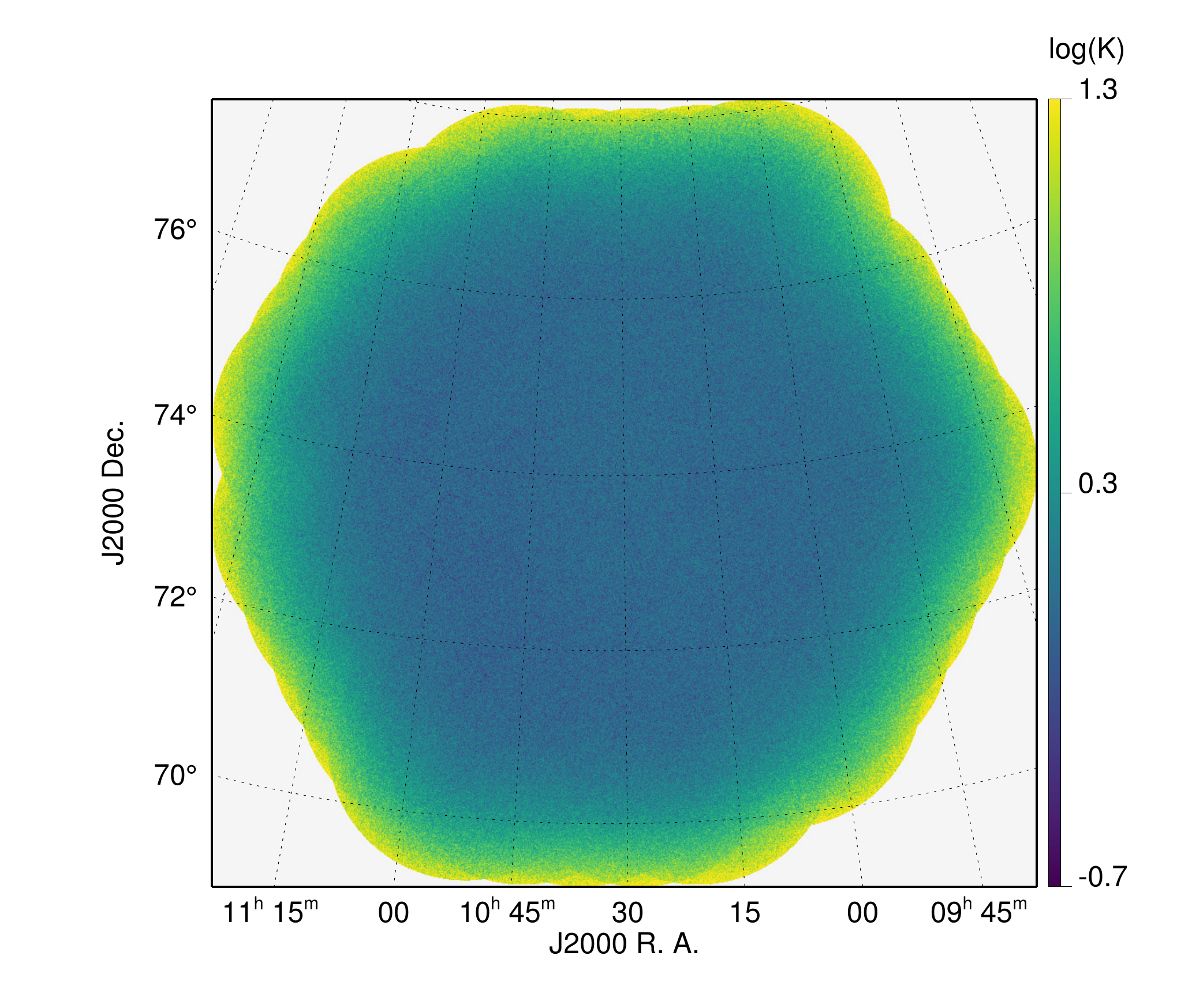}
\includegraphics[clip=true,trim=10 20 10 10 ,angle=0,width=0.8\linewidth]{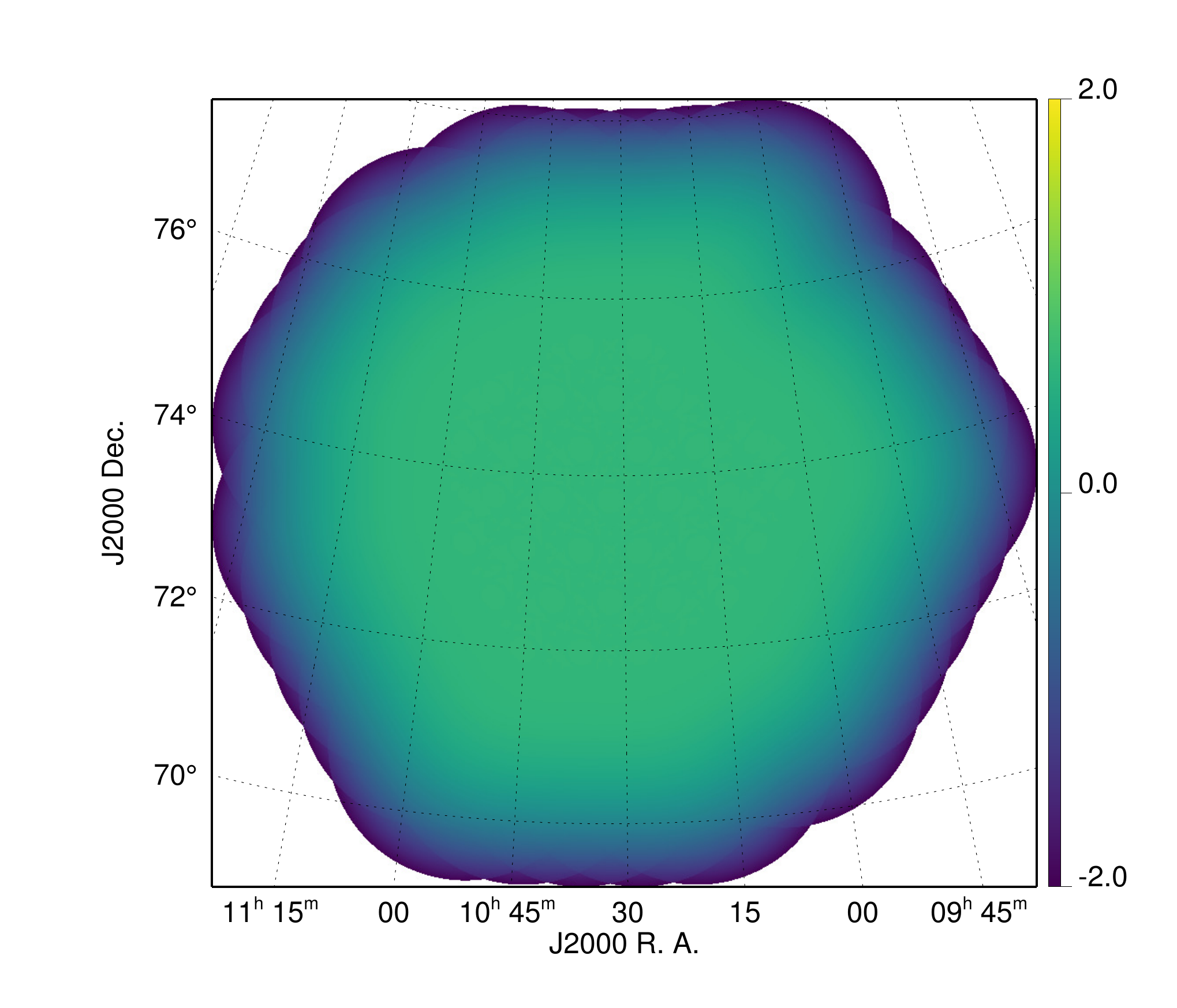}
\caption{
Maps displaying the noise (upper panel) and the relative weight (lower
panel) across the \DFi\ mosaic.  
Adopting the same logarithmic colorbar scales as in
Figure~\ref{noiseDF20} emphasizes the improvement compared to the
corresponding values for the single synthesis DF20.
}
\label{noiseDFdrao} 
\end{figure}

\subsection{Noise and Weight Maps for Mosaics}
\label{mosaicnoise}

The channel maps of the mosaicked \hi\ spectral line data are
contained in a data cube.  We can produce a noise map from
emission-free channels as we did for a single synthesis in
Section~\ref{singlenoise}.
An example is shown in the upper panel of Figure~\ref{noiseDFdrao} for
the \draost\ \DFi\ mosaic.  The corresponding weight map produced in
assembling the mosaic is shown in the lower panel. 

Figure~\ref{noiseDFdrao} shows the distinctive pattern of increased
noise around the perimeter as a result of the lower coverage and
effect of the primary beam there.  Across most of the field the noise
is 1.0~K, clearly reduced compared to \noisecenter~K for a single
synthesis because of the chosen dense packing of pointing centers (and
the typical non-unity scale factor $F_{\rm DRAO}/F_{\rm NVSS}$ 
applied to the single syntheses prior to mosaicking).

We note that because of the weighting scheme used in {\tt SUPERTILE},
the weight map can be used to create a theoretical noise map by
multiplying its inverse square root by the typical noise for a single
synthesis, $\noisesingledeltacorrected \sin \delta$~K 
(modified by the typical $F_{\rm DRAO}/F_{\rm NVSS}$ scale factor for
the syntheses in the mosaic).  Therefore, in figures for other regions
below we display only the noise map.
 
The noise map is used to create three contours.  Two contours, e.g.,
black outer dashed and solid contours in Figure~\ref{coverageDF},
indicate where the noise of the mosaic is identical to the noise in a
single synthesis at the FWHM of the primary beam and at the truncation
radius, respectively.  A third contour, e.g., white dashed in
Figure~\ref{coverageDF}, indicates where the noise is only a factor of
two larger than in the central, lowest-noise region, thus selecting a
low-noise portion of the larger mosaic
(inside the outer black dashed contour because the central noise in
the mosaic is lower than for a single synthesis).  The white dashed
contour is of particular significance in selecting data for analyses
discussed below.

\begin{figure}
\centering
\includegraphics[clip=true,trim=10 20 10 10 ,angle=0,width=0.8\linewidth]{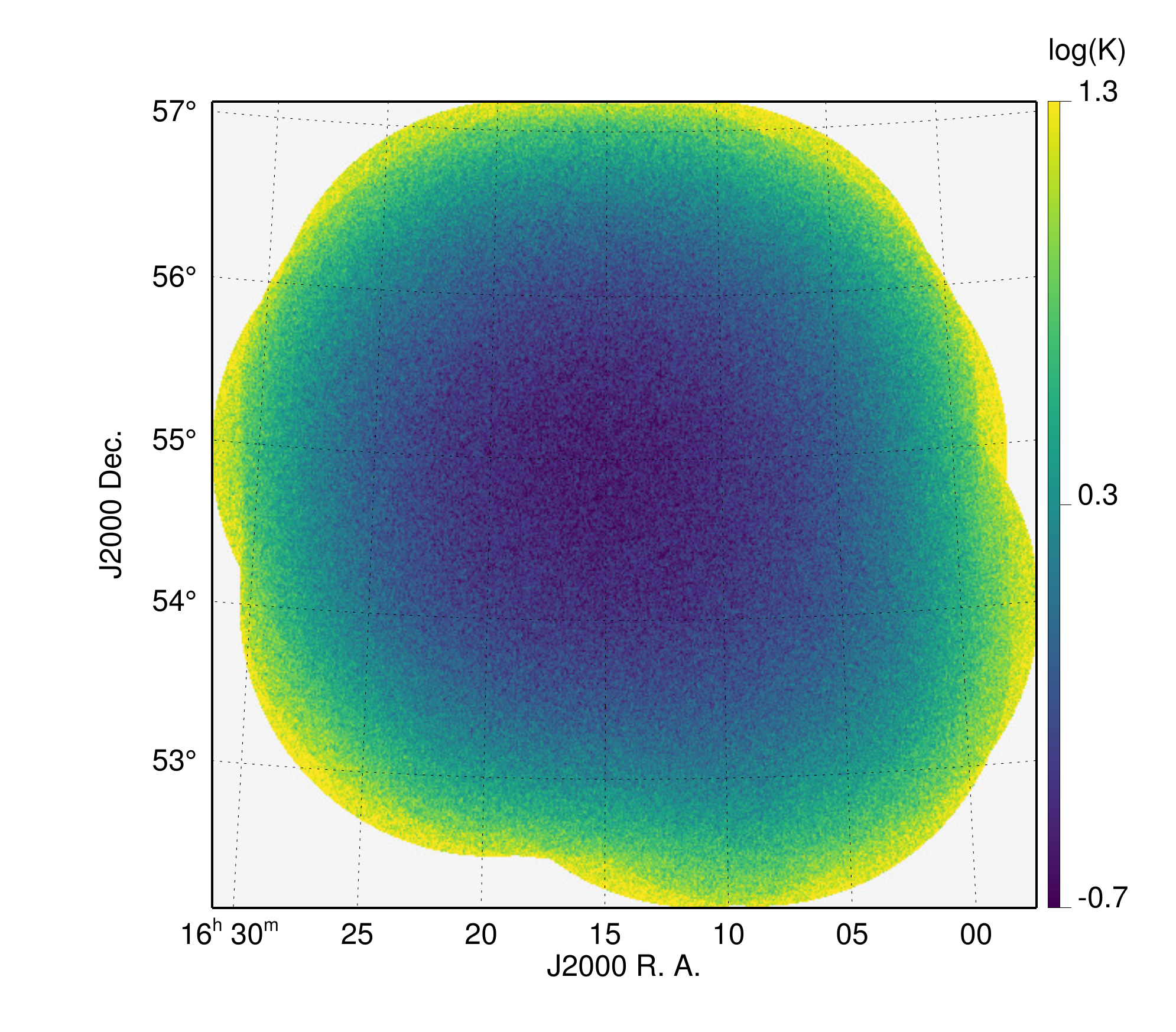}
\caption{
Noise map as in upper Figure~\ref{noiseDFdrao}, for \ENi\ mosaic.  
}
\label{noiseENdrao} 
\end{figure}

We have carried out a similar analysis for the \ENi\ mosaic.  The
noise map is shown in Figure~\ref{noiseENdrao}.  This clearly shows
that the denser packing of pointing centers of the syntheses making up
the mosaic (Figure~\ref{coverageEN}) has significantly reduced the
noise in the central region to 0.3~K, about a factor of 3 lower than
in \DFi.  The denser packing also reduces the relative extent of the
low-noise region marked by the dashed white contour in
Figure~\ref{coverageEN}.

Likewise, noise maps for the \UMi\ mosaic and the smaller \DRi\ and
\POi\ mosaics are presented in Figures~\ref{noiseUMdrao},
\ref{noiseDRdrao}, and \ref{noisePOdrao} in
Appendix~\ref{additionalfields}.

\subsection{\HI\ Signal Maps}
\label{mosaicps}

\begin{figure}
\centering
\includegraphics[clip=true,trim=10 20 10 10 ,angle=0,width=0.8\linewidth]{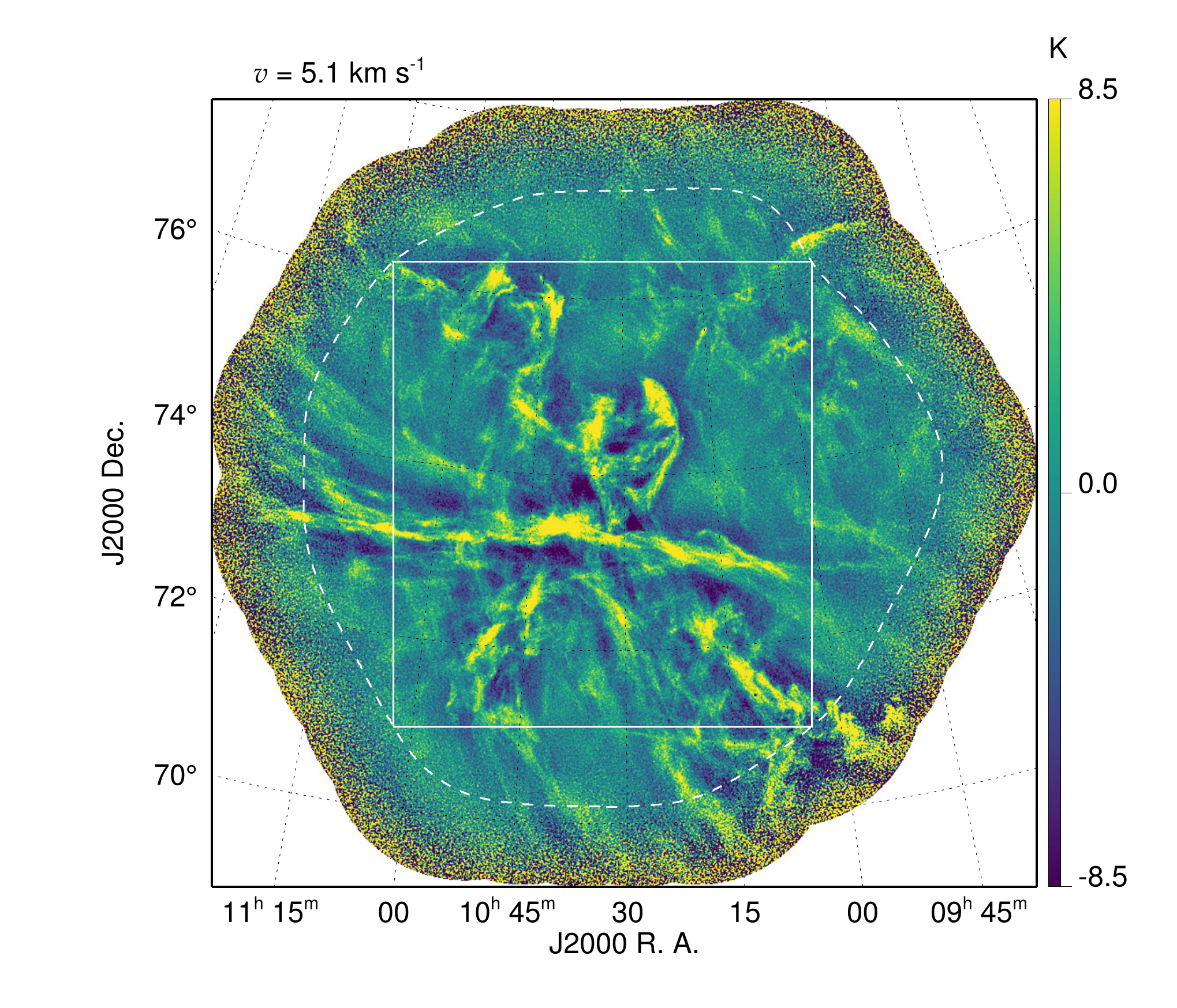}
\caption{
Channel map from \DFi\ mosaic ($v = 5.1$~\kms, in the LVC range).
Region used in calculation of power spectrum in
Figure~\ref{noiseDFdrao_ps} is indicated by a white rectangle (see
Section~\ref{mosaicps}).
}
\label{noiseDFdraohi} 
\end{figure}

\begin{figure}
\centering
\includegraphics[clip=true,trim=10 20 10 10 ,angle=0,width=0.8\linewidth]{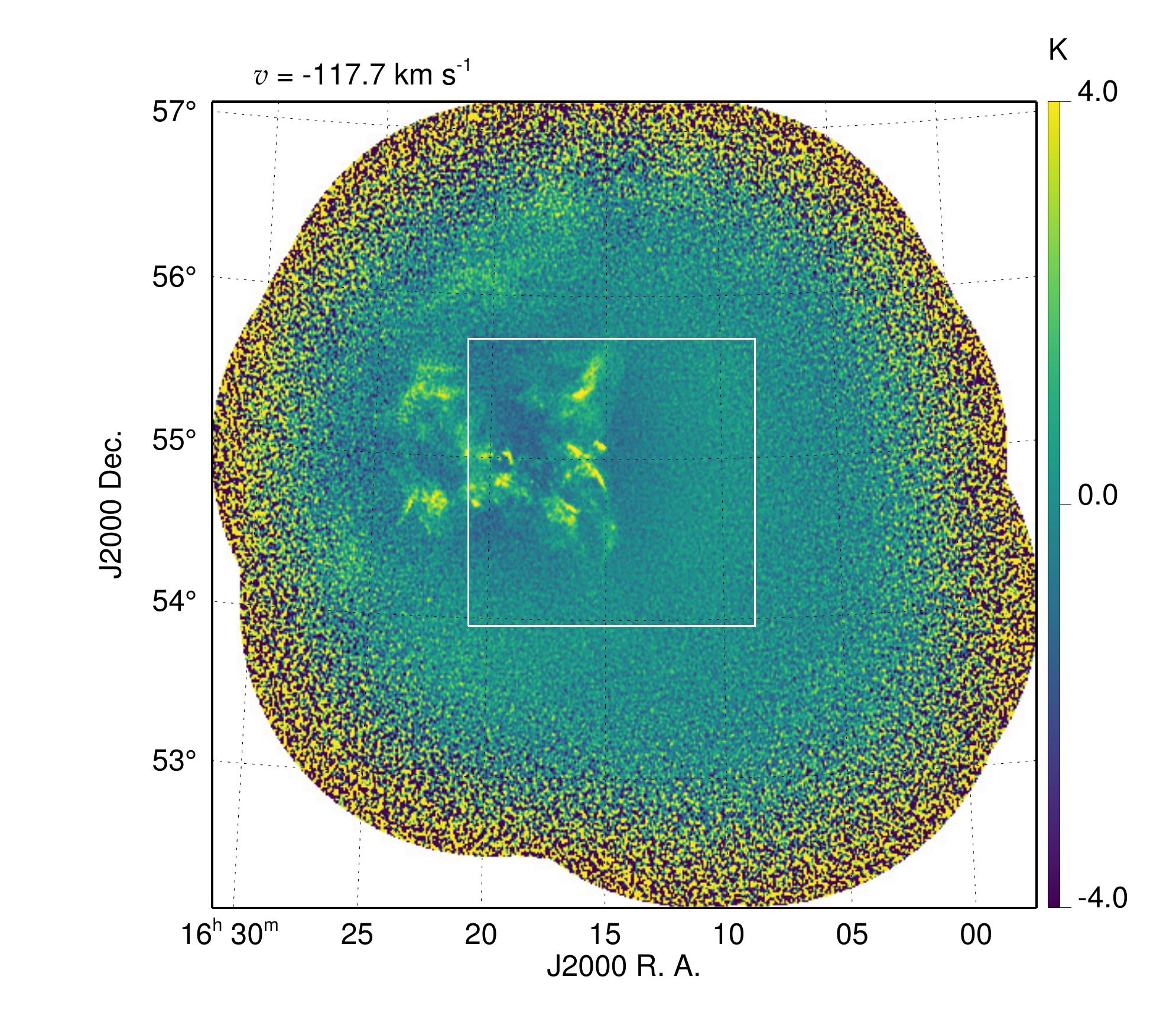}
\caption{
As in Figure~\ref{noiseDFdraohi}, for \ENi\ mosaic and $v =
-117.7$~\kms\ in the HVC range.  White rectangle indicates region used
for power spectrum analysis in Figure~\ref{noiseOTdrao_ps}.
}
\label{noiseENdraohi} 
\end{figure}

Figure~\ref{noiseDFdraohi} shows a representative channel map of the
\hi\ signal in the \draost\ \DFi\ mosaic near the peak of the average
spectrum.
Below we analyze the power spectrum in a low noise (and high signal to
noise) region shown in Figure~\ref{noiseDFdraohi} as the ``white
rectangle."  This is the rectangle of largest area that is aligned
with the pixel grid and fits inside the ``white dashed contour"
defined in Figure~\ref{coverageDF}, which is the region of least
noise.

Similar data are shown in Figure~\ref{noiseENdraohi} for the \ENi\
mosaic and in Figures~\ref{noiseUMdrao}, \ref{noiseDRdrao}, and
\ref{noisePOdrao} for the \UMi, \DRi, and \POi\ mosaics in
Appendix~\ref{additionalfields}.

\subsection{Insight from the Angular Power Spectrum}
\label{enmnoise} 

It is of interest to relate the signal and noise in \hi\ maps to their
manifestations in the Fourier domain (harmonic or \uv\ domain) and the
angular power spectrum.  In addition to enabling the science
application discussed in Section~\ref{interpretation}, this provides a
means of exploring the sensitivity of synthesis telescope observations
to various spatial scales and more immediately allows for direct
comparisons with (and integration of) observations from single-dish
telescopes, as will be described in Section~\ref{gbt_observations}.
Facilitating such analyses is the power spectrum model development
including the ``noise template" and fitting presented in
Appendix~\ref{appdfmps} building on earlier work described in
\citet{mamd2007} and \citet{mart15}.

The 2D angular power spectrum of a map $f(x,y)$ (image of the sky) is
the square of the modulus of the Fourier transform
$\tilde{f}(k_x,k_y)$ (the visibility) where $k$ is the spatial
frequency (wavenumber) in the \uv\ domain:
\begin{equation}
P(k_x,k_y)  = \vert \tilde{f}(k_x,k_y) \vert^2 \,. 
\label{amplitude}
\end{equation}

The collapsed one-dimensional (1D) power spectrum $P(k)$ is formed by
azimuthal averaging of $P(k_x,k_y)$ in annuli of constant
$k=(k_x^2+k_y^2)^{1/2}$.  We use this for both analysis and display
purposes (see Appendix~\ref{ps2D} for exceptional images with
directional structure).

The 2D Fourier transform reflects the sampling in the \uv\ domain,
which is quite uniform for the \draost\ syntheses, with noise
increasing at angular frequencies with little or no coverage.
Depending on the symmetry of the Fourier transform, differing amounts
of this structure will remain in $P(k)$.  In the limit where the
synthesized beam is circularly symmetric, there is a direct mapping of
the features in the 2D power spectrum to the 1D $P(k)$.  Of the two
deepest mosaics, \DFi\ is much closer to this limit than is \ENi.

\begin{figure}
\centering
\includegraphics[angle=90,width=1.0\linewidth]{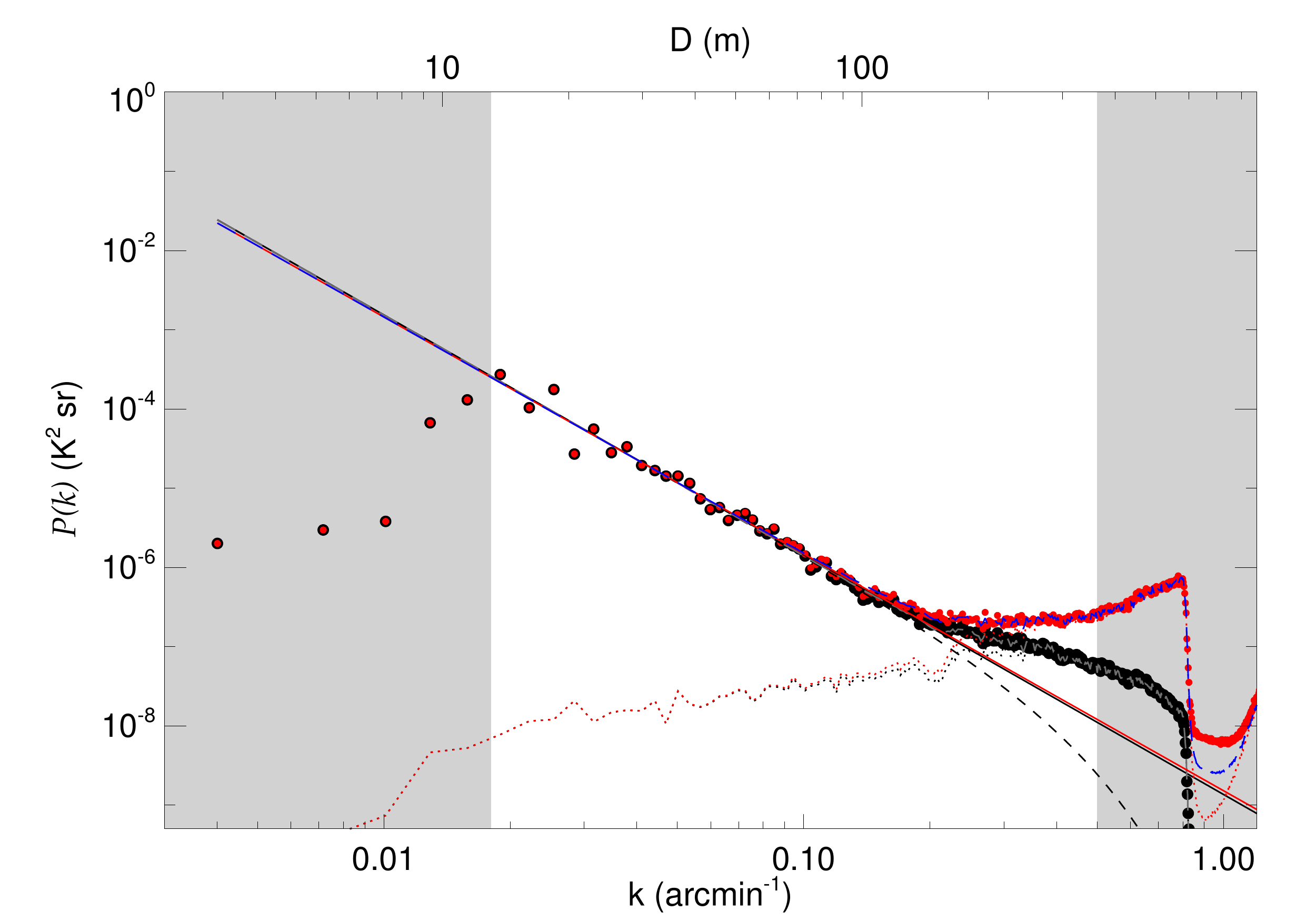}
\caption{
Power spectra of the \DFi\ mosaic for signal channel in the LVC range
and white rectangle region shown in Figure~\ref{noiseDFdraohi}.  Data
in red result from deconvolution by the 2D synthesized beam.  Data in
black are without deconvolution.  Dotted curves are the scaled noise
component, solid lines are the power-law component and long dashed
curves are the total power spectrum model (Equations~(\ref{plawd}) and
(\ref{plaw})).  Black dashed curve shows the modification of the
power-law by the synthesized beam.  Data in shaded ranges of $k$ were
excluded from the model fit (see Appendix~\ref{psfitting}).
}
\label{noiseDFdrao_ps}
\end{figure}

Figure~\ref{noiseDFdrao_ps} shows $P(k)$ for the LVC \hi\ channel map
in Figure~\ref{noiseDFdraohi} for \DFi.  There the data have been fit
to the 1D power spectrum formed after deconvolution in 2D using the
asymmetrical synthesized beam.  The fit to the power law model of
Equation~(\ref{plawd}) has power law exponent $\expon = -2.99 \pm
0.04$ and scales the noise template by $\eta = 1.59 \pm 0.03$.
We also show an alternative representation in which the modelling is
carried out without deconvolution, requiring accounting for the
lowering of power by the effective beam at high spatial frequencies
(see Equation~(\ref{plaw})).  These parameters are consistent with
those found above: $\expon = -3.03 \pm 0.03$ and $\eta = 1.60 \pm
0.03$.

\begin{figure}
\centering
\includegraphics[angle=90,width=1.0\linewidth]{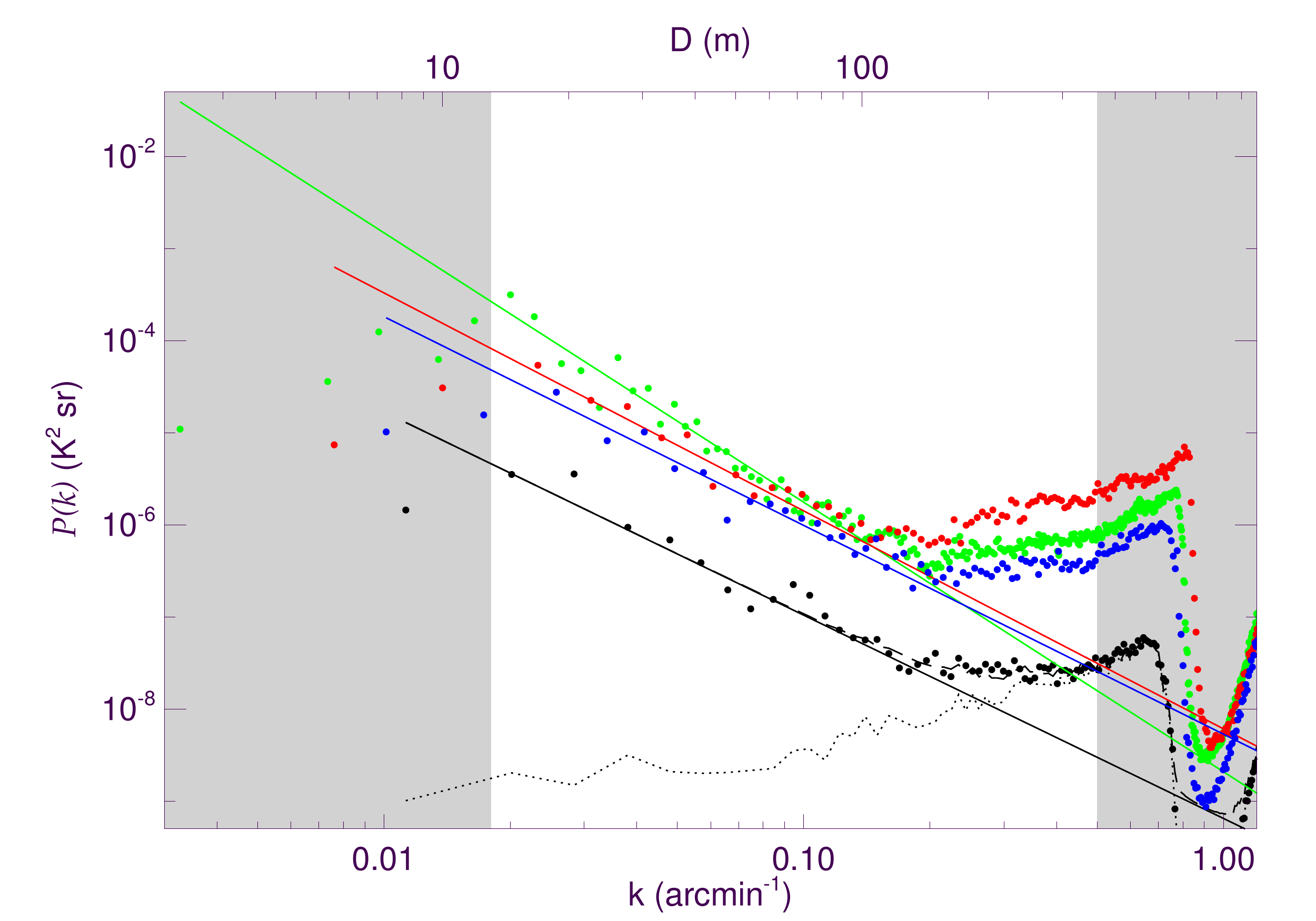}
\caption{
Power spectra from the \ENi\ mosaic (black, HVC), \UMi\ (green, LVC),
\DRi\ (blue, IVC), and \POi\ (red, IVC) for channel maps and white
rectangle regions shown in Figures~\ref{noiseENdraohi},
\ref{noiseUMdrao}, \ref{noiseDRdrao}, and \ref{noisePOdrao},
respectively, at same scale as for \DFi\ in
Figure~\ref{noiseDFdrao_ps}.  For \ENi\ the noise (dotted line) is
lower than in \DFi\ and so despite the signal being even lower the
power spectrum of the signal can still be analysed over a significant
range of $k$.  For the others, both signal and noise are higher than
in \DFi.
}
\label{noiseOTdrao_ps}
\end{figure}

To assess the level of noise achieved and the ability 
to investigate the variety of spatial structure using the \draost\
\hi\ data, we have modelled the power spectra in the other regions
mosaicked, in Figure~\ref{noiseOTdrao_ps} for the single channel maps
shown in Figures~\ref{noiseENdraohi}, \ref{noiseUMdrao},
\ref{noiseDRdrao}, and \ref{noisePOdrao}.

As in \DFi, the denser coverage in \ENi\ also lowers the power
spectrum of the noise, so that as seen in Figure~\ref{noiseOTdrao_ps}
the power spectrum of the signal can still be analyzed over a
considerable range in $k$, even though the \hi\ signal in this channel
is much weaker than in the channel selected for \DFi.  Here the noise
template is scaled by $\eta = 1.22 \pm 0.09$.  Generally $\eta$
decreases toward 1.0 as the \hi\ signal diminishes, and vice versa.
The power law for this HVC channel is significantly shallower ($\expon
= -2.21 \pm 0.12$) than that for the LVC channel in \DFi\ ($\expon =
-2.99 \pm 0.04$).

For the remaining \UMi, \DRi, and \POi\ power spectra in
Figure~\ref{noiseOTdrao_ps}, at high $k$ where the noise dominates,
the different levels are in accord with expectation based on the noise
maps presented in Section~\ref{mosaicnoise} and
Appendix~\ref{additionalfields}.  Similarly, the values of $\eta$ show
the expected differences with \hi\ signal strength, being $1.40 \pm
0.03$, $1.04 \pm 0.05$, and $1.27 \pm 0.05$ for the selected channels
in \UMi, \DRi, and \POi, respectively.  Despite the different noise
levels and signal strengths, the power spectrum of the signal can
still be analysed over a significant range of $k$.

Extending the comparison made between \DFi\ and \ENi, the power
spectrum falloff for the \UMi\ LVC channel is similar to that for
\DFi, but visibly steeper than for the HVC channel in \ENi; this is
quantified in the power-law exponent, $-2.93 \pm 0.08$.  On the other
hand the data for the IVC channels in \DRi\ and \POi\ appear more
compatible with the behaviour in \ENi.  Their exponents are $-2.27 \pm
0.14$ and $-2.4 \pm 0.2$, respectively.  

Thus the data are of sufficiently high quality 
that some differences can be discerned.  However, because these power
spectra are for individual channels and are thus responding to a
mixture of velocity and density fluctuations \citep{lazarian2000} the
exponents cannot be linked simply to the structure of the atomic gas.
This will be followed up in Section~\ref{interpretation}.

\section{Incorporation of Data at Low Spatial Frequency}
\label{gbt_observations}

A synthesis telescope with moveable antennas provides ample coverage
at discrete frequencies in the angular frequency domain, but is not
sensitive to angular frequencies below that corresponding to the
closest antenna spacing (some foreshortening of the baselines during a
synthesis observation provides some sampling of lower spatial
frequencies, but not much sensitivity).  In the case of a \draost\
synthesis, this is \bminst~m (Section~\ref{drao_observations}) or
\kminvalst\ (Figure~\ref{noiseDFdrao_ps} and
Appendix~\ref{psfitting}).

To fill in this short-spacing information, observations using a filled
aperture are required.  Many of the issues related to how this is
accomplished are discussed in \citet{stan2002}.

Historically, for the \draost, data from the DRAO 26~m telescope have
been used most often \citep[e.g.,][]{higgs2000,taylor2003,higgs2005}.
However, as discussed in Appendix~\ref{appen:short}, that telescope is
not ideally sized to maintain sensitivity through the transition
region in angular frequency where the coverage of the single filled
aperture and the interferometer overlap.

By contrast, in \dhigls\ we combined the \draost\ data with
complementary data from the GBT, whose 100~m diameter is 7.8 times
larger than the shortest baseline in the \draost.
As illustrated in Figure~\ref{overlapfig} in
Appendix~\ref{appen:short} and discussed below, using GBT data greatly
improves both the width of and the sensitivity in the transition
region and allows more choice in optimizing the addition of the
short-spacing data.

\subsection{\ghigls\ Observations and Cross Calibration}
\label{gbtobs}

We adopted observations made with the GBT as part of \ghigls\
\citep{mart15}.
On-the-fly observations of the fields produced Nyquist-sampled maps
with an effective \ghigls\ beam slightly elongated in the scanning
direction. The \ghigls\ beam can be modelled accurately in 2D and when
approximated by a 2D Gaussian is $9\parcm55 \times 9\parcm24$ FWHM
\citep{mart15}.
The data reduction pipeline for calibration of these \hi\
measurements, the correction for stray radiation, and baseline removal
is described in \citet{boothroyd2011} and \citet{mart15}.
The spectrometer channel spacing in the final \ghigls\ data cubes is
0.80~\kms\ with velocity coverage well over $\pm200$~\kms, i.e.,
beyond that covered by the \draost.  For use with the higher velocity
resolution data in the \POi\ mosaic and the MC and MG single syntheses
from the \draost\ (Table~\ref{beamsize}), we reprocessed the archived
GBT raw data to create POLFINE and G86FINE data cubes with channel
spacing 0.32~\kms.

The large SPIDER field was observed in 2\degree $\, \times$ 2\degree\
segments and then assembled into a final 10\degree $\, \times$
10\degree\ cube.  The nine central sections of SPIDER were measured
twice.  Other relevant \ghigls\ fields (Table~\ref{beamsize}) are N1
(measured 2 times), G86 (3), DRACO (3), POL(1), and UMA (1).  The
typical end-channel (emission-free) rms noise is 105~mK per
observation, reducing to 75~mK for two observations (appropriate to
the central region of SPIDER).  For the reprocessed G86FINE and
POLFINE fields, the noise is typically 96~mK and 170~mK, respectively.
Note that for the \draost\ \UMi\ mosaic we actually used the \ghigls\
\ncpl\ combined field to achieve slightly greater spatial coverage.

Each synthesis pointing used in the assembly of a given \draost\
mosaic has been registered and scaled to the NVSS flux scale as
described in Section~\ref{register}.  However, here we are interested
in having a consistent calibration between the \hi\ data from the
\draost\ mosaic and \ghigls.  To this end, as described in
Appendix~\ref{crosscalib}, we determined a cross calibration scale
factor for the brightnesses, $f_{\rm cc} \equiv I_{\rm GBT}/I_{\rm
DRAO} = \ccal \pm \estccal \pm \esyccal$, that is to be applied to the
\draost\ mosaicked \hi\ data to bring them to the \ghigls\ scale.

\subsection{Combining Data from \ghigls\ Fields and \draost\ Mosaics}
\label{combinegbtdrao}

As described below, we combined the two data sets by weighting their
Fourier transforms.  We refer to the \dgm\ product as the ``combined"
(or ``merged") image.  This combined product is the one contained in
the \dhigls\ data cube and the spectra therein are on the \ghigls\
scale.

Our approach to weighting is close to that of the {\tt AIPS} task {\tt
IMERG}.\footnote{
\url{http://www.aips.nrao.edu/cgi-bin/ZXHLP2.PL?IMERG}} 
Using the {\tt madr} software, the transformed data below some \kmin\
are taken from the \ghigls\ data and above some \kmax\ from the
\draost\ data, and within this ``image combination overlap range" the
data are weighted by complementary functions summing to unity.  We
used a cubic polynomial in $k$ with zero slope at the inner and outer
edges of the range.
Note that the overlap ranges for the cross calibration and the image
combination need not be the same.
However, because many of the same considerations apply (see
Appendix~\ref{overrange})
we did adopt the same range.  We note again that this \kmax\ is well
below the range affected by the \draost\ synthesized beam.  Refer to
Figure~\ref{overlapfig} in Appendix~\ref{appen:short}.

This approach is different than the {\tt CASA} task {\tt
feather}\footnote{
\url{http://casa.nrao.edu/docs/TaskRef/feather-task.html}}
and {\tt IMMERGE} \citep{stan1999,sault2011} in two ways.  First,
these use \kmin$\, = 0$~arcmin$^{-1}$ and so despite the
weighting function applied, the interferometric data are used below the
range actually sampled in the \uv\ domain.  Second, the weighting
function applied to the deconvolved single-dish data is simply the
Fourier transform of the single-dish beam and so with no restriction
on \kmax\ other than implied by the decreasing weight, data strongly
affected by the noise are incorporated.  Our adopted limits to the
range address both of these issues.

\begin{figure}
\centering
\includegraphics[clip=true,trim=20 20 10 10,angle=0,width=1.0\linewidth]{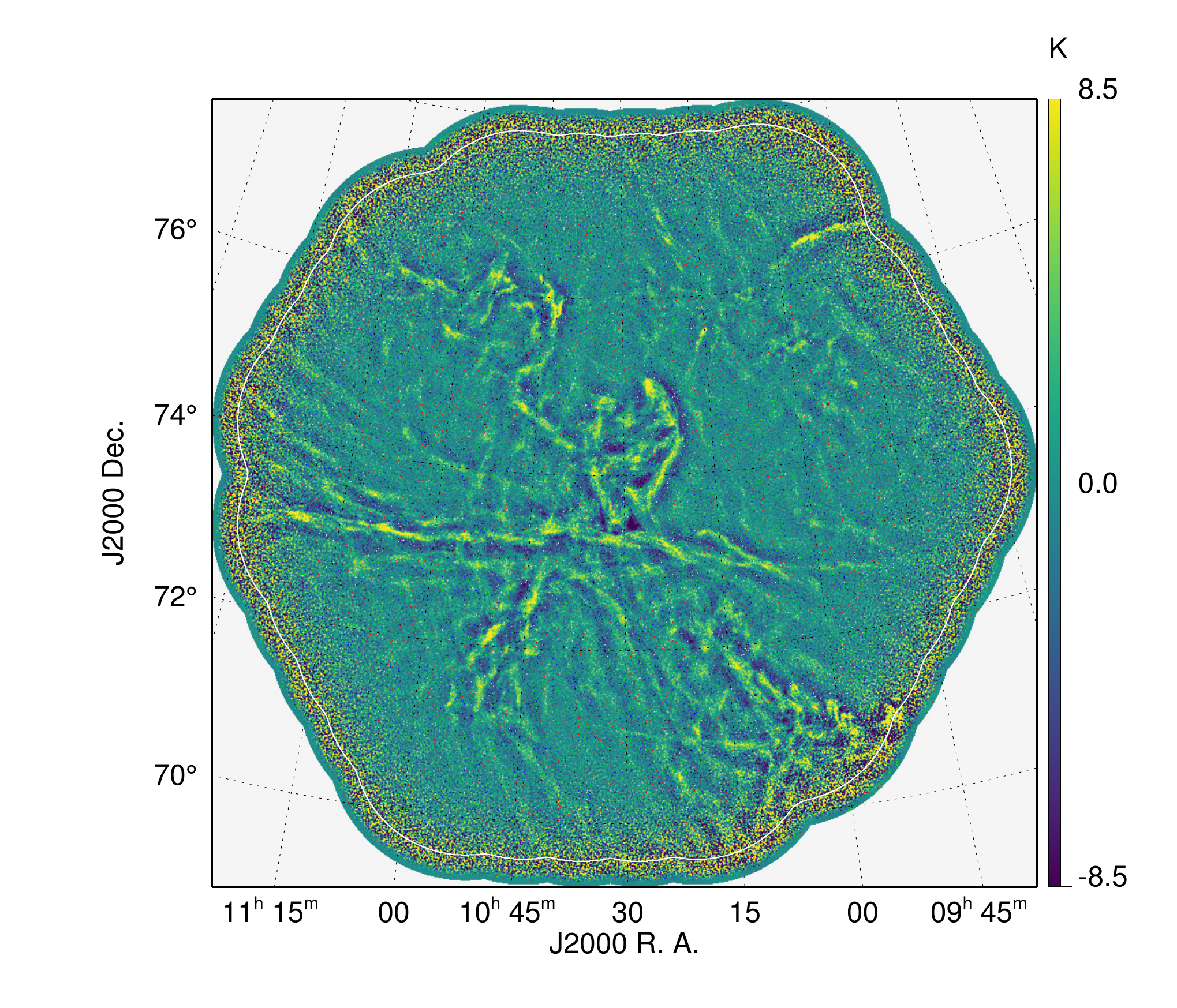}
\includegraphics[clip=true,trim=20 10 10 10,angle=0,width=1.0\linewidth]{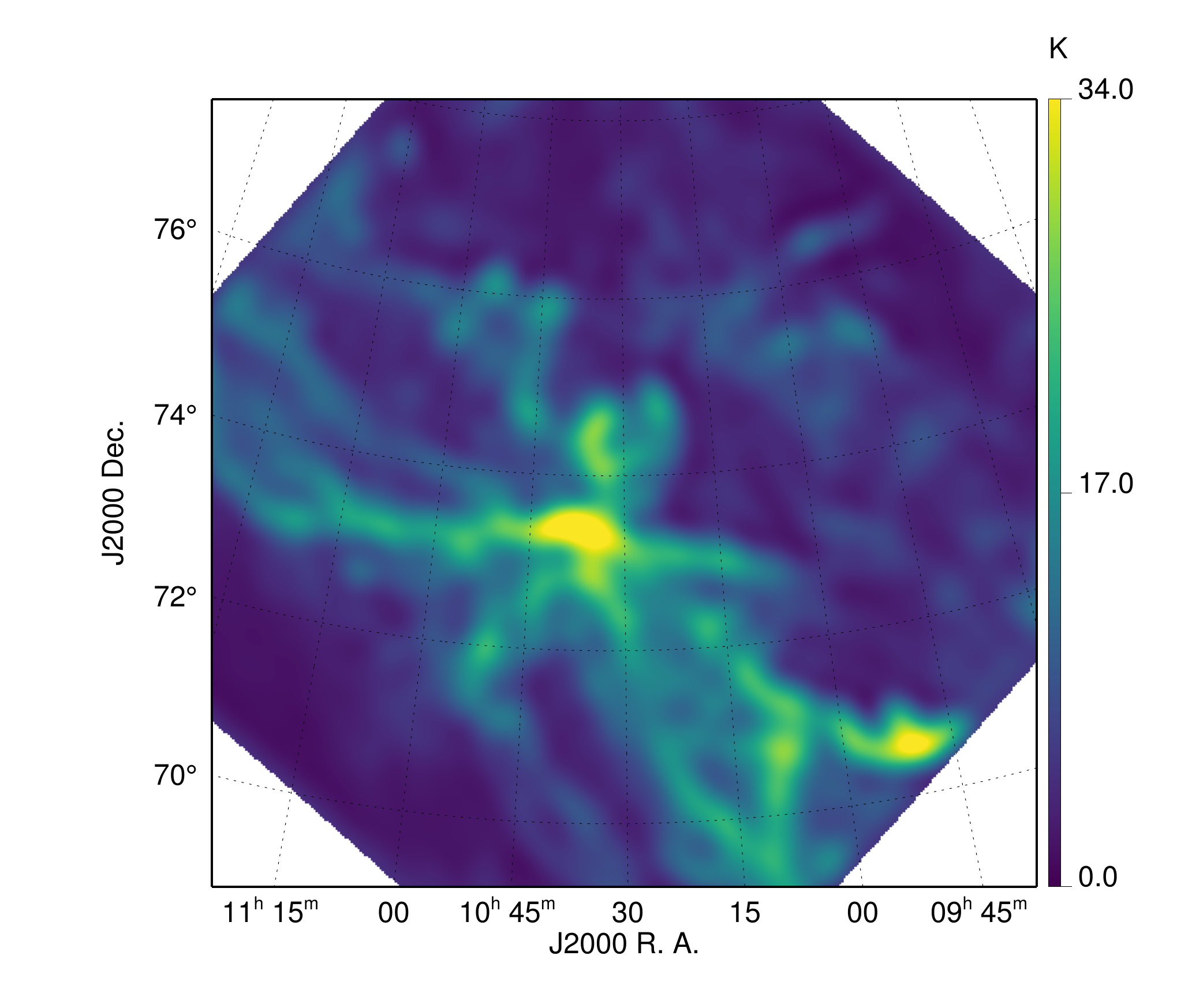}
\includegraphics[clip=true,trim=20 10 10 10,angle=0,width=1.0\linewidth]{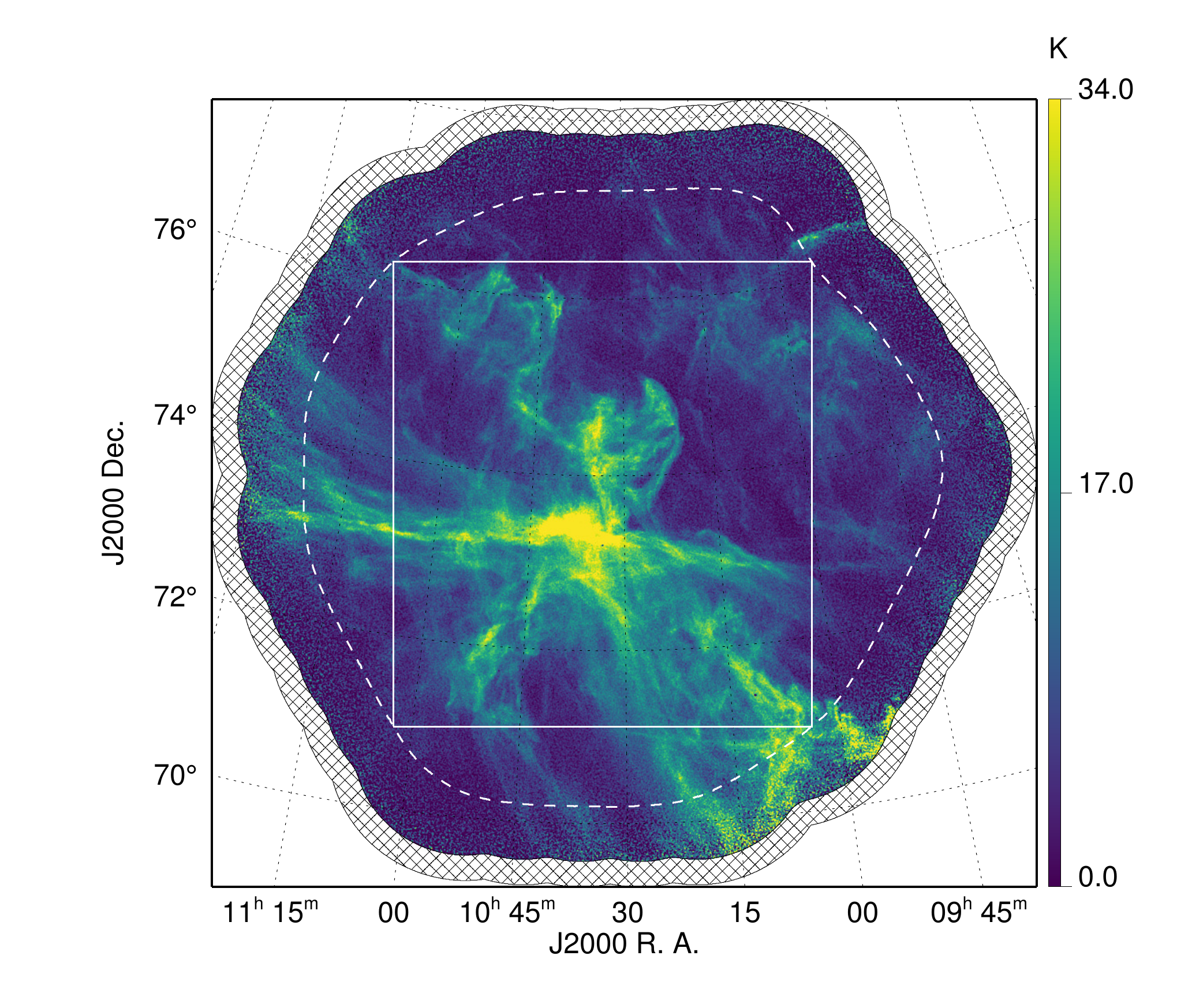}
\caption{
Channel maps for \DFG\ at $v =5.1$~\kms\ as in
Figure~\ref{noiseDFdraohi}.  Upper and middle: high and low resolution
filtered images, respectively.  Lower: Their addition, the final
\dhigls\ product.  See Section~\ref{combinegbtdrao} for details of the
annotations and Section~\ref{assessgbtdrao} for discussion.
}
\label{noiseDF} 
\end{figure}

In the combined image we are interested in preserving the high
resolution information in the \draost\ mosaic and so we use that
mosaic, scaled by $f_{\rm cc}$, in its original ICRS coordinate
system.  To reduce edge effects from the conversion to the Fourier
domain, the image is apodized with a taper beginning 10\farcm9 from
the edge and ending at the outer contour of the mosaic.
For details on the apodization of non-rectangular fields, refer to
Appendix~\ref{powerspecnonrect}.  The apodized and zero-padded image
was Fourier transformed, the weighting function applied in 2D, and
then the data were back transformed, producing a ``high resolution"
filtered image. 
An example for a single channel of \DFi\ is shown in the upper panel
of Figure~\ref{noiseDF}.
Pixels exterior to the inner edge of the apodized region
(see contour in this panel)
were flagged for exclusion in the combined cube.

The original \ghigls\ data were interpolated to the velocity channels
of the \draost\ data and initially to a 1\farcm5 ICRS grid centered on
that of the \draost\ mosaic.  It is sufficient to have a rectangular
region somewhat larger than the mosaic.  This image was
median-subtracted, apodized, Fourier transformed, and then deconvolved
with a 2D Gaussian approximation of the \ghigls\ beam.  Subsequently,
the weighting function was applied and the data back transformed to
produce a second ``low resolution" filtered image, which was finally
regridded to the finer grid of the \draost\ mosaic.
An example corresponding to the same channel is shown in the middle
panel of Figure~\ref{noiseDF}.
As done for the \draost\ data, pixels affected by the apodization of
the \ghigls\ data are flagged for exclusion.

The high and low resolution filtered data
were combined (added) in the image plane.  
To complete the 
example for the single channel in the \DFG\ region, this final product
is presented
in the lower panel of 
Figure~\ref{noiseDF}.  Pixels in hatched area (affected by the
apodization of the low resolution filtered image) and beyond are
encoded as zero in a modification of the weight map provided as an
extension to the \dhigls\ cube (Section~\ref{conclusions}).

This procedure was carried out for each velocity channel to produce
the \dhigls\ cube.
The final data products are cropped in the x- and y-directions to
exclude the hatched region and the remaining blank and hatched region
pixels are filled with the regridded original \ghigls\ data for
context.  Before any analysis these can be removed (masked) using the
modified weight map supplied with the data cube.
\dhigls\ data cubes are available on a public archive, as described at
the end of Section~\ref{conclusions}.  For any processing of these
data cubes (e.g., convolving to the native resolution of another
instrument such as \Planck), take note of the synthesized beam
parameters in Table~\ref{beamsize}.

\begin{figure}
\centering
\includegraphics[angle=90,width=1.0\linewidth]{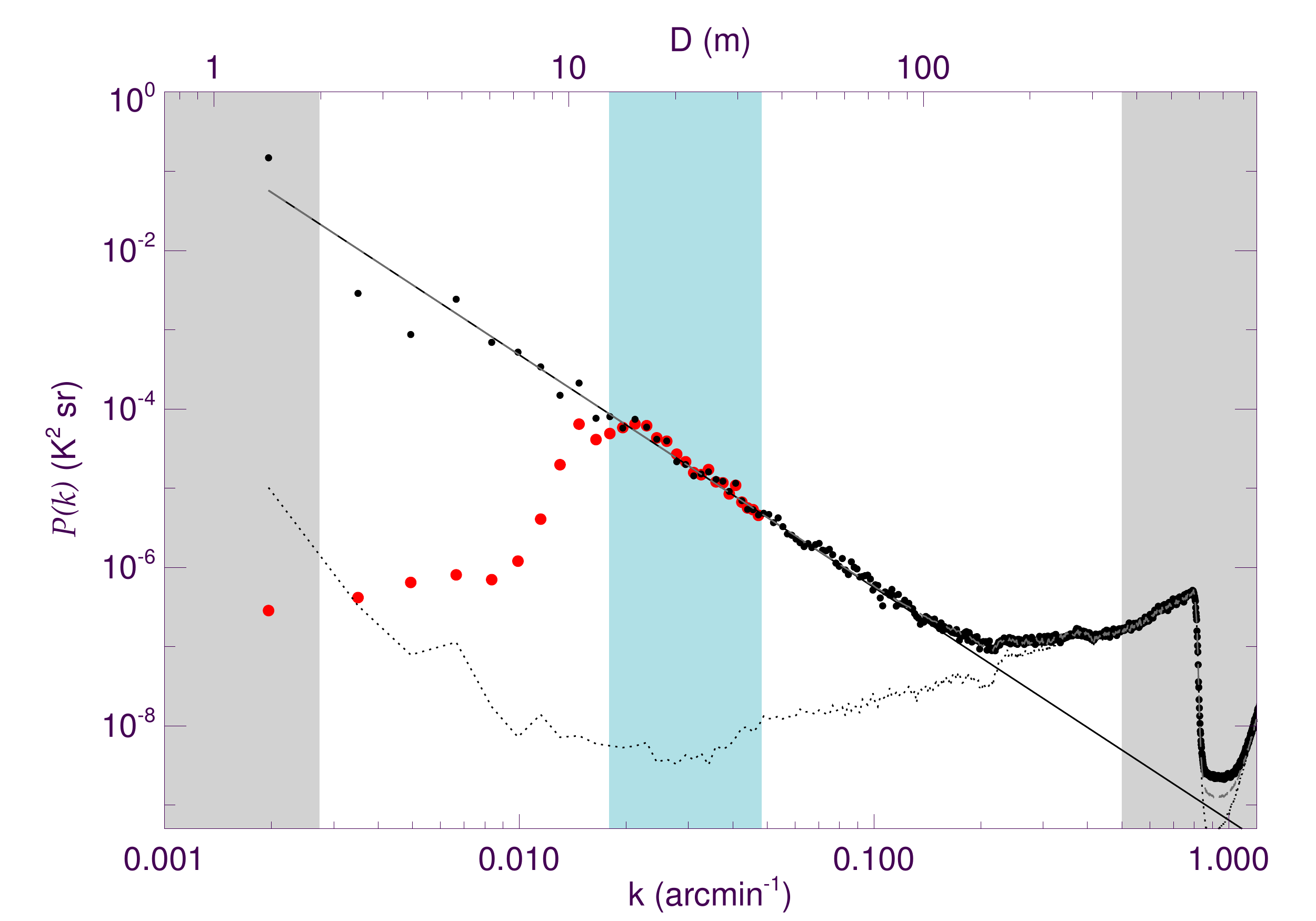}
\caption{
Power spectrum of single channel at $v = 5.1~$\kms\ as in
Figure~\ref{noiseDFdrao_ps}, but for the combined \dgm\ \DFG\ image
and for data within the white dashed contour (Figure~\ref{noiseDF}).
Model fit is to deconvolved data using Equation~(\ref{plawd}).
The blue region highlights the overlap range through which with
decreasing $k$ there is a transition from \draost\ to \ghigls\ data.
Shown for comparison (Section~\ref{assessgbtdrao}) is the power
spectrum of the scaled interferometric data alone (red, only for $k <
0.05$\,arcmin$^{-1}$).
}
\label{noiseDF_ps}
\end{figure}

\subsection{Assessment of the Short-spacing Correction}
\label{assessgbtdrao}

\hi\ emission has structure on all scales for which a power spectrum
in the \uv\ domain provides a useful statistical quantification.
The power spectrum of the combined channel map in
Figure~\ref{noiseDF}, for the data within the white dashed contour, is
given in Figure~\ref{noiseDF_ps}.
Given how the combined map was constructed, the power spectrum for
values of $k$ higher than the overlap range (in blue) is identical to
the power spectrum of the scaled interferometric data alone (shown in
red, but for clarity only for $k < 0.05$\,arcmin$^{-1}$).  Because of
the broad overlap in \uv\ coverage the latter power spectrum actually
agrees quite well through most of the overlap region, before
eventually falling off at low $k$, as in Figure~\ref{noiseDFdrao_ps}.
With the addition of the \ghigls\ data the power-law rise in the power
spectrum of the combined map extends to lower $k$ (with no
discontinuity as in Figure~\ref{df_ps_idl-eps-converted-to.pdf}).
The power-law exponent found within the white dashed contour, $-2.94
\pm 0.02$, is consistent with that found for the white rectangle
($-2.95 \pm 0.03$) and with those in Figures~\ref{noiseDFdrao_ps} and
\ref{df_ps_idl-eps-converted-to.pdf}.

The power spectrum shows that spatial fluctuations in the structure of
the \hi\ emission are smaller on small angular scales (high $k$)
compared to those on large scales.  Thus when the data are filtered to
contain only those small scales, the amplitudes in the image domain
are relatively small (and both positive and negative) as shown in the
upper panel of Figure~\ref{noiseDF}.  In contrast to this, when the
data are filtered to contain only large scales, the amplitudes in the
image domain are relatively large as shown in the middle panel of
Figure~\ref{noiseDF} (note the larger range on the colorbar).

We note that the high resolution filtered image has less dynamic range
than the channel map of the \DFi\ mosaic in
Figure~\ref{noiseDFdraohi}, because the weighting applied to the \uv\
data from the interferometer (Section~\ref{combinegbtdrao}) cuts off
the power quickly across the overlap region toward lower $k$ (see
Figure~\ref{overlapfig}), thus removing the brighter fluctuations on
the larger scales.  Those fluctuations are captured in the low
resolution filtered image.  Likewise, the low resolution filtered
image lacks the finer spatial detail in the original \ghigls\ data,
because the weighting applied cuts off faster than the effect of
primary beam of the GBT (Figure~\ref{overlapfig}). 
A corollary is that the overlap in \uv\ coverage of the \draost\ and
the GBT is so good that the spatial structure at scales covering the
central part of the overlap range could equally well be represented by
data of high quality derived from either instrument.

From the comparison in Figure~\ref{noiseDF} it can be appreciated that
short-spacing ``correction" is actually a contribution of fundamental
importance, laying out the basic structure of the map, which the
interferometric data adjusts to reveal the finer detail.  

\begin{figure}
\centering
\includegraphics[angle=90, width=1.0\linewidth]{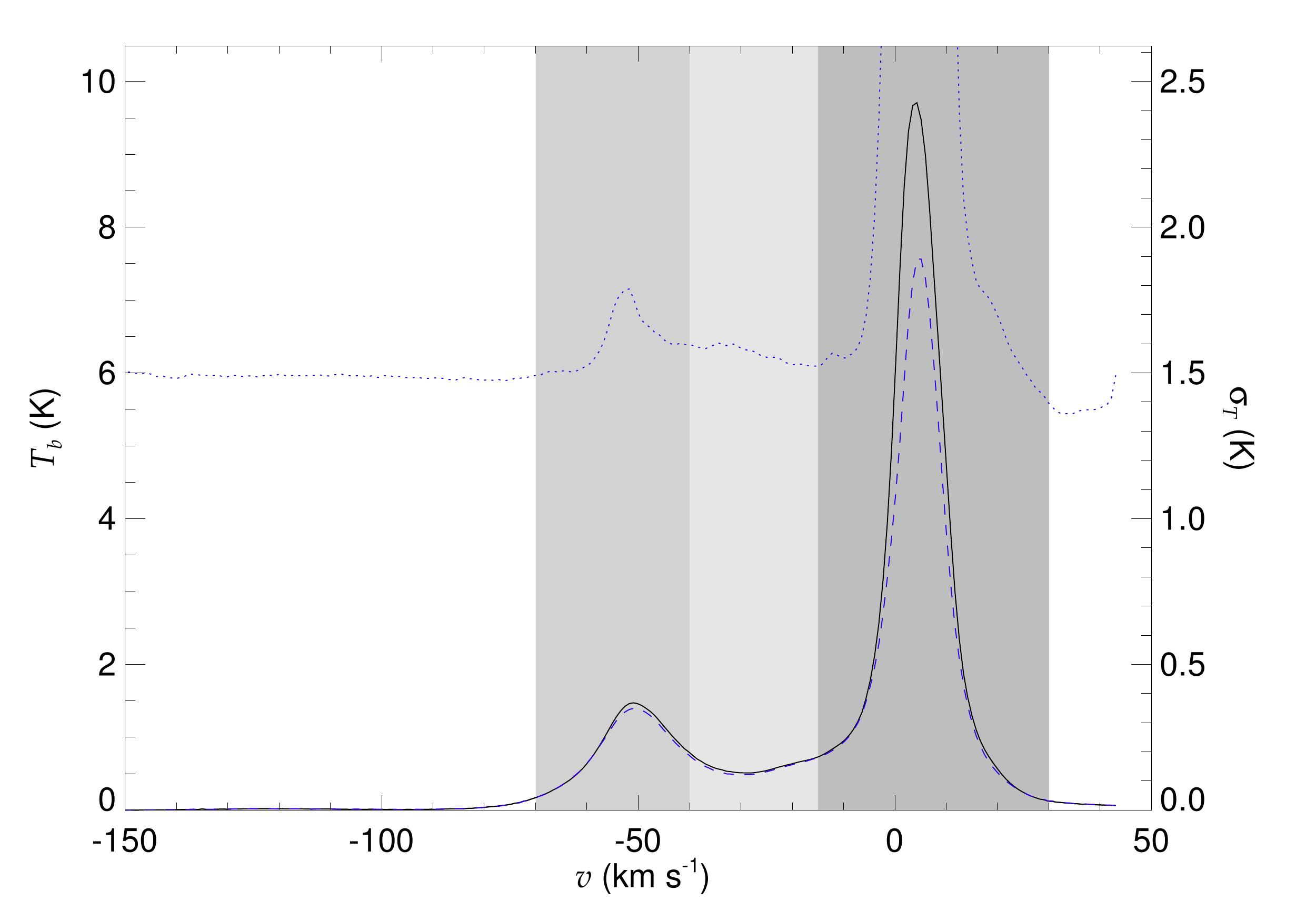}
\caption{
Spectral profiles within white dashed contour for the \dhigls\ \DFG\
cube.  Mean and median values of $\Trb$ (solid and dashed lines,
respectively, left axis).  Standard deviation about the mean of
$\Trb(\vrl)$, $\sigma_T$ (dotted line, right axis).
From right to left the shaded bands indicate the velocity ranges of
three velocity components: LVC and
an exploratory
subdivision of IVC into IVC1 and IVC2.
}
\label{sigmavelDF}
\end{figure}

\begin{table}
\caption{Velocity Component Boundaries (km~s$^{-1}$)}
\centering
\begin{tabular}{lcccc}
\hline \hline
Name & \multicolumn{4}{c}{LVC{\hskip 2.5em}IVC{\hskip 2.5em}HVC}\\
\hline
\emph{SPIDER\tablenotemark{a}} & \emph{39.9} & $-$\emph{14.9} & $-$\emph{88.1} & $-$\emph{159.7} \\
\DFG\tablenotemark{a}\tablenotemark{b} & $  30.0$ & $ -15.0$ & $ -70.0$ & \nodata \\
\rule{0pt}{3ex}\emph{N1} & \emph{57.6} & $-$\emph{10.8} & $-$\emph{59.9} & $-$\emph{151.6} \\
\ENG\tablenotemark{c}  & $  25.0 $ & $ -30.0$ & $ -80.0$ & $-145.0$ \\
\rule{0pt}{3ex}\emph{G86} & \emph{23.8} & $-$\emph{26.9} & $-$\emph{64.8} & $-$\emph{137.2} \\
\MGG & $  18.0$ & $ -15.0$ & $ -55.0$ & \nodata \\
\rule{0pt}{3ex}\emph{DRACO} & \emph{23.8} & $-$\emph{8.4} & $-$\emph{72.8} & $-$\emph{209.5} \\
\DRG & $  10.0$ & $  -12.0$ & $ -57.0$ & \nodata \\
\rule{0pt}{3ex}\emph{POL} & \emph{31.8} & $-$\emph{48.7} & $-$\emph{97.0} & $-$\emph{140.4} \\
\POG & $  15.0$ & $ -8.0$ & $ -40.0$ & \nodata \\
\rule{0pt}{3ex}\emph{POL} & \emph{31.8} & $-$\emph{48.7} & $-$\emph{97.0} & $-$\emph{140.4} \\
\MCG & $  10.0$ & $ -9.0$ & $ -55.0$ & \nodata \\
\rule{0pt}{3ex}\emph{UMA} & \emph{50.4} & $-$\emph{26.1} & $-$\emph{84.9} & $-$\emph{193.5} \\
\UMG & $15.0$ & $ -25.0$ & $ -65.0$ & \nodata \\
\hline
\end{tabular}
\tablenotetext{1}{Pair of rows: first row for \ghigls\ field
(italics); second row for gas in \dhigls\ region, a subset of the
\ghigls\ field.}
\tablenotetext{2}{IVC range subdivided in Figure~\ref{sigmavelDF}.} 
\tablenotetext{3}{Ranges shown in Figure~\ref{sigmavelEN}.} 
\label{compvel_table}
\end{table}

\begin{figure}
\centering
\includegraphics[angle=90,width=1.0\linewidth]{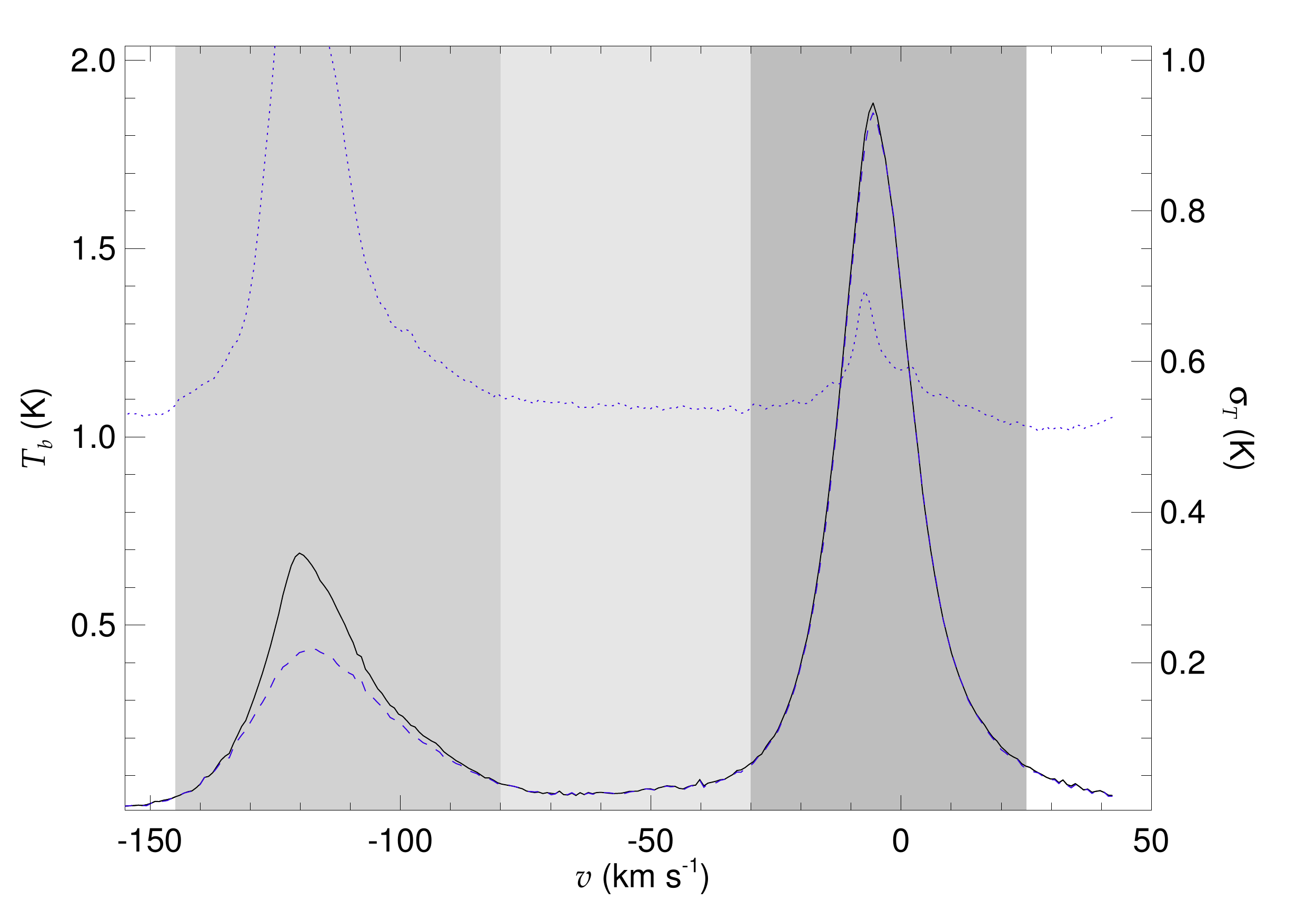}
\caption{ 
Spectral profiles as in Figure~\ref{sigmavelDF}, for \ENG.
There is no significant IVC in the area covered by the \draost\
mosaic.  LVC and HVC velocity ranges are indicated by the darker
shading.
}
\label{sigmavelEN}
\end{figure}

\section{Properties of Distinct Velocity Components}
\label{pvcs}

\subsection{Specification of VCs}
\label{svcs}

Within the cube distinct velocity structures can be identified.  To
quantify the division of the cube into VCs, we computed 
the mean, median and
standard deviation about the mean of each channel (see
Figure~\ref{sigmavelDF}), following the method in
\citet{planck2011-7.12} and \citet{mart15}.
Components often appear blended in the mean and median spectra, but
become more distinct in the standard deviation spectrum,
which depends on the fluctuations across the field, not just the
presence of signal, and so takes advantage of the rich structure
within the cube.  Peaks in the standard deviation spectrum were noted
and each was taken to indicate a separate velocity component.  For
\DFG, these are at about $3$~\kms, $-30$~\kms\ and $-50$~\kms.
Minima at about $-15$~\kms\ and $-40$~\kms\ suggest velocities at
which the cube can be divided into components.  For exploratory
purposes we divided the IVC range into two separate components, which
we label IVC1 and IVC2.
The separation here can be compared with the velocity cuts made in
\citet{planck2011-7.12} for the \ghigls\ data in the SPIDER field (of
which \DFG\ is a subset): LVC/IVC separation at $-14.9$~\kms\ and
IVC/HVC separation at $-88.1$~\kms\ (see Table~\ref{compvel_table}).
Here there is no HVC component readily identifiable, unlike for the
\ghigls\ SPIDER field where the lower resolution provides higher
sensitivity.  Also, the coverage in \DFG\ does not fully overlap with
the HVC seen in the \ghigls\ SPIDER data, and where it does is along
the noisy edge of the mosaicked region.

A second example is given in Figure~\ref{sigmavelEN} for \ENG.
Results are tabulated in Table~\ref{compvel_table} along with those
for the other \dhigls\ regions and the corresponding \ghigls\ fields
identified in Table~\ref{beamsize}.

\subsection{Maps of Column Density for VCs}
\label{integratemaps} 

\begin{figure}
\centering
\includegraphics[angle=0,width=0.85\linewidth]{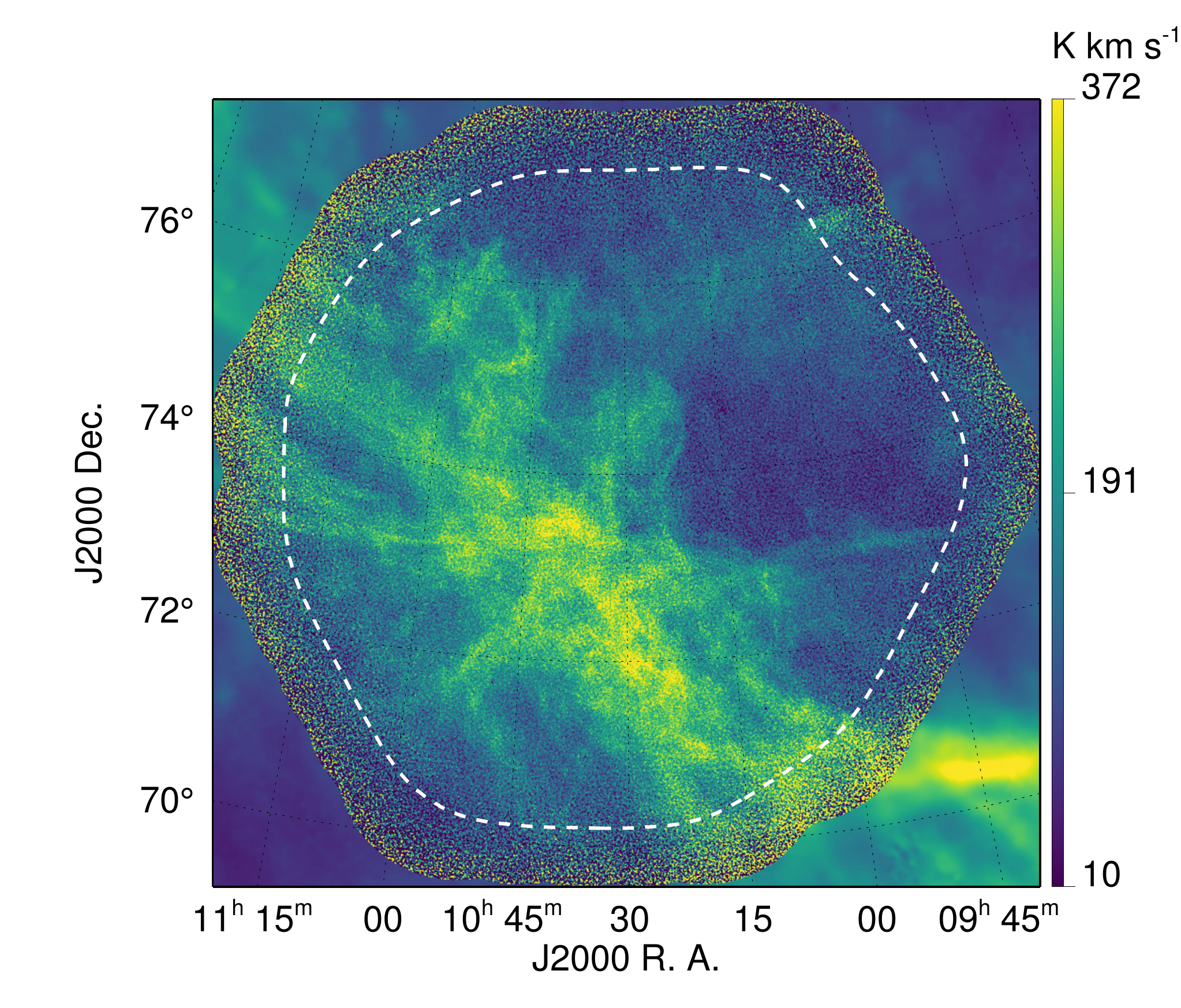}
\caption{
Integrated emission, \wh, in \DFG.  White dashed contour as in
Figure~\ref{coverageDF}. 
The area beyond where data have been combined has been filled in in
the cube using interpolated lower-resolution \ghigls\ data
(Section~\ref{combinegbtdrao}).
}
\label{fig:DFtvc}
\end{figure}

The data cubes collapsed along their velocity axis produce maps of
integrated emission, \wh$~=~\sum_{i}\Trb^{i} \Delta v$.  The
integrated emission is limited to channels where there is a measurable
signal.
An example of such a map is shown in Figure~\ref{fig:DFtvc} for \DFG\
with emission integrated between $-80$ and $+30$~\kms\ including both
IVC and LVC (Figure~\ref{sigmavelDF}).

\begin{figure}
\centering
\includegraphics[angle=0,width=0.8\linewidth]{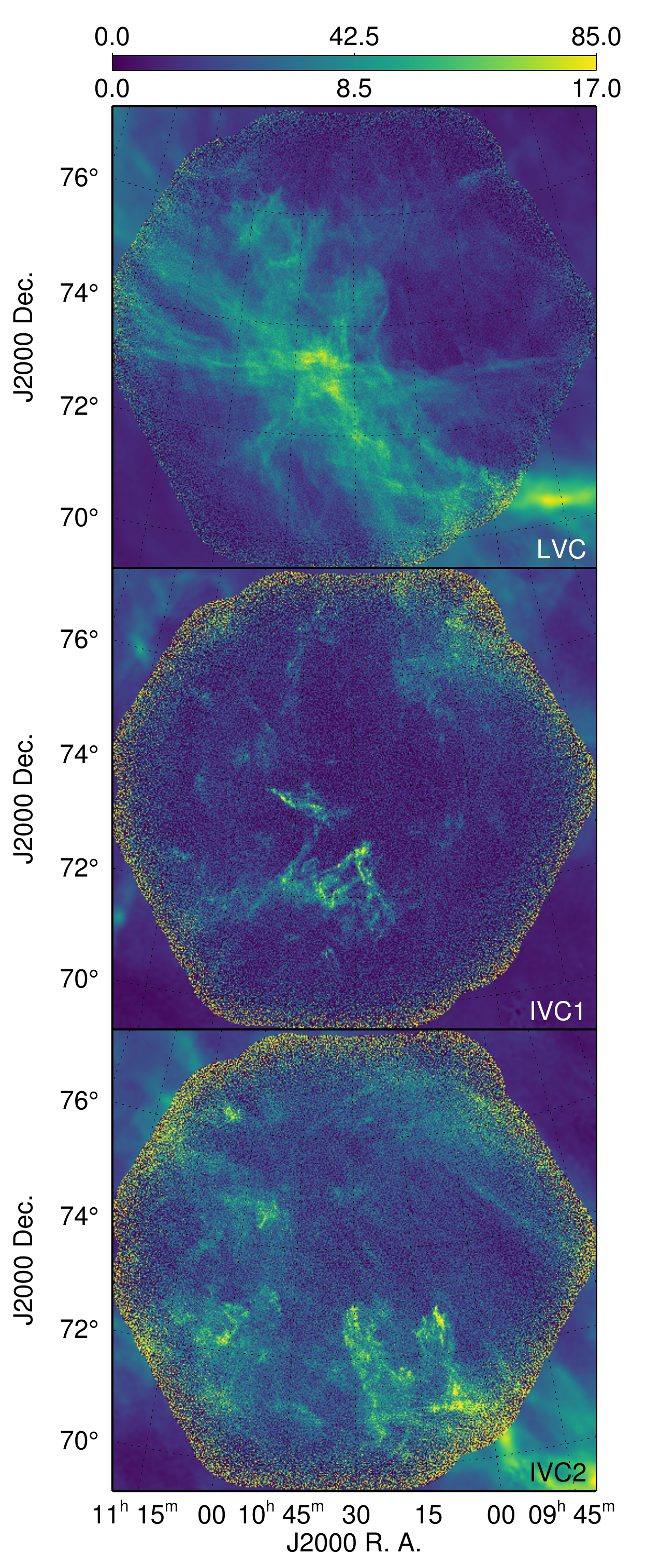}
\caption{
Maps of \nh\ for \DFG\ VCs centered at (top to bottom): $7.5$~\kms\
(LVC), $-27.5$~\kms\ (IVC1), and $-55$~\kms\ (IVC2) for velocity
ranges in Figure~\ref{sigmavelDF}.  Colorbar for lower two \nh\ maps
scaled by a factor of 5.  Units are \colunits.  Outer parts of the
maps are filled in with \ghigls\ data as in Figure~\ref{fig:DFtvc}.
}
\label{DFcomponents}
\end{figure}

\begin{figure}
\centering
\includegraphics[angle=0,width=0.8\linewidth]{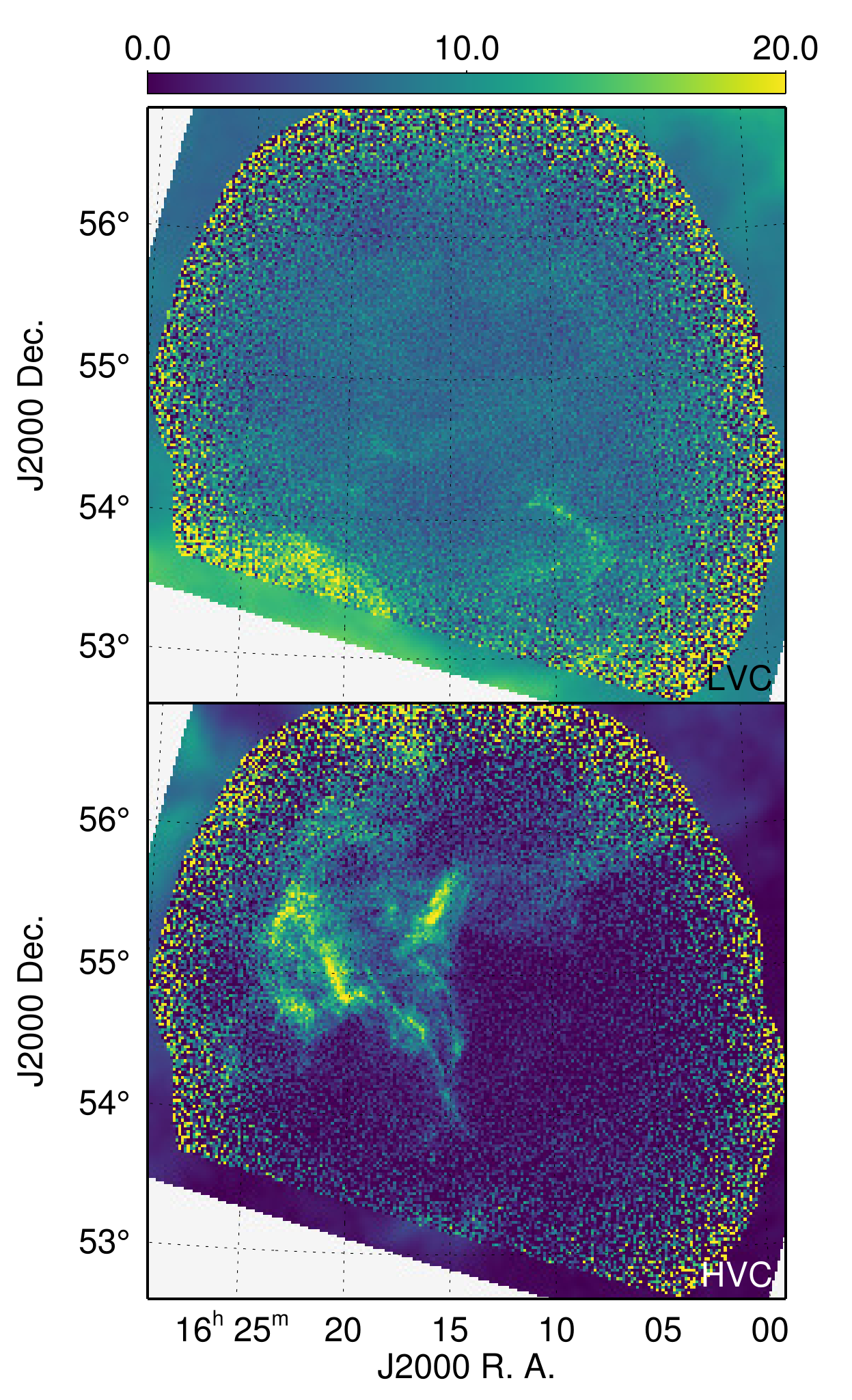}
\caption{
Like Figure~\ref{DFcomponents}, for \ENG\ VCs centered at $-2.5$~\kms\
(LVC, top) and $-112.5$~\kms\ (HVC, bottom)
for velocity ranges in Figure~\ref{sigmavelEN} and
Table~\ref{compvel_table} (no significant IVC emission in area covered
by \draost\ mosaic).  Areas not covered by \ghigls\ N1 survey are
white.
}
\label{ENcomponents}
\end{figure}

The quantity \wh\ is of particular interest because in the optically
thin regime the column density \nh$ = C\,$\wh\ cm$^{-2}$, where $C =
0.1823 \times 10^{\colnum}$~cm$^{-2}$ (K \kms)$^{-1}$.
Although of some significance for the \ghigls\ data, an optical depth
correction correction might seem less important for the \dhigls\ data
presented here because they are noisier.  Nonetheless, the correction
is a systematic one and for consistent treatment of the data in both
surveys an optical depth correction was made assuming a single spin
temperature $\Trs = 80$~K \citep{mart15} in Equation~(\ref{taucorrect}).
The related issue of the dependence of optical depth corrections on
the angular resolution of the observations is explored in
Appendix~\ref{resdep}.

Some resulting \nh\ maps produced from the \dhigls\ \HI\ cubes are
shown in Figure~\ref{DFcomponents} and Figure~\ref{ENcomponents}.
Others for the remaining regions are included in
Appendix~\ref{additionalfields}.  These \nh\ products are available
for download from the \dhigls\ archive (see end of
Section~\ref{conclusions}).

As presented in Appendix~\ref{prepdata} and discussed further in
Section~\ref{discresultsthick}, the optimal S/N in an \nh\ map is
obtained by integrating only a few channels near the peak emission,
about 10 in the case of \DFG.  However, for the above \nh\ maps of a
full VC integration is carried out over many more channels to cover
the range in which emission is seen anywhere in the map.  Therefore,
the \nh\ maps in Figures~\ref{DFcomponents} and \ref{UMAcomponents}
clearly appear noisier than color maps based on single channels
with strong emission (see Figures~\ref{DFthree} and \ref{DFthreeIVC}
below).
The spectrum at each pixel could be integrated separately, keeping
only data above some noise threshold, but unlike for the centroid
velocity this would bias \nh.  Fitting Gaussian components and then
integrating them over velocity is an alternative that accounts
implicitly for noisy empty channels.

\subsection{Maps of Centroid Velocity for VCs}
\label{vcmaps}

\begin{figure}
\centering
\includegraphics[angle=0,width=0.8\linewidth]{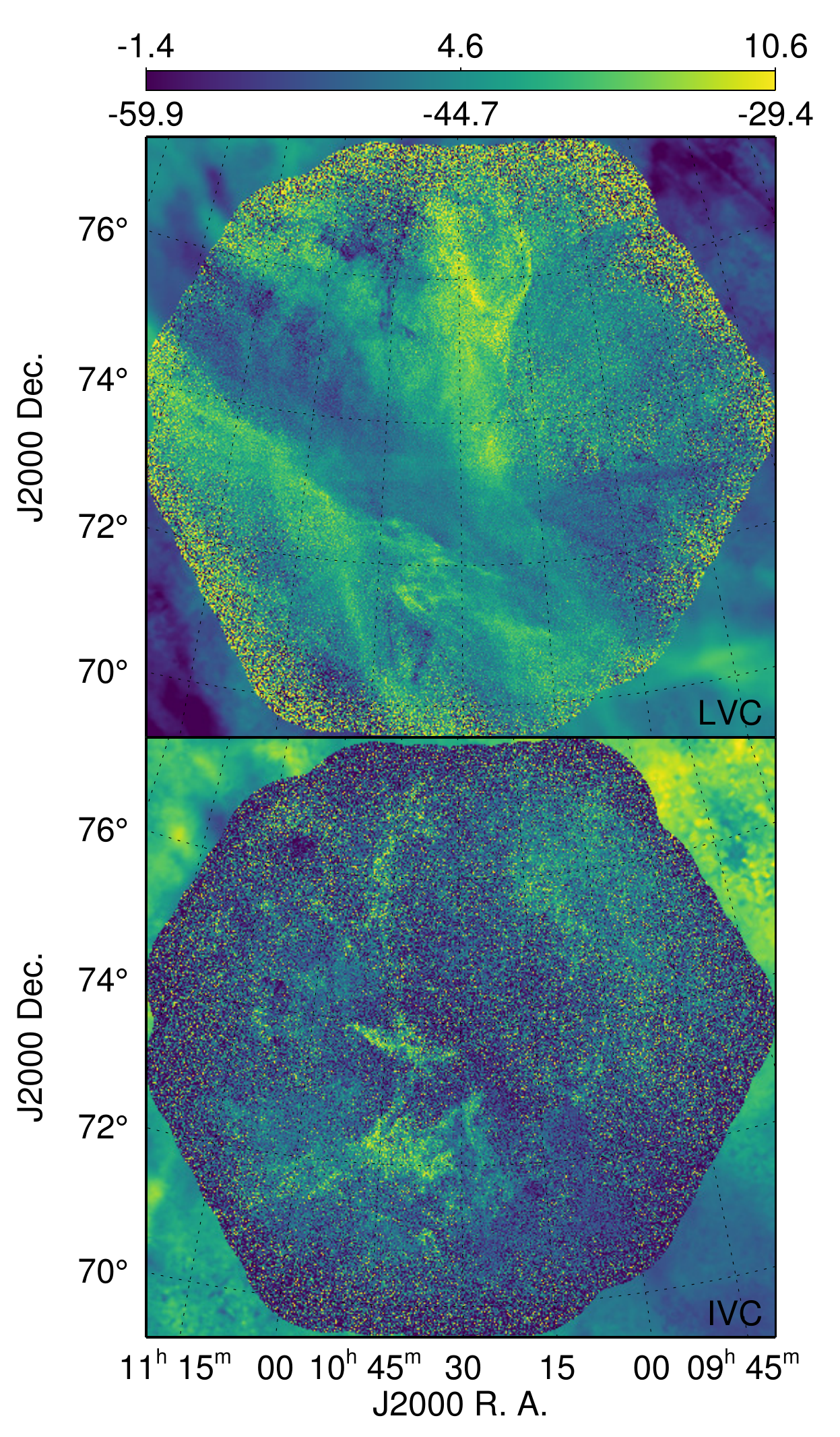}
\caption{
Centroid velocity maps for \DFG\ 
for VCs centered at $7.5$~\kms\ (LVC, top) and $-42.5$~\kms\ (IVC
including IVC1 and IVC2, bottom) for velocity ranges in
Table~\ref{compvel_table}.
Units are \kms.
}
\label{componentsDFv_ps}
\end{figure}

\begin{figure}
\centering
\includegraphics[angle=0,width=0.8\linewidth]{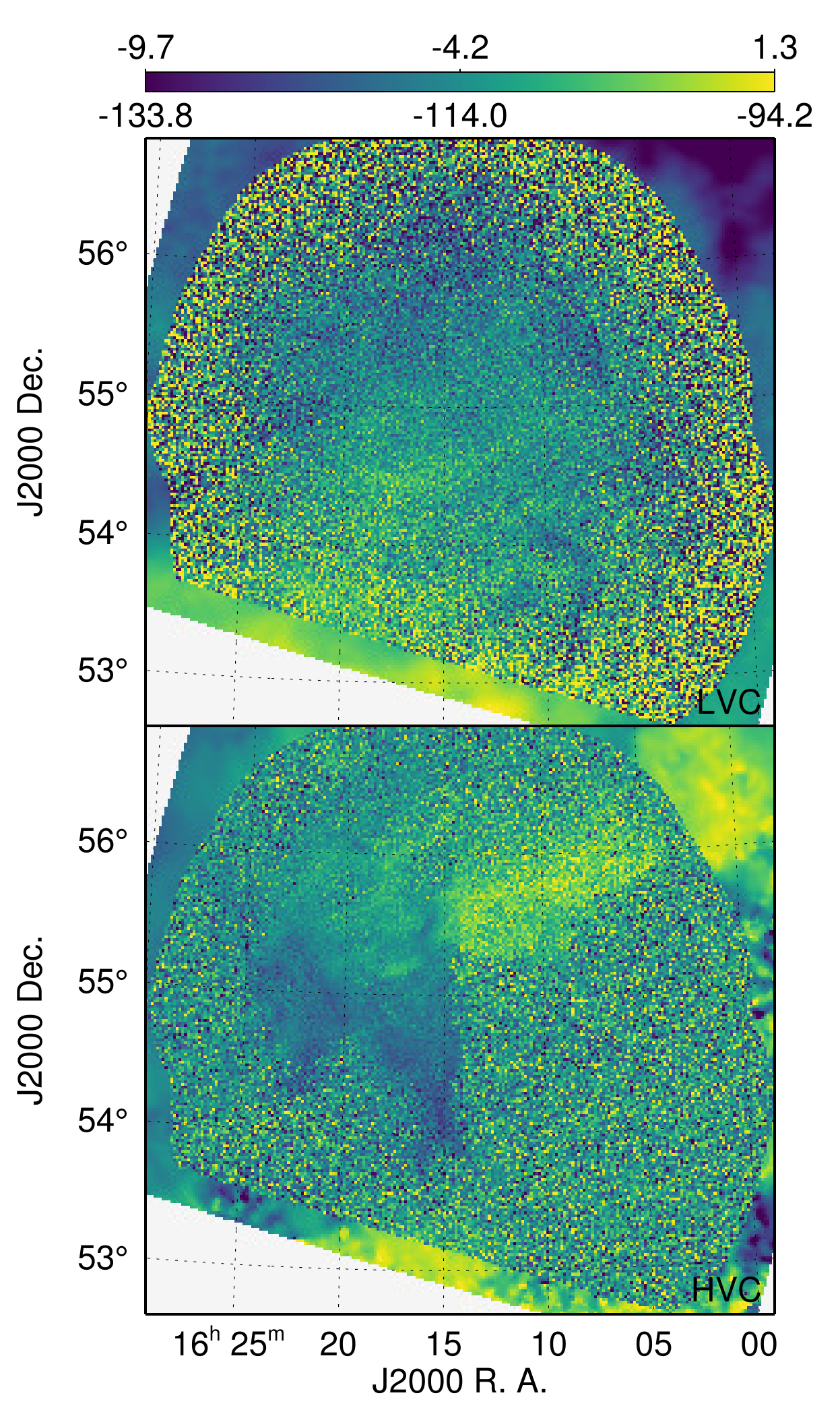}
\caption{
Like Figure~\ref{componentsDFv_ps}, for \ENG\ VCs centered at
$-2.5$~\kms\ (LVC, top) and $-112.5$~\kms\ (HVC, bottom) for velocity
ranges in Figure~\ref{sigmavelEN} and Table~\ref{compvel_table}.
}
\label{componentsENv_ps}
\end{figure}

On the basis of work on fractional Brownian motion (fBm) simulations,
\citet{mamd2003b} suggested a direct mapping between the power
spectrum of the centroid velocity map of a VC and the 3D velocity
field.
This motivated our computation of centroid velocity maps for all of
the VCs in the \dhigls\ regions.  The \dhigls\ cubes have not been
filtered and so noise is a factor in determining a reliable centroid
velocity power spectrum.  In order to reduce some of this uncertainty,
each spectrum in the cube is first filtered using a 5-channel-wide
Hanning window (i.e., a Hanning function with 5 non-zero values).  The
centroid velocity field, $v_{c}$, is then determined using only those
channels in the given component's velocity range with signal above
twice the emission-free end channel noise, $\sigma_{\rm ef}$:
\begin{equation}
v_{c} = \frac{\sum_{i} v^{i} \Trb^{i} \Delta v}{\sum_{i} \Trb^{i} \Delta v} {\hskip 1 em} ;  {\hskip 1 em}\Trb^{i} > 2 \sigma_{\rm ef} \, .
\end{equation}

Centroid velocity maps for \DFG, \ENG, and \UMG\ are shown in
Figures~\ref{componentsDFv_ps}, \ref{componentsENv_ps}, and
\ref{componentsUMv_ps}, respectively.  
These images show structures on all scales and sometimes systematic
gradients, most visible at large scales (e.g., the HVC in \ENG\ in
Figure~\ref{componentsENv_ps}).
The centroid velocity maps look somewhat less noisy than the \nh\ maps
because of the spectral smoothing and clipping performed.

\section{Power Spectrum Analysis of \hi\ Structure}
\label{interpretation}

\subsection{Power Spectrum Analysis of Maps of \nh}
\label{nhanalysis}

As described in Appendix~\ref{psmodel} we deconvolved the data with
the appropriate 2D synthesized beam (Table~\ref{beamsize}) before
computing the 1D power spectrum and fitting it with the model in
Equation~(\ref{plawd}).  As in Appendix~\ref{prepdata} and
Appendix~\ref{powerspecnonrect}, we selected data within the
non-rectangular white dashed contour.
The model has three parameters to be fitted: the power law exponent,
the amplitude normalization, and the scale factor of a noise template.

\begin{figure}
\centering
\includegraphics[clip=true,trim=40 0 0 0, angle=0,width=1.1\linewidth]{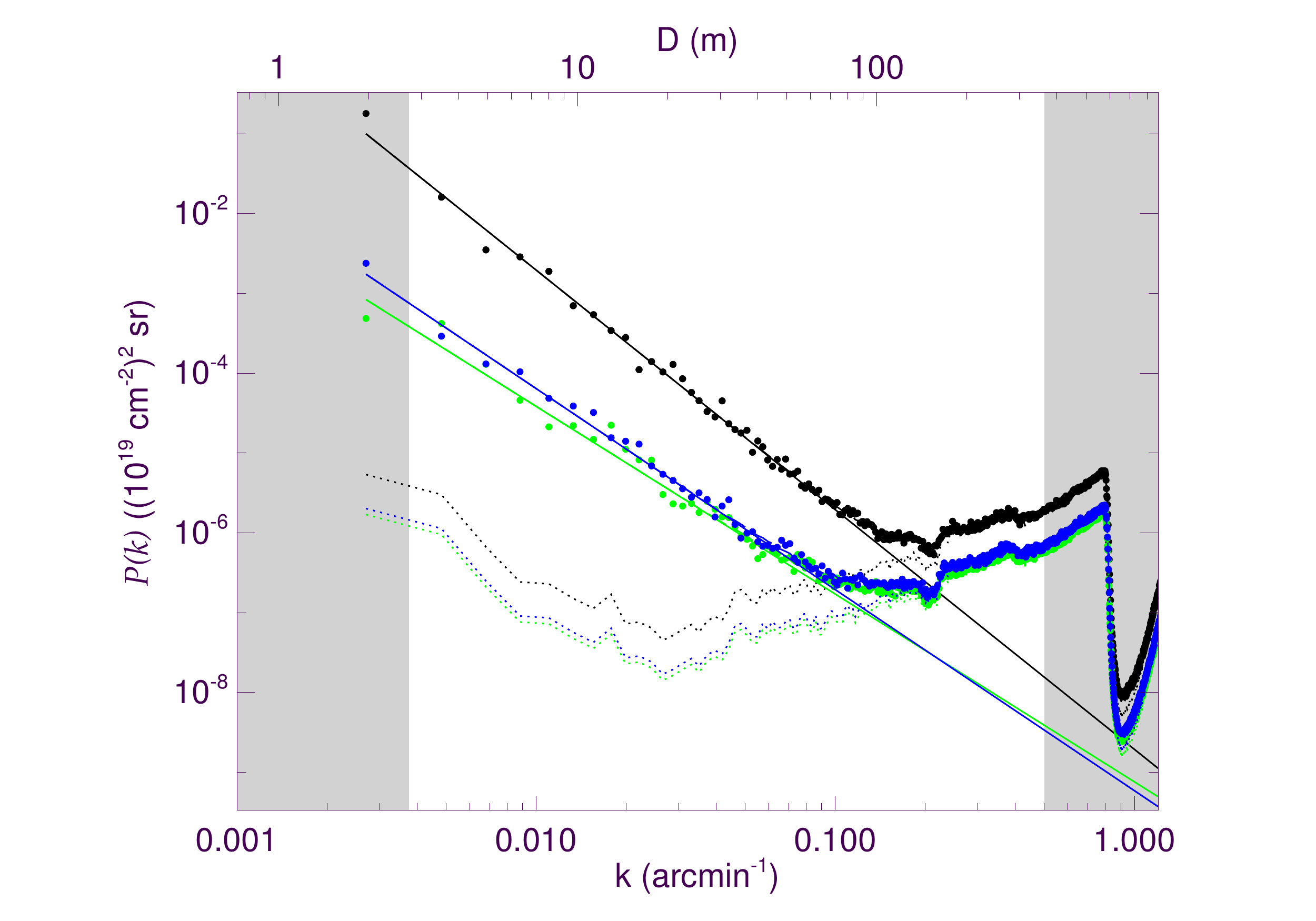}
\includegraphics[clip=true,trim=40 0 0 0, angle=0,width=1.1\linewidth]{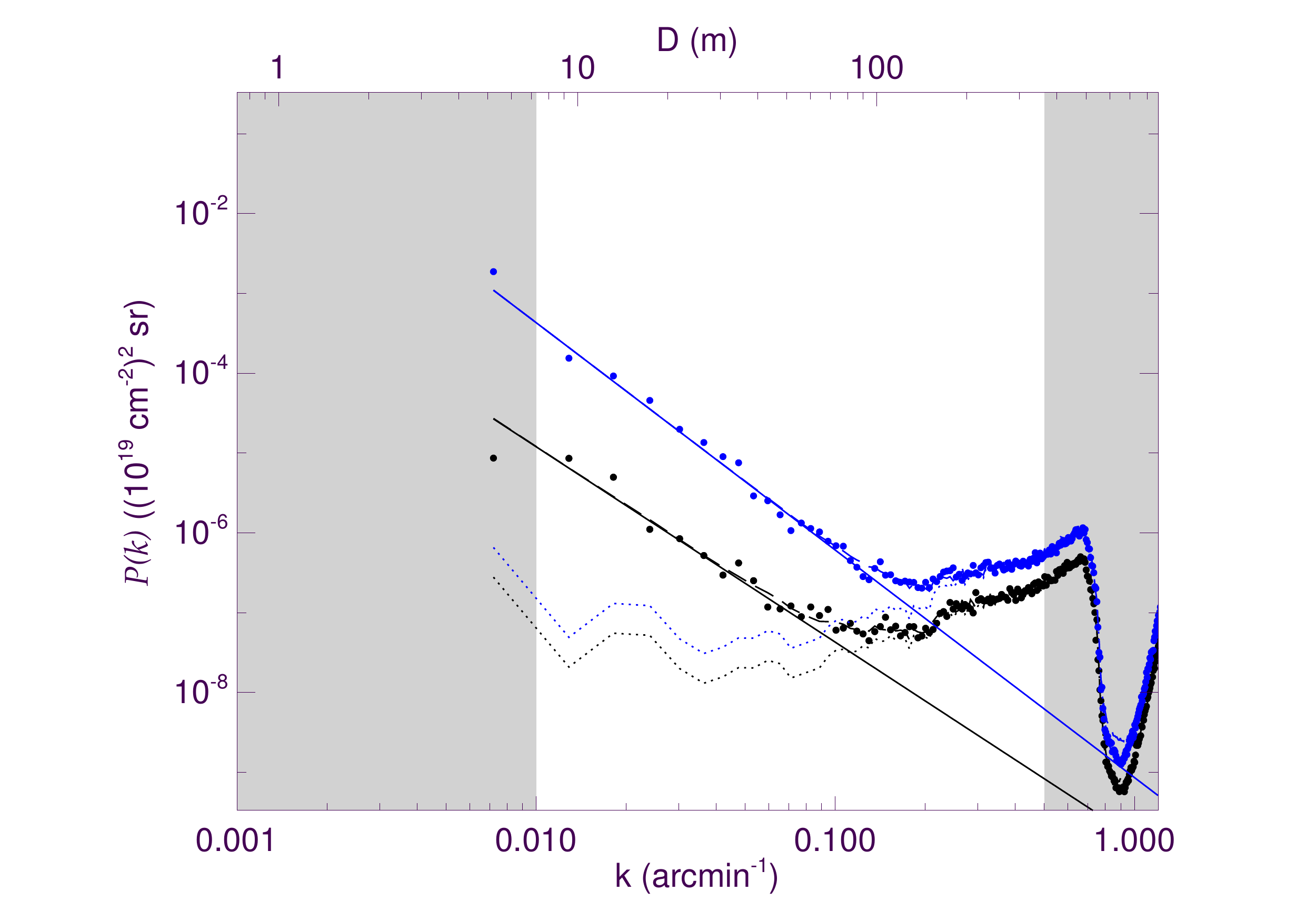}
\includegraphics[clip=true,trim=40 0 0 0, angle=0,width=1.1\linewidth]{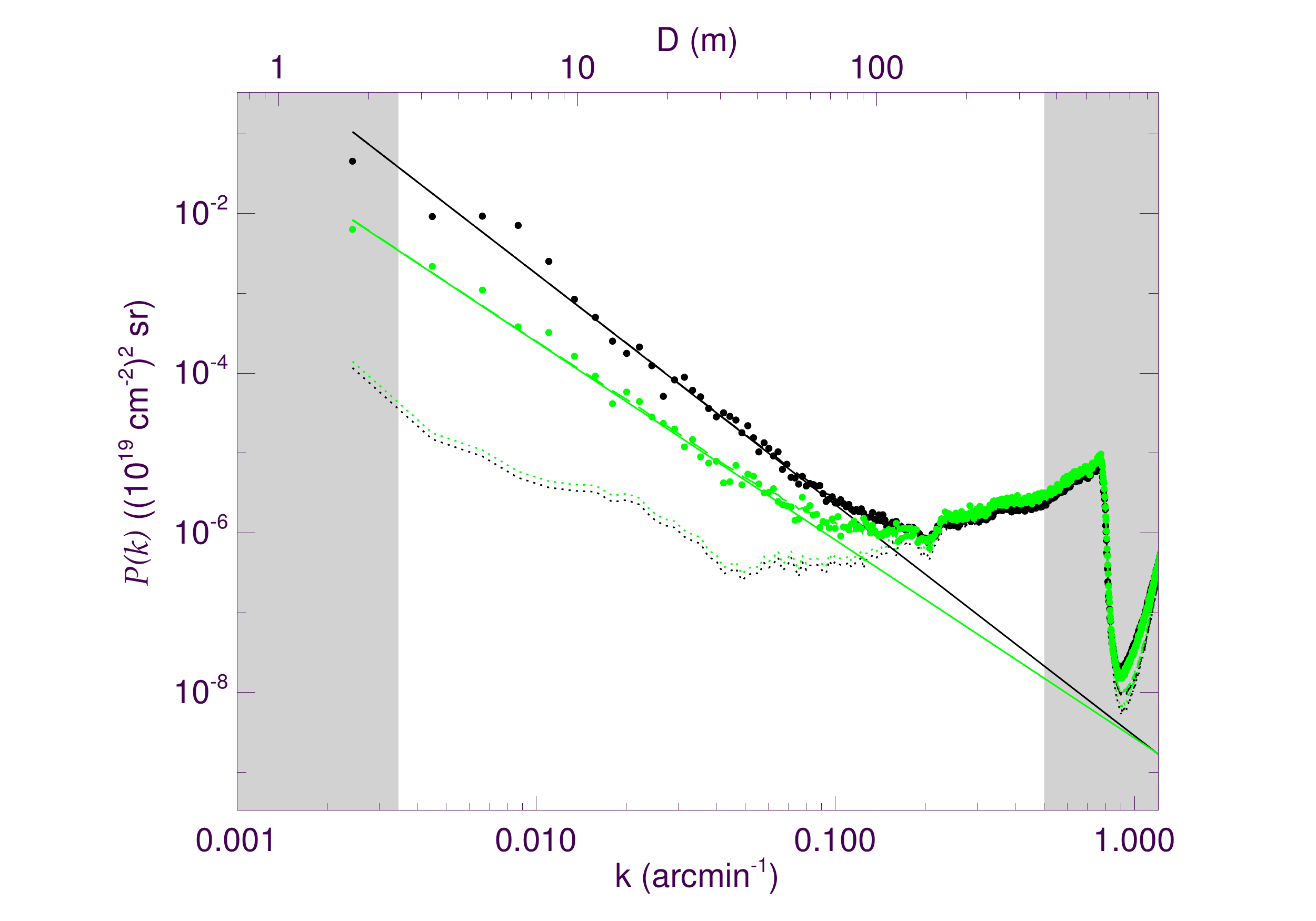}
\caption{
Power spectra for images of emission integrated over distinct velocity
ranges.
Upper: for \DFG\ for the ranges
in Figure~\ref{DFcomponents}: LVC (black), IVC1 (green), and IVC2
(blue).
Middle: for \ENG\ for the ranges
in Figure~\ref{ENcomponents}: LVC (black) and HVC (blue).
Lower: for \UMG\ for the ranges
in Figure~\ref{UMAcomponents}: LVC (black) and IVC (green).
}
\label{componentsDF_ps}
\end{figure}

The power spectra of the \nh\ maps of three VCs in \DFG\
(Figure~\ref{DFcomponents}) are shown
in the upper panel of
Figure~\ref{componentsDF_ps}, where it is readily apparent that the
spectra for IVC1 and IVC2 are consistently shallower than for the LVC,
indicative of \emph{relatively} more power on smaller scales.  This is
quantified by the power-law fits, also shown in the figure.
The power spectrum of the \nh\ map of the combined IVC
(Table~\ref{compvel_table}) is similarly shallower.  The exponents for
LVC and IVC
tabulated in Table~\ref{powertable} are significantly different.  Fits
to data within the more restricted white rectangle give consistent
results within the uncertainties.  We also note that while there are
systematic uncertainties ($< 0.1$) in the derivation of the exponents,
assessment of their relative differences is not impacted
(Appendix~\ref{psfitting}).

\ENG\ has very little IVC emission.  Power spectra for the other two
VCs (Figure~\ref{componentsDF_ps}, middle) indicate that the HVC
spectrum is marginally steeper than that for the LVC.  Additionally,
the HVC exponent in Table~\ref{powertable} can be compared to the
exponent for the rectangular field in the single \ENi\ HVC channel at
$-117.7$~\kms\ used in Section~\ref{enmnoise}, $-2.21 \pm 0.12$, or
more appropriately here, the exponent for the same channel in the
\dhigls\ cube within the white dashed contour, $-2.30 \pm 0.09$.  This
suggests a steepening of the power spectrum with the inclusion of
additional channels (see the discussion in
Section~\ref{discresultsthick}).

The \UMG\ data are not as deep as in \DFG\ or \ENG, but nevertheless
the \hi\ signal in the power spectrum is well defined.  For the \nh\
maps of the LVC and IVC components (Figure~\ref{UMAcomponents}) we
again found a steeper exponent for the LVC component (see
Figure~\ref{componentsDF_ps}, lower, and Table~\ref{powertable}).  The
value of the LVC VC is comparable to that for the single LVC channel
value, $-2.93 \pm 0.08$, from Section~\ref{enmnoise} for the white
rectangle region in Figure~\ref{noiseUMdrao}.
\UMG\ is immediately adjacent to and overlaps slightly \DFG\
(Figures~\ref{field_locations} and \ref{DFthree}) for which we found a
similar pair of exponents for the LVC and IVC components
(Table~\ref{powertable}).
We note that the \UMi\ mosaic contains a deeper observation of the
\URSA\ subregion for which a steeper power spectrum (exponent
$-3.6\pm0.2$) was found in previous work \citep{mamd2003}.  This is
revisited in Appendix~\ref{discdetailsursa}.

Results for \DRG\ and \POG\ are presented in
Appendix~\ref{additionalfields}.

\begin{table}
\caption{Power-law Model Results for \DFG, \ENG, and \UMG}
\centering
\begin{tabular}{lccc}
\hline
\hline
Map & \DFG & \ENG & \UMG \\
\hline
LVC \nh\ &$-3.00\pm0.03$ 	& $-2.45\pm0.17$  	& $-2.89\pm0.04$ \\
   {\hskip 2.5 em} $v\tablenotemark{a}$ & $-2.78\pm0.08$ & $-2.43\pm0.16$ & $-3.03\pm0.09$ \\
IVC \nh\ & $-2.60\pm0.04$ 	& \nodata              	& $-2.48\pm0.06$ \\ 
 {\hskip 2.5 em} $v$  & $-2.17\pm0.10$ & \nodata &  $-2.59\pm0.15$ \\
HVC \nh\ & \nodata 			& $-2.85\pm0.07$		& \nodata  \\
	  {\hskip 2.5 em} $v$  & \nodata & $-2.78\pm0.30$  & \nodata \\
\hline
\HI,  LVC+IVC\tablenotemark{b}       &  $-3.00\pm0.04$ 	& $-2.45\pm0.17$ & $-2.85\pm0.04$ \\
\hline
E(B-V)$_{\rm xgal}$\tablenotemark{c,d} & $-2.69\pm0.07$ 	& $-1.55\pm0.11$ 	& $-2.66\pm0.08$ \\
$\tau_{\rm 353}$\tablenotemark{c} & $-2.39\pm0.08$ 	& $-1.46\pm0.15$ 	& $-2.43\pm0.08$ \\
857\,GHz\tablenotemark{c,d} & $-2.64\pm0.07$ & $-1.19\pm0.17$ & $-2.52\pm0.07$ \\
857\,GHz\tablenotemark{c} & $-2.53\pm0.07$  & $-1.10\pm0.19$  & $-2.48\pm0.08$ \\
\hline
\end{tabular}
\tablenotetext{1}{Centroid velocity map}
\tablenotetext{2}{Proxy for dust emission: combined \nh\ 
components
weighted by
$\tau_{\rm dust}$/\nh\ of the correlated dust components from
\citet{planck2011-7.12}}
\tablenotetext{3}{From \citet{planck2013-p06b}, 5\arcmin\ resolution
}
\tablenotetext{4}{No point sources}
\label{powertable}
\end{table}

\subsubsection{Comparison with Power Spectra of Thermal Dust Emission}
\label{discresultsthermal}

Because dust and gas are well correlated in the intermediate latitude
ISM (e.g., \citealp{planck2011-7.12, planck2013-XVII}), it is of
interest to examine also the power spectra of dust maps to see how they
compare with those from \HI.

A dust map records all dust emission along the line of sight,
independent of velocity.  First, we compute a dust-proxy map from the
combination of \nh\ maps of the LVC and IVC weighted by the opacity
$\tau_{\rm dust}$/\nh\ for those two VCs \citep{planck2011-7.12}.  HVC
is excluded because it is observed to have little correlated far-IR or
submillimeter emission.  This proxy is expected to map the dust
correlated with \HI.  A further consideration in these comparisons is
that this proxy does not account for dust associated with molecular
and/or ionized hydrogen.  Exponents for this proxy are recorded in a
central row of Table~\ref{powertable}.

The thermal dust maps that we use for comparison are from the \Planck\
thermal dust model as described in \citet{planck2013-p06b}
and available on the Planck Legacy
Archive.\footnote{\url{http://www.cosmos.esa.int/web/planck/pla}/.}
They have a 5\arcmin\ beam and a 1\farcm7 grid.
For the regions of interest here, the product most relevant for
comparison with the dust proxy is E(B-V)$_{\rm xgal}$ scaled from the
radiance derived using data with point sources removed.
To assess systematic effects we also present results from maps of 
$\tau_{\rm 353}$ (dust optical depth at 353~GHz) from data
including point sources
and maps of the 857\,GHz
intensity used to determine the thermal dust model,
both without and with point sources.
The intensity maps relate to column density with further modulation by
spatial variations in the dust temperature and in the exponent of the
frequency dependency of the opacity.

For each of these dust maps, we selected regions corresponding to the
\dhigls\ white dashed contours, removing a median and applying an
apodization, thus ensuring that all maps have the same masked area and
edge-smoothing.  The deconvolved power spectra of the \dhigls\
dust-proxy maps were fit, as before, with a power law and scaled noise
template.  The power laws from thermal dust products were similarly
fit, first deconvolving by the 5\arcmin\ beam.  Our initial choice of
a white noise model did not appear to apply to all cases.  Instead we
limited the fitted $k$-range to $k < 0.08$~arcmin$^{-1}$, where the
noise is significantly lower than the power law, eliminating the noise
model from the fit altogether.
From an empirical assessment of the effects of such adjustments to the
fitting procedure, we deduce
that there could be a systematic uncertainty of 0.1 in the exponent in
comparison to that from analysis of the corresponding proxy \nh\ map.

The results of the modelling the dust maps are summarized in the lower
part of Table~\ref{powertable}.
In \DFG, the exponents for the submillimeter dust products are
reasonably consistent with that for the \nh\ (LVC+IVC) dust proxy,
although the latter power law appears marginally steeper.
The results for \UMG\ are similar.
On the other hand, there is a much more significant deviation relative
to the exponent for the exponent for the dust proxy in the case of
\ENG.  This arises because this field is an ``extragalactic window"
and so has a very low column density of LVC ($\langle \nhm \rangle =
6\times$\colunits) and correspondingly low Galactic dust emission.
Fluctuations of the cosmic infrared background radiation, which have a
power law exponent about $-1$ \citep{planck2011-7.12}, seriously
contaminate the dust maps, causing the flattening of the power
spectrum ($\sim-1.5$) in this specially selected region.  Increasing
the size of the dust maps of \ENG\ by a factor of 2, thus sampling
brighter Galactic dust emission, results in a steeper power law with
exponent $-2.9$.

\subsection{Dependence on Thickness: Velocity Channel Analysis}
\label{discresultsthick}

Angular power spectra of the ISM gas (and dust) such as discussed
above reflect a self-similarity that has long been associated with a
turbulent cascade \citep{chan49}.  It is not necessarily
straightforward to quantify all aspects of this turbulence directly
from observations.
For example, as seen in Figure~\ref{DFthree}, the CNM structure in the
channel maps changes quickly with velocity, but a channel map does not
necessarily represent a slice of space/distance.  Likewise a spatial
slice need not exhibit the same velocity throughout the map.
In general the pattern of emission in individual channels in the \HI\
velocity cube can reflect a combination of both the density and the
velocity structure of the ISM.
Attempts have been made to disentangle the effects of these two
fields.

For optically thin gas, \citet{lazarian2000} predict that for typical
velocity field exponents the power law exponent of the integrated \hi\
emission should decrease (the spectrum steepens) as the spatial
thickness of the sampled region increases, approaching the power law
exponent of the underlying 3D density field.  The important condition
is that the physical depth ($\delta d$) associated with the VC must
exceed the largest transverse scale $s$ of the map, which can be
obtained from Equation~(\ref{trans1}) below.  The intermediate
Galactic latitude fields \DFG, \UMG, and \ENG\ span about 7\deg,
5\deg, and 2\deg, respectively.  Adopting a distance of about 100~pc
for the LVC in these fields, then from Equation~(\ref{trans1}) below
the transverse extent $s$ of the entire field would be of order 12,
8.5, and 3.5~pc, respectively.

In general, the LVC is composed of WNM, CNM, and thermally unstable
components \citep{saur14}.  The scale height of the WNM, $H_{z} \sim
400$~pc \citep{kalb03}, supporting the view that the range $\delta d$
of gas contributing to the integrated column density significantly
exceeds $s$ for intermediate-latitude LVC gas over regions of the size
observed in \dhigls.  Furthermore, CNM is thought to form within the
WNM via a thermal phase transition.  There is a pressure threshold for
this transition and so $\delta d$ might be somewhat smaller for the
CNM.  Nevertheless, the spatially-thick condition $\delta d > s$ seems
likely to be met here.  This in turn suggests that the measured LVC
power laws ($-3.00\pm0.03$, $-2.45\pm0.17$, $-2.89\pm0.04$ in row 1 of
Table~\ref{powertable}) represent the 3D density power spectrum of the
\HI\ gas.

Using fBm simulations, \citet{mamd2003b} found that increasing the
number of LVC channels being integrated does indeed steepen the power
law as the relative contribution of power on the smallest scales is
reduced.  They find that the exponent decreases by of order unity and
interpret this phenomenon as a measure of the change from 2D to 3D in
the captured topology.

The large velocity range needed to capture a ``complete" VC could
often be associated with a large thickness, although as discussed
velocity does not map uniquely to distance.
To explore such potential variations in the power spectra of the data,
we started with the channel map with peak emission in the standard
deviation spectrum (e.g., Figure~\ref{sigmavelDF}) and added channels
around these peaks, on alternating sides, eventually building up
component maps as in Figures~\ref{DFcomponents} and
\ref{ENcomponents}.  At each step we used the model in
Equation~(\ref{plawd}) to determine the power-law exponent.
As the number of channels in the slice integrated increases, the S/N
first increases and then decreases.  Estimating the relative S/N as a
function of the slice width by $P_0/\eta$ with
$k_0=0.01$\,arcmin$^{-1}$ we found the width at which the S/N was at a
maximum.

\begin{figure}
\centering
\includegraphics[angle=90,width=1.0\linewidth]{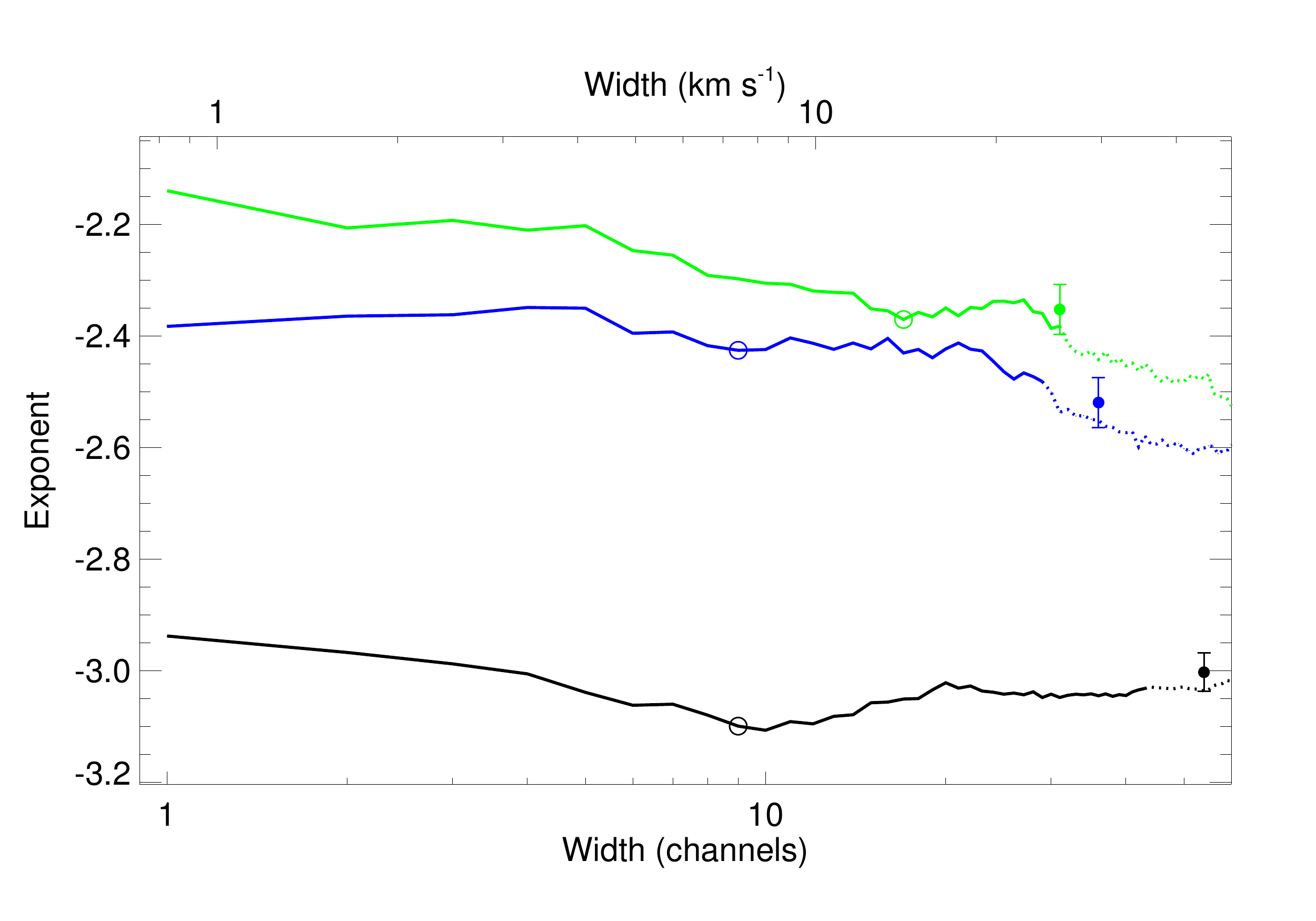}
\caption{
Modelled power-law exponents as a function of width of the velocity
slice for \DFG\ emission within the LVC, IVC1, and IVC2 ranges.
Velocity slices are centered at $3.0$~\kms\ (black), $-27.5$~\kms\
(green), and $-52.0$~\kms\ (blue), respectively
(Figure~\ref{sigmavelDF}).
Trend line is solid where gas is entirely within the range of the VC.
Filled circle, for exponent found fitting the complete VC
(Figure~\ref{componentsDF_ps}), falls slightly off the trend line at
the same width, because the slice centers and centers of the
VC velocity ranges do not quite coincide.
Unfilled circle indicates exponent for the velocity slice for which
S/N of power spectrum is maximum.
}
\label{exponentDF}
\end{figure}

The trend lines in Figure~\ref{exponentDF} show the results for \DFG\
for the three velocity ranges identified in Figure~\ref{sigmavelDF}.
The exponents for IVC1 and IVC2 are similar to one another but
different than for LVC.  We also marked the exponent for the slice
width at which the S/N was maximum and plotted the exponents found in
the power spectrum analysis of \nh\ of the complete VCs
(Figure~\ref{componentsDF_ps}).

Our main result is that there is no significant decrease in the
exponent with width of the slice (a possible surrogate for spatial
thickness).
This is the case for \ENG\ and \UMG\ as well and in contrast to a
decrease of the exponent with slice width by 0.45 found by
\citet{mamd2003} in \URSA\ (see also Appendix~\ref{discdetailsursa}).
However, for the HVC in \ENG\ there appears to be a steepening of the
power spectrum with the inclusion of additional channels, from
exponent $-2.30 \pm 0.09$ for a single channel to $-2.85\pm0.07$ for
the full VC, which is in the sense expected from theory and
simulations.

\subsection{Geometrical Effects Relating to Distance}
\label{discresultsgeom}

Given the angular extent of \DFG, \UMG, and \ENG, the power spectrum
samples well down to about $k$ of 0.005, 0.007, and
0.02\,arcmin$^{-1}$, respectively.
At the other extreme, the S/N is sufficient in each underlying mosaic
to probe the power spectrum readily up to $k$ of about
0.2\,arcmin$^{-1}$, corresponding to a scale of 5\arcmin, somewhat
larger than the synthesized beam.

For an adopted distance of LVC gas of 100~pc, the \dhigls\ data probe
down to scales of 0.14~pc and up to 10~pc in the case of \DFG\ and
\UMG.

The IVC, on the other hand, might lie in the more distant Galactic
halo, say 1~kpc.  The power spectrum in that case would be probing
transverse scales about 10 times larger, perhaps up to 100~pc in the
case of \DFG\ and \UMG.  Noting that the transition from probing 3D to
probing 2D turbulence occurs when $\delta d < s$ 
(here, $s\sim100$~pc),
the relatively shallow spectra seen for the IVC VCs
might indicate that these data are probing such a 2D topology, in
which case the observed exponent would depend on the exponent of the
velocity field as well (also relatively shallow, see
Section~\ref{centroid}).  However, not much is known about the depth
of the IVC.
A typical IVC column density is $3\times$\colunits\ and so volume
density $\vnh = 0.1$~cm$^{-3}$ would translate to depths of 100~pc
using Equation~(\ref{density1}) below.  In Section~\ref{showoff} we
present evidence for significant CNM in the IVC that would indicate
higher $\vnh$ and lower depths within these structures.  However, the
relevant $\delta d$ is the amount by which gas across the region is
spread out along the line of sight, not the depth of individual
features.

The Draco nebula, estimated to be at a distance of about 500~pc
\citep{gladders98}, is an interesting IVC because it shows an
interaction with the embedding Galactic halo (Section~\ref{nemi},
\citealp{mamd2016}).  In the interaction region the gas has become
dense enough to be molecular, and so the depth there might be of order
a pc, smaller than the angular extent of 15~pc.  The power spectrum is
again relatively shallow ($-2.68 \pm 0.07$, Appendix~\ref{drsummary}),
perhaps suggestive of a 2D topology, but $\delta d$ across the region
is not readily known.  Alternatively a shallower behaviour compared to
typical LVC might simply reflect different conditions in the driving
of the turbulence in the two components.

The HVC gas in N1 is part of Complex C at an estimated distance of
$\sim10\pm0.3$~kpc \citep{thom2008}.  Adopting this distance, the
power spectrum would be probing transverse scales up to 300~pc for the
2\deg\ \ENG\ field.  The power spectrum exponent is not atypical, and
so if revealing a 3D topology would imply $\delta d > 300$~pc and a
mean density $<~0.01$~cm$^{-3}$.  From the line profiles, it appears
that the HVC component could be divided into a broad (WNM-like) and
narrow (CNM-like) component.  The narrow-line components are quite
compact, some about 3\arcmin\ or 10~pc at the adopted distance.  These
individual features have \nh\ about $7\times$\colunits\ and so a
typical $\vnh$ of 3~cm$^{-3}$, perhaps greater locally if clumpy below
the beam scale.
This is much larger than the average density, but does not necessarily
imply $\delta d < s$ because again the relevant $\delta d$ is the
amount by which the gas, including the condensed structures
(filaments, nuggets), is spread out along the line of sight.
Furthermore, the turbulent conditions in this infalling gas could be
quite different from Galactic LVC gas, making interpretation more
challenging.

\subsection{Power Spectra of Maps of Centroid Velocity}
\label{centroid}

We also analysed the centroid velocity maps for all VCs in \DFG, \ENG,
and \UMG\ (Figures~\ref{componentsDFv_ps}, \ref{componentsENv_ps}, and
\ref{componentsUMv_ps}, respectively), with the goal of determining
the power spectrum of the 3D velocity field \citep{mamd2003b}.  To
quantify the power spectrum the centroid velocity data were
deconvolved by the \draost\ beam and modelled with a power law.
Interestingly, the noise component at high $k$ is similar to the noise
template used for the \nh\ power spectrum models up to $k\sim0.8$;
however, the shape is not as consistent and so we chose to fit a power
law model without a noise template.  This precludes the use of high
$k$ power in the model fit, limiting the range to $k <
0.06$~arcmin$^{-1}$.  An associated systematic uncertainty in the
exponent in comparison to analysis of \nh\ maps of the same VC could
amount to 0.1.

As summarized in Table~\ref{powertable}, the power law exponents for
the centroid velocity maps are for the most part the same as those
found for the \nh\ maps, within the uncertainties.  This interesting
empirical result is not a general expectation of theory and implies
that the 3D velocity field and the 3D density field (see discussion in
Section~\ref{discresultsthick}) have rather similar statistical
properties.  An exception might be the centroid velocity map of \DFG\
IVC, which being a combination of IVC1 and IVC2 with distinct
velocities and spatial structure is more complex.

\begin{figure}
\centering
\includegraphics[clip=true,trim=29 0 0 0,angle=0,width=1.045\linewidth]{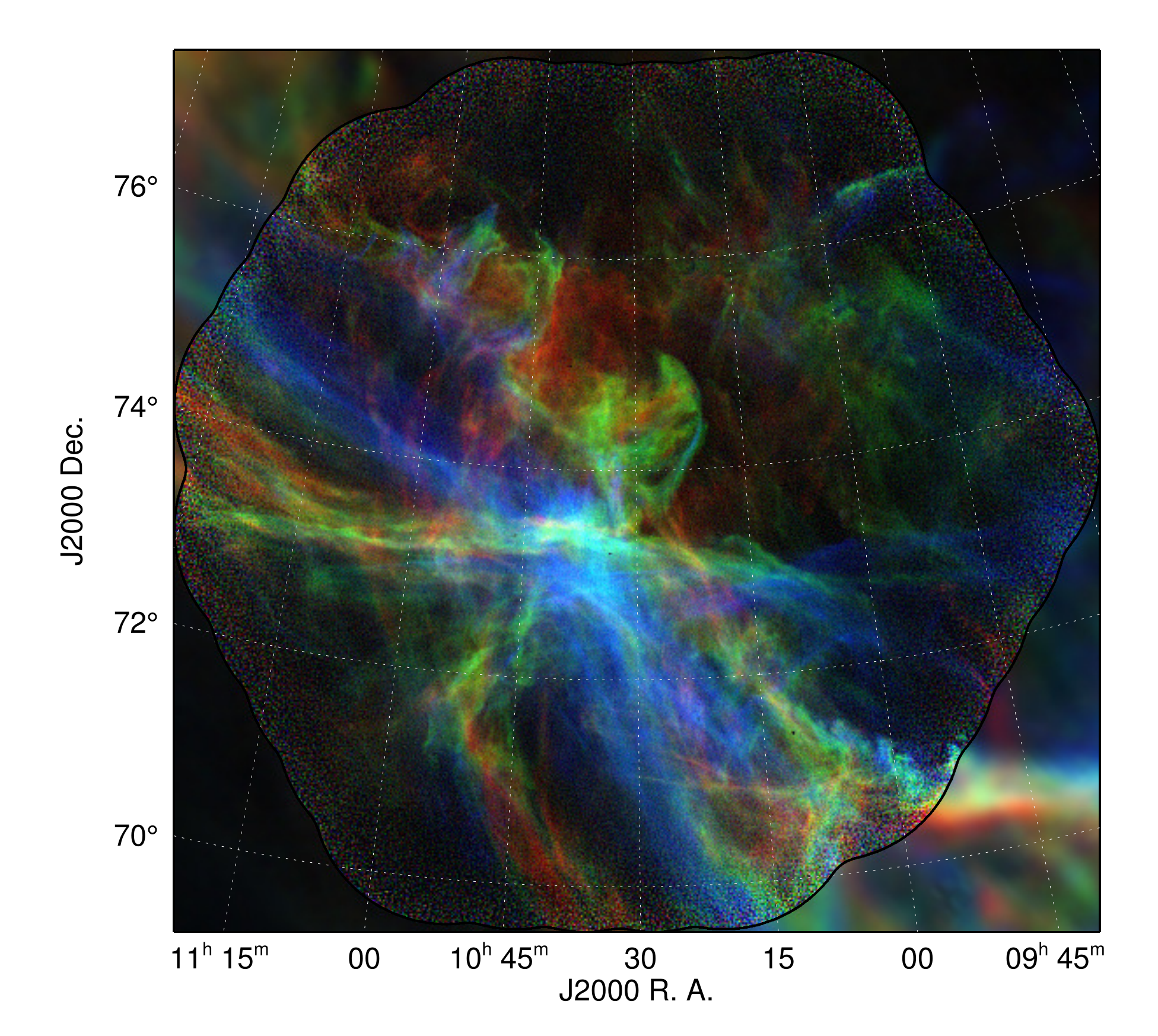}
\includegraphics[angle=0,width=0.8585\linewidth]{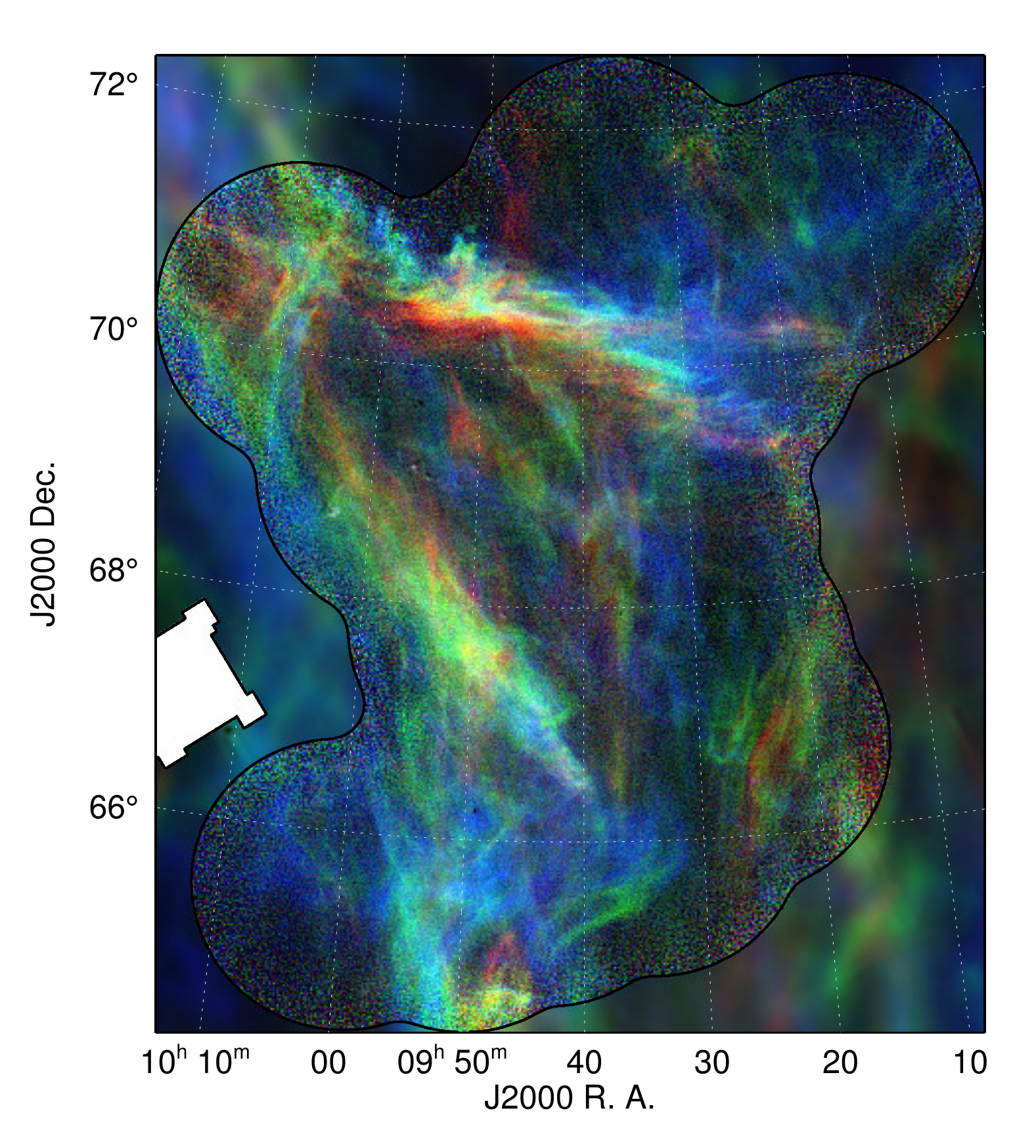}
\caption{ 
Illustration of dramatic changes in channel maps with velocity in
\DFG\ (upper) and \UMG\ (lower).  RGB images were made using three
distinct LVC channels from the cubes at 7.59~\kms\ (red), 5.12~\kms\
(green), and 2.64~\kms\ (blue); for each color the identical intensity
range was used, from 0~K (black) to 40~K (color saturation).
Upper left portion of \UMG\ image joins onto lower right of \DFG\
image (see also Figure~\ref{field_locations}).  In parts of the
overlap the different angular resolutions of the \dhigls\ and the
\ghigls\ data can be appreciated (a black curve separates the two).
}
\label{DFthree}
\end{figure}

\begin{figure}
\centering
\includegraphics[clip=true,trim=29 0 0 0,angle=0,width=1.045\linewidth]{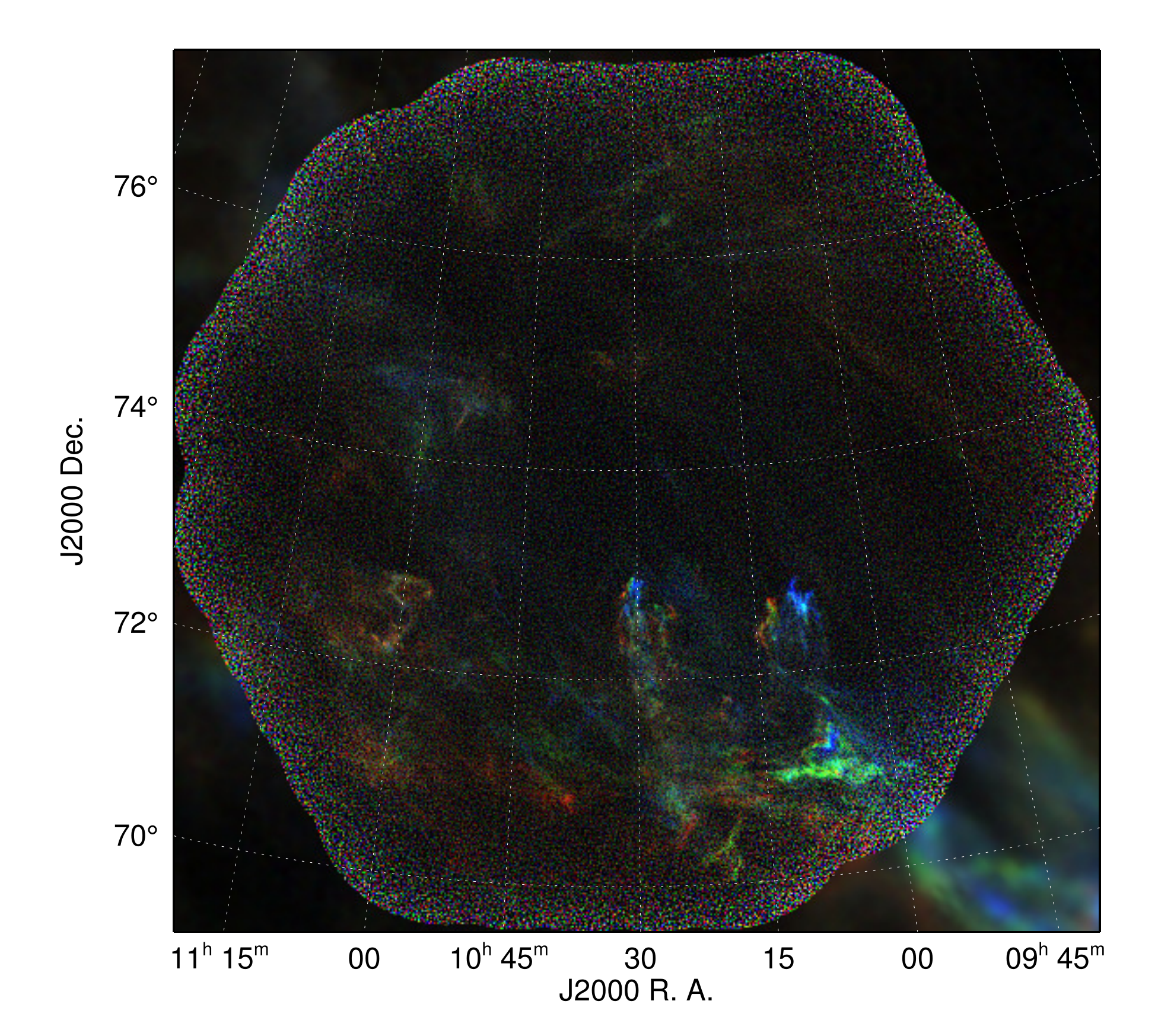}
\includegraphics[angle=0,width=0.8585\linewidth]{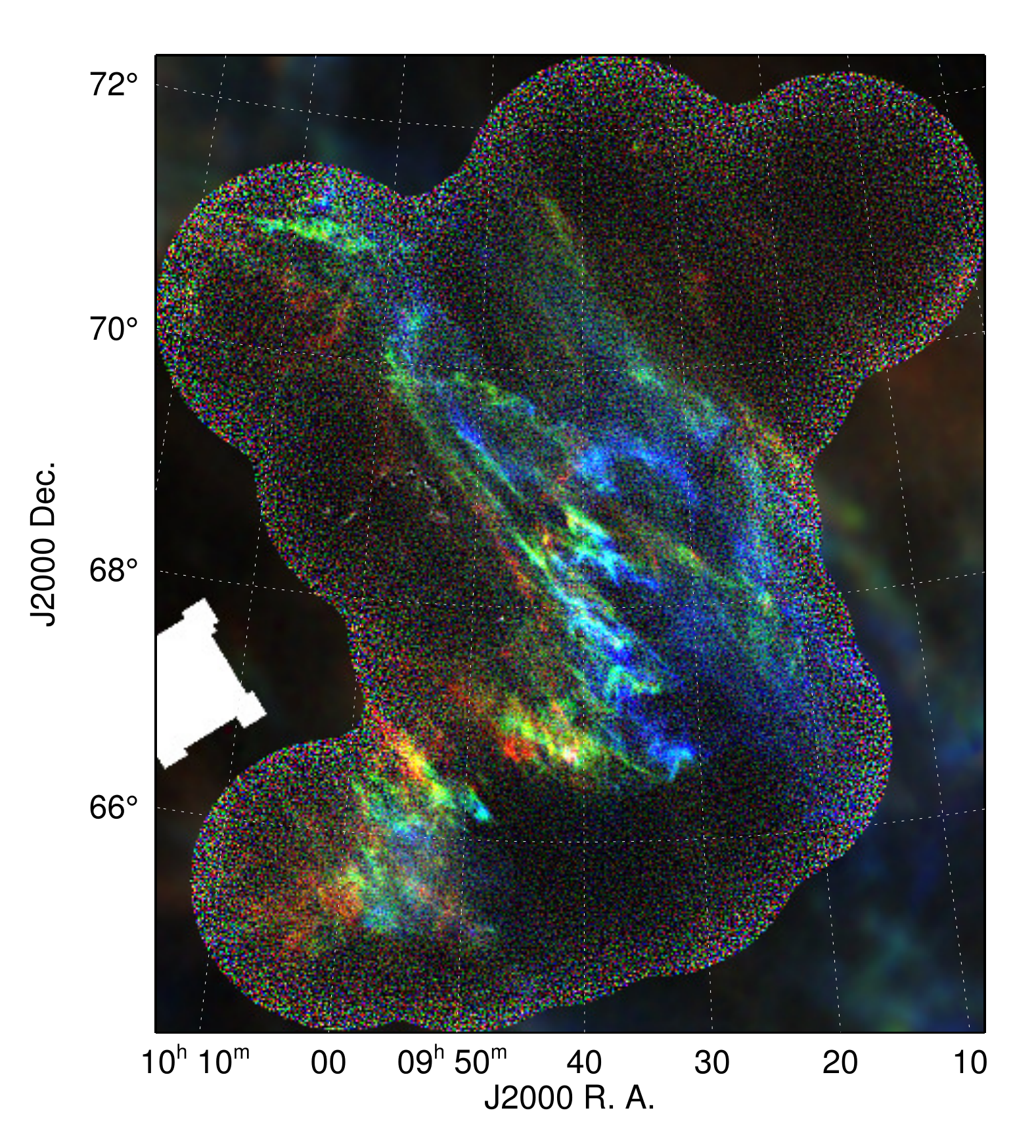}
\caption{ 
Similar to Figure~\ref{DFthree} but for IVC channels with intensity
extremes at 0~K and 15~K because the IVC emission is fainter.  The
three distinct channels used from the cubes were at $-49.28$~\kms\
(red), $-51.76$~\kms\ (green), and $-54.23$~\kms\ (blue).
}
\label{DFthreeIVC}
\end{figure}

\begin{figure}
\centering
\includegraphics[angle=0,width=1.0\linewidth]{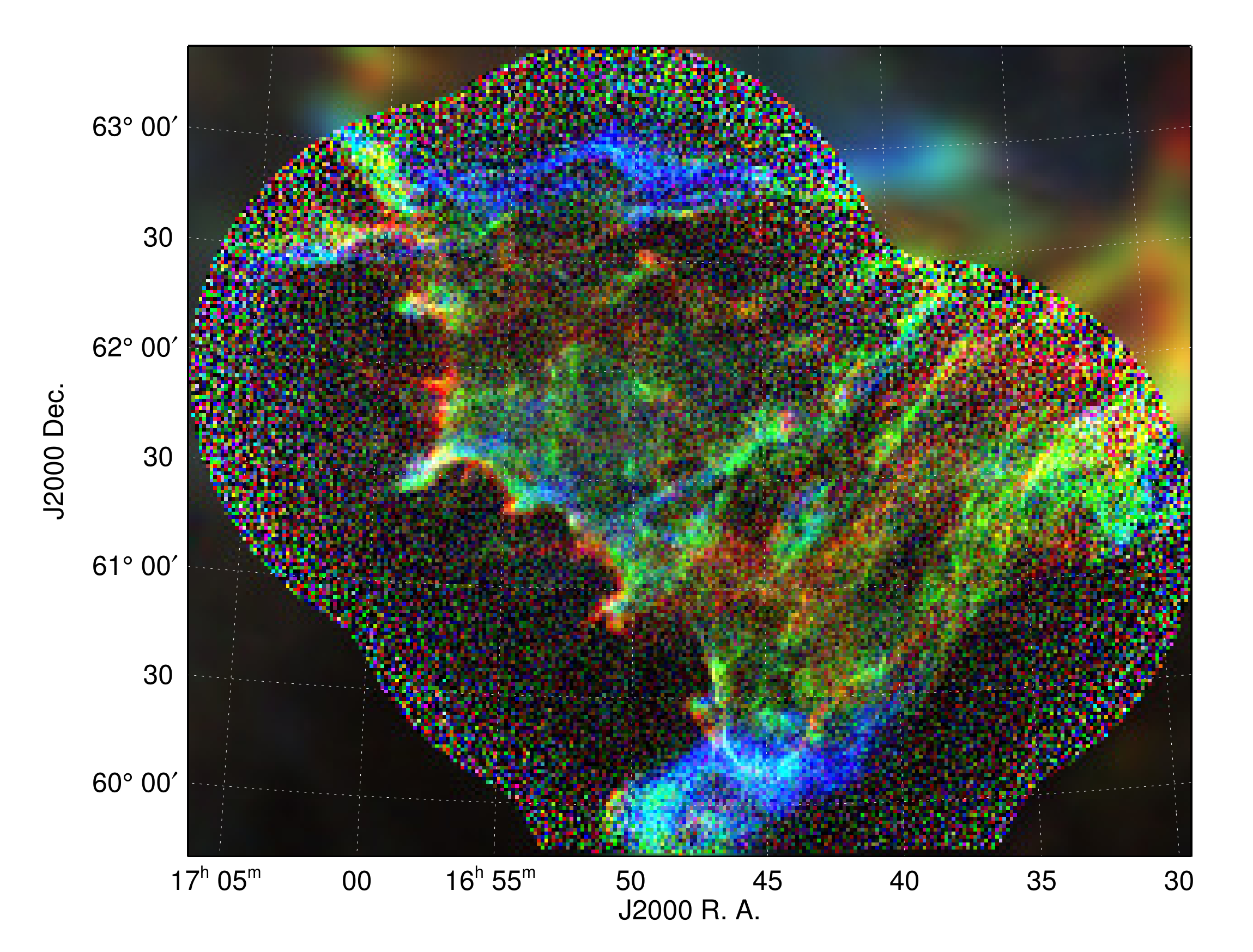}
\caption{ 
Similar to Figure~\ref{DFthree}, for \DRG\ IVC channels with intensity
extremes at 0~K and 15~K as for the IVC in Figure~\ref{DFthreeIVC}.
The three distinct channels used from the cube were at $-19.61$~\kms\
(red), $-22.08$~\kms\ (green), and $-24.56$~\kms\ (blue).
}
\label{DRthree}
\end{figure}

\begin{figure}
\centering
\includegraphics[angle=0,width=1.0\linewidth]{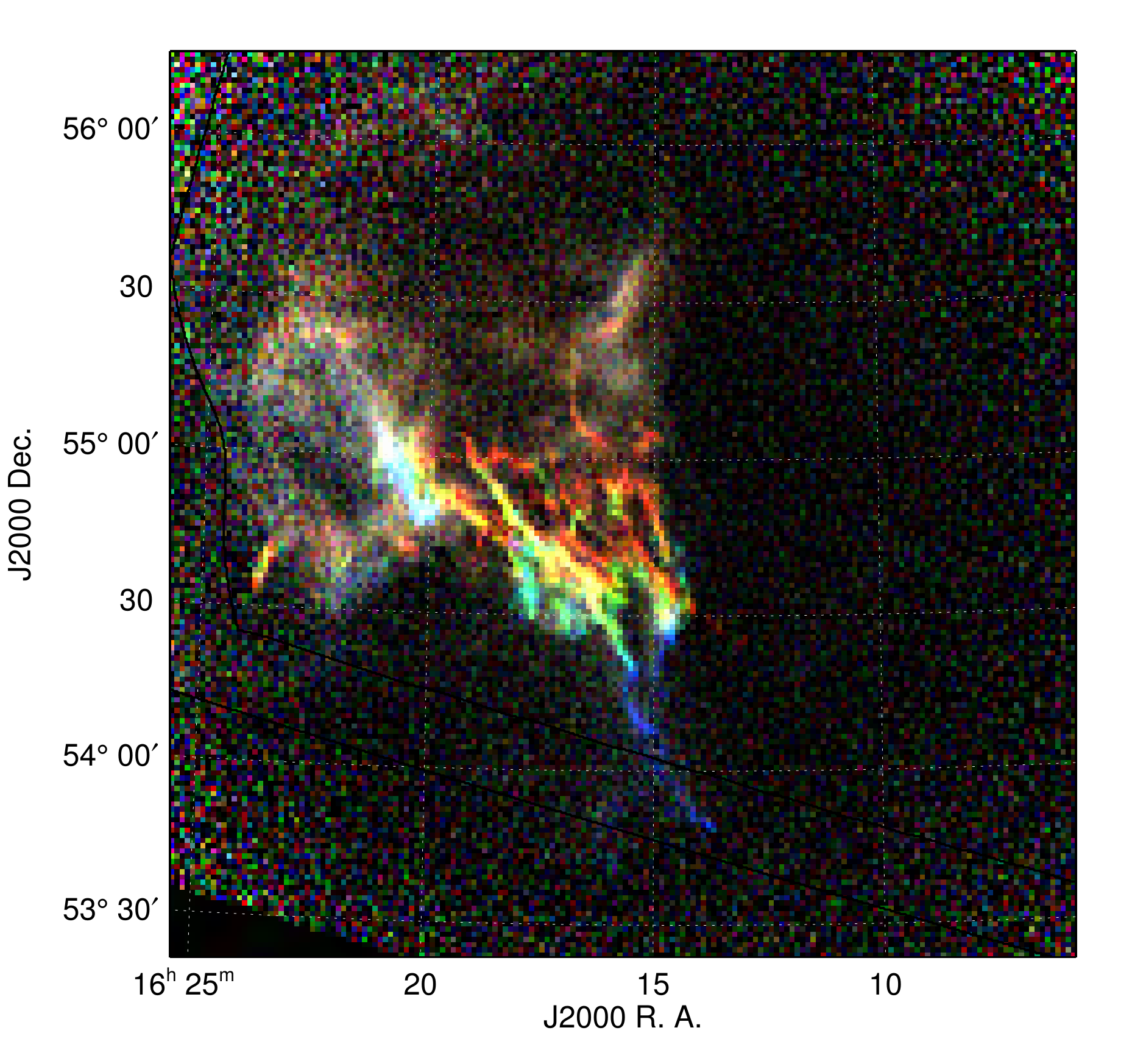}
\caption{ 
Similar to Figure~\ref{DFthree}, for \ENG\ HVC channels with intensity
extremes at 0~K and only 5~K because the HVC emission is relatively
weak.  The three distinct channels used from the cube were at
$-120.17$~\kms\ (red), $-122.64$~\kms\ (green), and $-125.12$~\kms\
(blue).
}
\label{ENthree}
\end{figure}

 \section{Structures Associated with Narrow Emission and Absorption Lines}
 \label{showoff}

For the intermediate latitude lines of sight targeted, the \dhigls\
data reveal the distribution of \hi\ in unprecedented detail, to
sub-pc scales for the LVC gas.  In this section we provide a glimpse
of this rich information to encourage further exploration of the
cubes.  The empirical relationship of the diffuse \hi\ emission and
the fine structure in both intensity and velocity should provide
valuable insight into the neutral atomic component of the multiphase
ISM \citep{cox05,wolfire2015} when combined with relevant numerical
modelling incorporating the essential physics of the turbulent
magnetized interstellar medium
\citep[e.g.,][]{bry14,dob14,hen14,kim14,kim15} and including the
influence of dynamics on the transition to molecular hydrogen
\citep{ste14,val16}.  Likewise, the new dynamical information
illustrated on the interaction of IVC and HVC gas in the Galactic halo
should provide important constraints for accurate numerical models of
these phenomena in 3D \citep[e.g.,][]{hp09,sch15}.

\subsection{Structure in Channel Maps}
\label{nemi}

Examination of the \hi\ data cubes shows that the structure in the
channel maps changes dramatically over even small changes of velocity.
At the spatial positions of these distinctive structures, the line
profiles have significant narrow components.
To illustrate this we made RGB color images of the \hi\ emission using
three distinct channels offset by 2.47~\kms.  Examples are given for a
variety of environments:
LVC gas in the overlapping \DFG\ and \UMG\ regions in the \ncpl\
(Figure~\ref{DFthree}); IVC gas in these same regions
(Figure~\ref{DFthreeIVC}) and in \DRG\ (the Draco nebula,
Figure~\ref{DRthree}); and HVC gas in \ENG\ (Figure~\ref{ENthree}).

For each figure, the intensity ranges for the channels represented by
different primary colors are identical.  Therefore, 
when the line profiles are sufficiently narrow 
and the emission at different velocities does not overlap spatially,
the image appears red, green, or blue.  With partial overlap at
similar intensity, the image appears yellow or cyan (other blending
possibilities resulting in magenta or white are rare in these
examples).

The channel offset used, 
2.47~\kms, corresponds to the FWHM of a thermally broadened \hi\ line
at a kinetic temperature of about 130~K, even without allowance for
instrumental and turbulent broadening.  The fine structure
that appears in a single colour in this visualization comes from
narrow line components of CNM gas, as can be corroborated by
independent estimates of the density and spin temperature of the gas
(Sections~\ref{vol} and \ref{nabs}).

The LVC \hi\ seen in Figure~\ref{DFthree} has a turbulent and
filamentary nature; angular power spectra of single channel maps are
illustrated in Figures~\ref{noiseDFdrao_ps}, \ref{noiseOTdrao_ps}, and
\ref{noiseDF_ps}.

In the IVC range the channel maps of \DFG\ and \UMG\ near $-50$~\kms\
(Figure~\ref{DFthreeIVC}) have a streaky appearance, suggestive of a
preferred orientation for the structures (approximately along constant
Galactic latitude).  Another distinguishing feature, compared with the
appearance in the LVC range, is higher contrast frothy small-scale
structure, also seen in \POG\ near $-21$~\kms
(Figure~\ref{noisePOdrao}).  These maps have slightly flatter power
spectra (see Figure~\ref{noiseOTdrao_ps} and also
Figure~\ref{componentsDF_ps} and Table~\ref{powertable}).  Power
spectrum analysis for images with such exceptional directional
structure is discussed in Appendix~\ref{ps2D}.

The color image of the Draco nebula, the IVC in \DRG\
(Figure~\ref{DRthree}), reveals some particularly interesting
features.  Relative to a diagonal running north east to south west
(upper left to lower right)
there is IVC gas to the upper right and a void, at those velocities,
to the lower left.  It has been suggested that infalling higher
velocity gas is interacting with more tenuous gas at high altitude in
the Galactic disk \citep{goerigk83}.  The \dhigls\ observations show
the complexity of this interaction.  Nevertheless, there are some
important systematics that will be important to model.  For example,
in the central portion along this boundary there is a change in the
characteristic velocity of the narrow line emission across the
``leading edge," from red to green (in radial motion approaching more
quickly) in Figure~\ref{DRthree} on an arc minute (0.15\,pc) scale, as
one passes from lower left to upper right.
Along this edge the density is enhanced locally to the extent that CO
is present and detectable in the \Planck\ CO maps
\citep{planck2013-p03a}.  The corrugated structure of the edge might
result from hydrodynamic instability in the interaction region
\citep{mamd2016}.

In \ENG\ the LVC has no prominent small-scale structure except for a
filament at $-7$~\kms\ at $16^{\rm h}10^{\rm m}, +54\deg10$\arcmin;
this filament was studied earlier with the GBT \citep{mamdmart2007}
and is seen much more clearly in our \dhigls\ data with the \draost\
resolution.
By contrast, the HVC image in Figure~\ref{ENthree} reveals an
intricate pattern of coherent narrow ribbons of emission at each
particular velocity, with the structure in this velocity range
shifting systematically with increasingly negative velocity to the
lower right, along the general direction of the ribbons.  The HVC gas
in \ENG\ is part of Complex C at an estimated distance of $\sim
10\pm2.5$~kpc \citep{thom2008}.

\subsection{Volume Density}
\label{vol}

It is useful to check that the volume density in these structures is
relatively high, as would be expected in the CNM as compared to the
WNM.

If a column density \nh\ arises in a physical depth $\delta d$, then
the volume density $\vnh$ is given by\footnote{
Quantities in square brackets define the units of the parameters in
the ratios in the following equations and are chosen so that the
ratios are not too far from unity.}
\begin{equation}
\label{density1}
\vnh = 3.2\, \frac{\nhm}{[10^{\colnum}\, {\rm cm}^{-2}]}\, \frac{[1\,{\rm pc}]}{\delta d} \,\, {\rm cm}^{-3} \, .
\end{equation} 

A structure of angular size $\psi$ at distance $d$ has a transverse
size
\begin{equation}
\label{trans1}
s  = 0.29\, \frac{\psi}{[10\arcm]}\, \frac{d}{[100\, {\rm pc}]} \,\, {\rm pc}\, .
\end{equation} 

Supposing that $\delta d$ can be estimated from the narrowest
dimension of the structure, $\psi_{\rm n}$, then
\begin{equation}
\label{density2}
\vnh = 11\, \frac{\nhm}{[10^{\colnum}\, {\rm cm}^{-2}]}\, \frac{[10\arcm]}{\psi_{\rm n}}\, \frac{[100\, {\rm pc}]}{d} \,\, {\rm cm}^{-3} \, .
\end{equation} 

For an optically thin \hi\ line component approximated by a Gaussian
specified by its peak brightness temperature $\Trp$ and FWHM,
\begin{equation}
\label{column1}
\frac{\nhm}{[10^{\colnum}\, {\rm cm}^{-2}]} = 16\,    \frac{\Trp}{[20\, {\rm K}]}\, \frac{{\rm FWHM}}{[4\, {\rm km\,s}^{-1}]} \, ,
\end{equation} 
and 
\begin{equation}
\label{density3}
\vnh = 174\, \frac{\Trp}{[20\, {\rm K}]}\, \frac{{\rm FWHM}}{[4\, {\rm km\,s}^{-1}]}\, \frac{[10\arcm]}{\psi_{\rm n}}\, \frac{[100\, {\rm pc}]}{d} \,\, {\rm cm}^{-3} \, .
\end{equation}

\begin{table}
\caption{Parameters of select structures and derived volume density}
\label{workedexamplesden}
\centering
\begin{tabular}{lccccc}
\hline \hline
Feature & Distance & FWHM & $T_p$ & $\psi$ & $\vnh$ \\
             &  (pc)         & (\kms) & (K)      & ($\arcmin$)  & (${\rm cm}^{-3}$)  \\
 \hline
{\DFG}a & 100-200\tablenotemark{a} & 4 & 20 & 15-30 & 30-120 \\
{\DFG}b & 100-200\tablenotemark{a} & 3 & 4 & 4 & 40 \\
{\ENG}a & 150\tablenotemark{b} & 4 & 10 & 6 & 100 \\
{\ENG}b & 10000\tablenotemark{c} & 4 & 15 & 3 & 4 \\
\hline
\end{tabular}
\tablenotetext{1}{\ncpl\ LVC distance, \citet{meyer91}}
\tablenotetext{2}{LVC distance, \citet{mamdmart2007}}
\tablenotetext{3}{HVC distance, \citet{thom2008}}
\end{table}

Parameters and derived volume densities for a few worked examples are
presented in Table~\ref{workedexamplesden}.  Consider first the green
``face'' and filaments in \DFG\ (Figure~\ref{DFthree}, upper), which
we tabulate as {\DFG}a.  From Equation~(\ref{density3}), we find a
volume density in the range $\vnh \sim 30 - 120\, {\rm cm}^{-3}$,
somewhat denser than the conditions for CNM gas modelled in
\citet{wolfire2003}.

\DFG\ also hosts some remarkably long narrow structures.  Among these,
in the left hand side of channel maps near 9~\kms, is one ({\DFG}b)
that runs over 3\deg\ along $\delta \sim +73\pdeg5$ but is barely
resolved by the \draost\ synthesized beam, indicating an axial ratio
of over 100.  From the parameters in Table~\ref{workedexamplesden},
$\vnh \sim 40\, {\rm cm}^{-3}$ if it is indeed a filament in 3D.

Likewise, for the above-mentioned LVC filament in \ENG\ ({\ENG}a),
we find $\vnh \sim 100\, {\rm cm}^{-3}$.

The HVC ribbons seen in {\ENG} are neutral condensed structures with a
density much higher than the estimated average, $\vnh \sim 0.03\, {\rm
cm}^{-3}$ \citep{thom2008}, with a low covering factor.  Furthermore,
there is clear substructure along the ribbons.  The brightest such
nugget ({\ENG}b) at $16^{\rm h}20^{\rm m}37^{\rm s}, +54\deg
55\arcmin\ 25$\arcsec\ in the $-125$~\kms\ channel map, has $\vnh \sim
4\, {\rm cm}^{-3}$.  The integrated column density of even this
brightest nugget is still quite low, \nh\ $\sim 10^{20}\, {\rm
cm}^{-2}$ from Equation~(\ref{column1}), requiring the angular
resolution and sensitivity of the \dhigls\ product to bring out the
fine detail across the region.

\subsection{Spin Temperature from Absorption against Radio Galaxies}
\label{nabs}

\hi\ absorption of bright background radio sources
\citep[e.g.,][]{roy13,mur15}
can be found in \dhigls\ spectra.  The absorption coefficient of \hi\
varies inversely with spin temperature $\Trs$, favouring the detection
of cold gas.

Thanks to the relatively small size of the synthesized beam ($\theta_{\rm sb}$ in
arc seconds in Table~\ref{beamsize}) some radio point sources are
detected with high continuum brightness temperatures $\Trc$.  For a
source of flux density $S_{\nu}$,
\begin{equation}
\Trc =  170\, \left(\frac{[60\arcsec]}{\theta_{\rm sb}}\right)^2 \, \frac{S_{\nu}}{[1 {\mathrm {Jy}}]}\, {\rm K}    \,.
\label{tcon}
\end{equation}
\hi\ absorption reduces this background emission to $\Trc
\exp(-\tau)$.  Because of 
continuum
removal in the processing of the spectra, this results in a
negative-going feature in the spectrum, of depth $\tau T_{\mathrm c}$
for small optical depth.

When $\tau$ is small, the net temperature $\Trn$ recorded in the
\dhigls\ spectrum is therefore
\begin{equation}
\Trn =  \Trb (1 - \Trc/\Trs)  \,,
\label{tnet}
\end{equation}
which becomes negative when $\Trc > \Trs$ or $S_{\nu} > 0.4$~Jy for
$\Trs = 80$~K.
From the NVSS catalog browser\footnote{
\url{http://www.cv.nrao.edu/nvss/NVSSlist.shtml}} 
we find about 0.25 such sources per square degree and although this is
small we have found instances of favorable alignments producing the
expected narrow absorption features.\footnote{
We also note that fitting the position and angular extent of the
absorption features in channel maps where they are prominent confirms
the registration of the mosaics and expected size of $\theta_{\rm
sb}$.}

\begin{table}
\caption{Parameters\tablenotemark{a} of select absorption features and derived $\Trs$}
\label{workedexamplests}
\centering
\begin{tabular}{lccrrcc}
\hline \hline
Region & NVSS Source & $\Trc$ & $\vlsr$  & $\Trn$ & $\Trb$ & $\Trs$\\
 \hline
\DFG  & J$101132+712440$ & 380 & $+5.1$ & $-70$ & 25 & 84 \\
\UMG & J$094912+661459$ & 444 & $+2.6$ & $-106$ & 25 & 85 \\
\UMG & J$094912+661459$ & 444 & $-53.0$ & $-58$ & 13 & 80 \\
\DRG & J$164829+600722$ &   41 & $-23.7$ &  $+5$ & 18 & 55 \\
\hline
\end{tabular}
\tablenotetext{1}{Velocity in \kms\ and temperatures in K}
\end{table}

In \DFG\ a prominent example is 4C~$+71.09$ (NVSS J$101132+712440$)
which can be seen as a dark dot against emission at 5.12~\kms\ (green)
in Figure~\ref{DFthree} (upper).  For this source, $\Trc = 380$~K.
The spectrum is most negative at 5.94~\kms\ with $\Trn = -70$~K.  The
FWHM of the absorption feature is about 3.5~\kms.  $\Trb$ interpolated
spatially in the channel map is 25~K, with about 20~K attributable to
narrow emission of FWHM 4~\kms, slightly broader than the absorption.
Rearranging Equation~(\ref{tnet}), the spin temperature can be
estimated from
\begin{equation}
\Trs  =  \Trc\, \Trb /(\Trb  - \Trn)    \,.
\label{tspin}
\end{equation}
For this example, using the numbers collected in
Table~\ref{workedexamplests}, this evaluates to $\Trs =84$~K and
confirms that the optical depth is appropriately small ($<0.25$) and
that this is CNM gas as suggested above on the basis of $\vnh$.  That
the line width is slightly broader than thermal is usually attributed
to turbulence, which also appears to play a role in the pressure
distribution \citep{wolfire2015}.

There are many more examples of absorption indicating cold CNM gas in
the LVC range.  In \UMG, 4C~$+66.09$ (NVSS J$094912+661459$) can be
seen deeply absorbed at 2.64~\kms\ (blue in Figure~\ref{DFthree},
lower) with FWHM about 3.8~\kms\ as for nearby emission.  Using the
values in Table~\ref{workedexamplests}, $\Trs = 85$~K.

Although lines of sight to bright radio sources rarely intersect
sufficiently bright IVC gas, which has a lower covering factor than
the LVC gas, there are a few examples in the \dhigls\ data.
Toward 4C~$+66.09$ in \UMG\ the deepest absorption in the IVC range is
at $-53.0$~\kms\ (cyan in Figure~\ref{DFthreeIVC}, lower), with FWHM
about 3~\kms\ (absorption and emission).  In this IVC feature, $\Trs =
80$~K.  Using $\psi_{\rm n} \sim 10\arcmin$ and taking $d <500$~pc, we
find $\vnh > 20\, {\rm cm}^{-3}$.  Thus the IVC gas that reveals the
frothy and streaky structure across the whole region is dense like CNM gas.

In \DRG, absorption of the weaker source NVSS J$164829+600722$ still
produces a relative deficit visible in the $-23.7$~\kms\ emission
(cyan) in Figure~\ref{DRthree} and we find $\Trs \sim 55$~K.  This gas
is part of the prominent higher column density region in the south
which is molecular \citep{herb93}.

For the HVC range, because of factors like insufficient spectral
coverage, faintness of emission, and low covering factor, there are no
examples of absorption in \dhigls\ data.

\section{Summary}
\label{conclusions}

We report on high resolution ($\sim1\arcmin$) \hi\ data from the
\draost\ acquired to complement the low resolution \ghigls\ data from
the GBT \citep{mart15} and higher resolution dust emission data such
as are available from \Herschel/SPIRE surveys and \Planck.  The
\dhigls\ data probe intermediate-latitude gas of moderate column
density over several distinct velocity components comprising local gas
in the Galactic disk, the ``Galactic fountain,'' and infalling \hi\
from outside the Galaxy.

Following calibration of all of the \draost\ data to a common NVSS
scale using point sources in each field, an additional single cross
calibration factor is found to adjust \draost\ data to the \ghigls\
scale.  Thanks to the sensitivity and resolution of the \ghigls\ data,
this factor can be found from a direct comparison of the Fourier
transformed data in the region where the spatial frequencies of the
two instruments overlap.\footnote{
As described in Appendix~\ref{appen:short} this combination of
facilities is unique in the significant overlap in spatial frequency.}
The cross calibration factor is 1.12.  This enables an accurate
combination of the short-spacing information from \ghigls\ with the
interferometric \draost\ data to produce the \dhigls\ products.
Disregarding this adjustment would result in artificially steep power
spectra in the combined data.

We present a power spectrum analysis of the \hi\ structure, based
first on maps of integrated emission or column density \nh\ for
several distinct ranges of velocity.  In the analysis we make use of a
power law model of the \hi\ signal in combination with models of the
noise and of the effects of the synthesized beam at high spatial
frequencies.  For the four large regions incorporating many \draost\
pointings, we find a range in power spectrum exponents from $-2.5$ to
$-3.0$, with IVC tending to lie toward the shallower end and LVC
somewhat steeper.  The HVC gas is well characterized in the \ENG\
region and has an exponent toward the high end of this range.
For LVC gas, increasing the number of channels integrated to form the
\nh\ map, from one to the number defining a full velocity component,
does not produce any large (i.e., order unity) decrease of the power
law exponents.
These results and trends are complemented by a power spectrum analysis
of the centroid velocity maps of these components, which implies that
the 3D velocity field and the 3D density field have similar
statistical properties.

The \dhigls\ regions contain distinctive structure at low,
intermediate, and high velocities down to the resolvable scale.
Illustrative figures are provided that demonstrate that this structure
can change dramatically even over a few channels.  This can be
appreciated in the movies of the data cubes on the DHIGLS archive as
well.  Estimates of the volume density associated with this
narrow-line emission are relatively high, consistent with this gas
being in the phase called the CNM.  Where this can be probed using
absorption against background radio galaxies, the spin temperature is
found to be low, again typical of CNM gas.  Examples of absorption are
found in the LVC and IVC ranges.  Because of several conspiring
factors, there are no examples of absorption in the HVC range in the
\dhigls\ data.

A digital archive at \url{\dhiglsarchive} contains the \dhigls\ \hi\
data cubes and derived products.  On the digital archive the FITS data
cube for each region has extensions as follows:
0, the cube of spectra, $\Trb$ in K; 
1, the modified weight map (closely related to the noise map,
Section~\ref{mosaicnoise}) of the \draost\ mosaicked data (e.g.,
Figure~\ref{noiseDFdrao} bottom), set to zero to indicate the area
in which the data are simply the lower-resolution \ghigls\
data, not the new \dhigls\ product (Section~\ref{combinegbtdrao});
and
2, a mask carried over from the \ghigls\ data recording the few pixels
in that data with no baseline fit removed from the spectrum (\dhigls\
data could be interpolated to replace these spectra if necessary).
For a preview of the \hi\ structure in the cube, we provide a movie
showing successive channel maps over the entire velocity range.
We also include FITS cubes whose planes are \nh\ maps ($\Trs = 80$~K)
for different velocity VCs (see Table~\ref{compvel_table}).  The
standard deviation spectra used to choose the velocity ranges, as in
Figure~\ref{sigmavelDF}, are presented as well.

\section*{Acknowledgments}

We acknowledge support from the Natural Sciences and Engineering
Research Council (NSERC) of Canada and thank the DRAO staff for their
outstanding support of this ambitious project.
We also thank the referee for constructive comments that have led
to improvements in the presentation of our results.


\begin{appendix}

\counterwithin{figure}{section}
\counterwithin{table}{section}

\section{Implementation of Power Spectrum Analysis}
\label{appdfmps}

Here we provide details of how we computed the power spectrum and
extracted quantitative information, building on work described in
\citet{mamd2007} and \citet{mart15}.

In practice in computing the 2D power spectrum, $P(k_x,k_y)$
(Equation~(\ref{amplitude})), we first altered the image $f(x,y)$ by
subtracting the median and then apodizing using a cosine function
along its rectangular boundary.  These both reduce edge effects
inherent to any Fourier transform of a finite nonperiodic function,
which otherwise would appear as excess power along the $u=0$ and $v=0$
axes \citep{MivilleDeschenes2002}.  Apodization in the image plane
over five pixels at each edge is normally sufficient to suppress this
``centered cross" in $P(k_x,k_y)$ in the \uv\ domain, but even after
apodization the potential artificial cross should be monitored and if
necessary masked in 2D.
For the collapsed 1D power spectrum, notionally formed by azimuthal
averaging, taking the median rather than the average is an effective
complement to apodization and mitigation against any residual cross,
and to be conservative we adopted that strategy.

Because of the asymmetry of the synthesized beam
(Table~\ref{beamsize}), we chose to deconvolve the power spectrum
moduli in 2D in the \uv\ domain before calculating the 1D power
spectrum, distinguished as $P_{\rm d}(k)$.

Figure~\ref{noiseDFdrao_ps} shows the 1D power spectrum for the \hi\
channel map in Figure~\ref{noiseDFdraohi} for \DFi, with and without
deconvolution.

\subsection{Power Spectrum of the Noise}
\label{noisetemplate}

The noise in a mosaic made using the interferometric data from the
\draost\ is quite complex.  The spectral noise is not uniform over the
map (see example in upper panel of Figure~\ref{noiseDFdrao}).  The
power spectrum of an emission-free channel is not flat (see example in
Figure~\ref{noiseDFdrao_ps}) but it has a consistent shape and scale
for a given mosaic.
For example, at values of $k$ corresponding to antenna spacings of
$34L$ to $38L$ ($D\sim146$ to $163$~m or $k\sim0.20$ to
$0.22$\,arcmin$^{-1}$), the noise is slightly reduced because the
strategy for the separation of the moveable antennas and their
discrete movements results in twice as much coverage at these antenna
spacings.  This can be seen clearly in the power spectrum of the
emission-free channel
in Figure~\ref{noiseDFdrao_ps}.
At higher values of $k$ the desired \hi\ signal is being reduced by
the synthesized beam and this example shows how, even in relatively
bright intermediate Galactic latitude regions like \DF, the noise
dominates at large $k$.
As described in Section~\ref{drao_observations}, a Gaussian taper was
applied in the Fourier domain, apodizing the visibilities at high $k$.
In Figure~\ref{noiseDFdrao_ps} it can be seen how this taper has
affected the level and shape of the power spectrum of the noise above
$k = 0.4$\,arcmin$^{-1}$.

We exploit the consistent shape to calculate a median of the 2D
power spectra for a set of emission-free channels for each mosaic.  
We call this the ``noise template."  After deconvolution by the
relevant beam, we form the 1D equivalent, $N_d(k)$.  At high $k$ this
is greatly amplified compared to the version without deconvolution,
$N(k)$.
Because the noise is always larger when there is an \hi\ signal
\citep[e.g.,][]{mart15}, this template needs to be scaled by a fitting
factor $\eta >1$, as in Equations~(\ref{plaw}) and (\ref{plawd})
below.

\subsection{Power Spectrum Models}
\label{psmodel}

The 1D power spectrum of deconvolved data can be described with a
parameterized model consisting of a power law representation of the
signal plus a noise component:
\begin{equation}
P_{\rm model,d}(k) = P_{0} \; (k/k_0)^{\expon} + \eta N_{\rm d}(k) \,,
\label{plawd}
\end{equation}
where $P_{0}$ is the amplitude of the power law at some representative
scale $k_0$ and $\expon$ is the scaling exponent.  The exponent is
alternatively called the spectral index or the slope (in a log$-$log
representation).

Without deconvolution, the original data are modified by the effective
beam decreased at high $k$ through $\psfk^2(k)$ so that the model
would be
\begin{equation}
P_{\rm model}(k) = \psfk^2(k) \; P_{0} \; (k/k_0)^{\expon} + \eta N(k) \,.
\label{plaw}
\end{equation}
For application in the above 1D model the effective $\psfk^2(k)$ can
be estimated directly from the 1D power spectrum formed from the 2D
power spectrum of a representative synthesized beam at the center of
the \draost\ mosaic.  As a check, fitting such a 1D power spectrum
with a Gaussian results in equivalent 1D Gaussian beams for \DF\ and
\EN\ that have quite similar FWHM, $56\parcs8\pm0\parcs7$ and
$58\parcs9\pm0\parcs9$, respectively, both reasonably consistent with
the 2D Gaussian beam statistics as summarized in Table~\ref{beamsize}.

Example power spectra of signal and noise are shown in
Figure~\ref{noiseDFdrao_ps} for the \DFi\ mosaic.  Generally a noise
template is calculated from many emission-free channels
(Appendix~\ref{noisetemplate}).  In this example for simplicity the
noise template is from a single representative emission-free channel
($v = 39.7$~\kms, as also used in Figure~\ref{noiseDF20}, left).
 
\subsection{Fitting the Power Spectrum}
\label{psfitting}

We adopt the model represented by Equation~(\ref{plawd}) to probe the
power spectrum of the signal to smaller spatial scales while properly
accounting for the noise.

We also mitigated against further uncertainty in model fits by
excluding data above a value \kmax; the results below are not
sensitive to the precise value so long as the noise can be adequately
assessed.  We adopted \kmaxvalst\ (which corresponds to baselines of
\bmaxst~m).

In the \draost\ observations there are no data corresponding to
spacings shorter than \bminst~m and so there is little power at $k<
\,$\kminvalst\ except from foreshortened baselines.  This is reflected
in the power spectrum of the signal in Figure~\ref{noiseDFdrao_ps},
where there is a precipitous departure below the rising power law at
\kminvalst.  The remedy for this is to add short-spacing data acquired
with a large single antenna (Section~\ref{gbt_observations}). For the
power spectrum fits to the \draost\ data alone, we excluded data below
a value \kmin, here taken to be \kminvalst.

The three best-fit model parameters were found using the IDL routine
{\tt mpfit.pro} \citep{markwardt2009}, with the described
uncertainties as weights.  The weights were further modified at high
$k$ using the prescription in \citet{mart15}, specifically
substituting fractional error $b = 0.07$ into their Equation~(4) to
account for the uncertainty of the effect of the beam.

The fits to the models for \DFi\ data are shown in
Figure~\ref{noiseDFdrao_ps} in Section~\ref{enmnoise}.  The exponent
is $\expon = -3.03 \pm 0.03$.  The uncertainties that we cite are the
formal 1$\sigma$ errors from the fits.  Alternative choices in the
fitting analysis could lead to systematic uncertainties in the
exponent, which we estimate to be no larger than 0.1. We also
found that relative differences between the exponents for different
channels or VCs (Section~\ref{nhanalysis}) are robust against the
systematic effects.

\subsection{Power Spectra of Non-Rectangular Regions}
\label{powerspecnonrect}

Apodizing a (median-subtracted) map prior to computing the Fourier
transform has the benefit of reducing edge effects.  Normally maps can
be made rectangular and so only a simple rectangular taper is
required.  The unique shapes of the \dhigls\ regions are distinctly
non-rectangular.  One alternative is to define an inscribed rectangle,
as we have done, but this inevitably leaves out high signal to noise
data.  This led us to an alternative approach aimed at incorporating
as much spatial information from each mosaic as possible, in this case
using data within the white dashed contour (e.g., Figure~\ref{coverageDF}).

We modified the shape of the taper so that rather than following a
straight edge of the map it followed a contour of constant weight (or
noise) in the \draost\ mosaic.  The median-subtracted maps were
tapered to zero orthogonal to the chosen contour.  The 35\arcmin\
taper was centred on the contour.  These tapered maps were then
cropped to a rectangular map aligned with the zeroed edges of the map.
Because of the unique shapes of the \dhigls\ regions, information is
missing between the zeroed edges of the map and the boundary of the
rectangle; this was padded with zeroes.  In practice the 2D taper was
obtained by convolving a unit mask defined by the contour with a
circularly-symmetric tapering function of the desired shape and width
and then cropping.

The drawback of any tapering (apodization) of the map is that the
resulting Fourier transform has also been modified: the true Fourier
transform has been convolved with the Fourier transform of the taper.
The Fourier transforms of the various non-rectangular zero-padded
apodization functions obtained as described above are all
centrally-peaked, suggesting that a convolution in the Fourier domain
will not have a significant effect on the measured power law exponent
of the original \dhigls\ map. (The amplitude is affected, but that is
not of interest here.)

We have tested this conjecture using simulations.  Fractional Brownian
motion (fBm) maps were created with a given power law exponent
\citep{mamd2003b}.  We apodized these maps with the custom-shaped
\dhigls\ taper, Fourier transformed them, and computed the power
spectrum.  Each power spectrum was then modelled to find the power law
exponent.  We found that the results were consistent with those used
to make the fBm maps within the model uncertainties.

\subsection{Power Spectra in 2D}
\label{ps2D}

Implicit in the construction of the power spectrum in 1D is that the
2D power spectrum is azimuthally symmetric.  For exceptional images
with a pronounced streaky appearance, the directionality affects the
2D power spectrum.  For example, for the single channel of IVC
emission in the \UMG\ region at $-51.76$~\kms, seen in green in
Figure~\ref{DFthreeIVC}, lower, the directionality of the structure is
roughly on the upper-left to lower-right diagonal, and in the image of
the 2D power spectrum there is more power in the \uv\ plane orthogonal
to this, roughly in two opposed quadrants.  This can be brought out
most effectively in the image by removing the radial dependence, i.e.,
by dividing the 2D power spectrum by the model fit to the 1D power
spectrum, which has exponent $-2.35 \pm 0.03$.

To quantify what is seen visually, as a simple approximation we made a
1D power spectrum by azimuthally averaging within these opposed
quadrants, and another 1D power spectrum for the other two
complementary quadrants.  The power law exponents were quite similar
($-2.28 \pm 0.04$ and $-2.27 \pm 0.05$, respectively).  Thus the
systematic effect is apparently not large.  However, as expected the
amplitudes were different, in this case by a factor of about 2.
Given the similar exponents, it is not surprising that the amplitude for 
the overall 1D power spectrum is close to the mean of the
amplitudes for the two separate 1D power spectra.

We are often interested in power spectra of images of integrated
emission, \nh\ (Section~\ref{nhanalysis}).  Because the \hi\ structure
changes over many channels, the directionality is less pronounced in
\nh\ (e.g., compare Figures~\ref{DFthreeIVC} and \ref{UMAcomponents}),
so that forming the 1D power spectrum is more suitable and less prone
to systematic effects.  For the case above, recall that the exponent
for the somewhat noisier \nh\ image of the full IVC VC in \UMG\ was
$-2.48\pm0.06$ (Table~\ref{powertable}).

Also, in comparing the power spectra of images of two independent tracers,
for example \nh\ and thermal dust (Section~\ref{discresultsthermal}), it
seems likely that any systematic effects will be similar given a
common directionality.

\section{Ranges in the \uv\  Domain for Adding Short-spacing Data}
\label{appen:short}

\begin{figure}
\centering
\includegraphics[angle=0,width=1.0\linewidth]{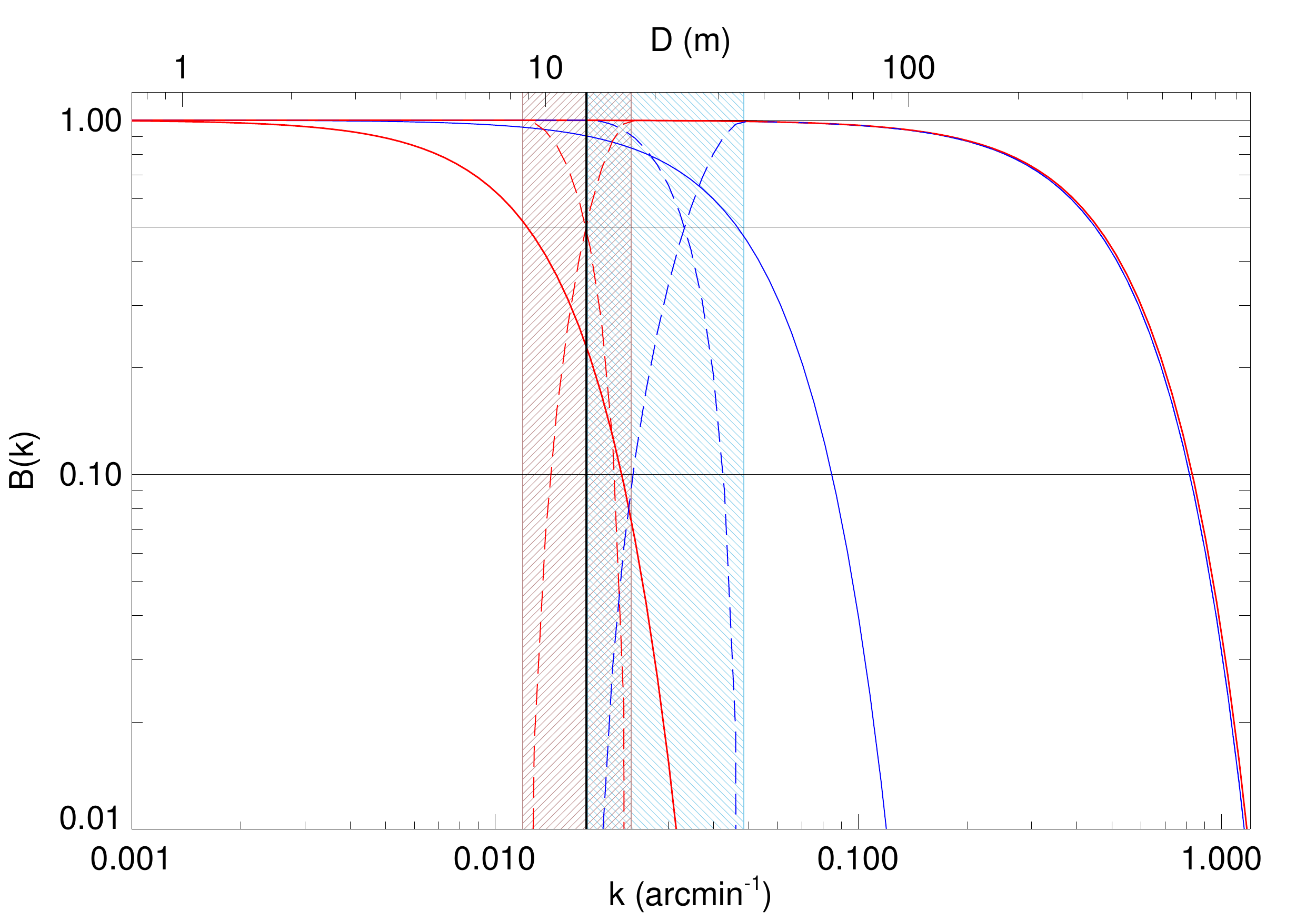}
\includegraphics[angle=0,width=1.0\linewidth]{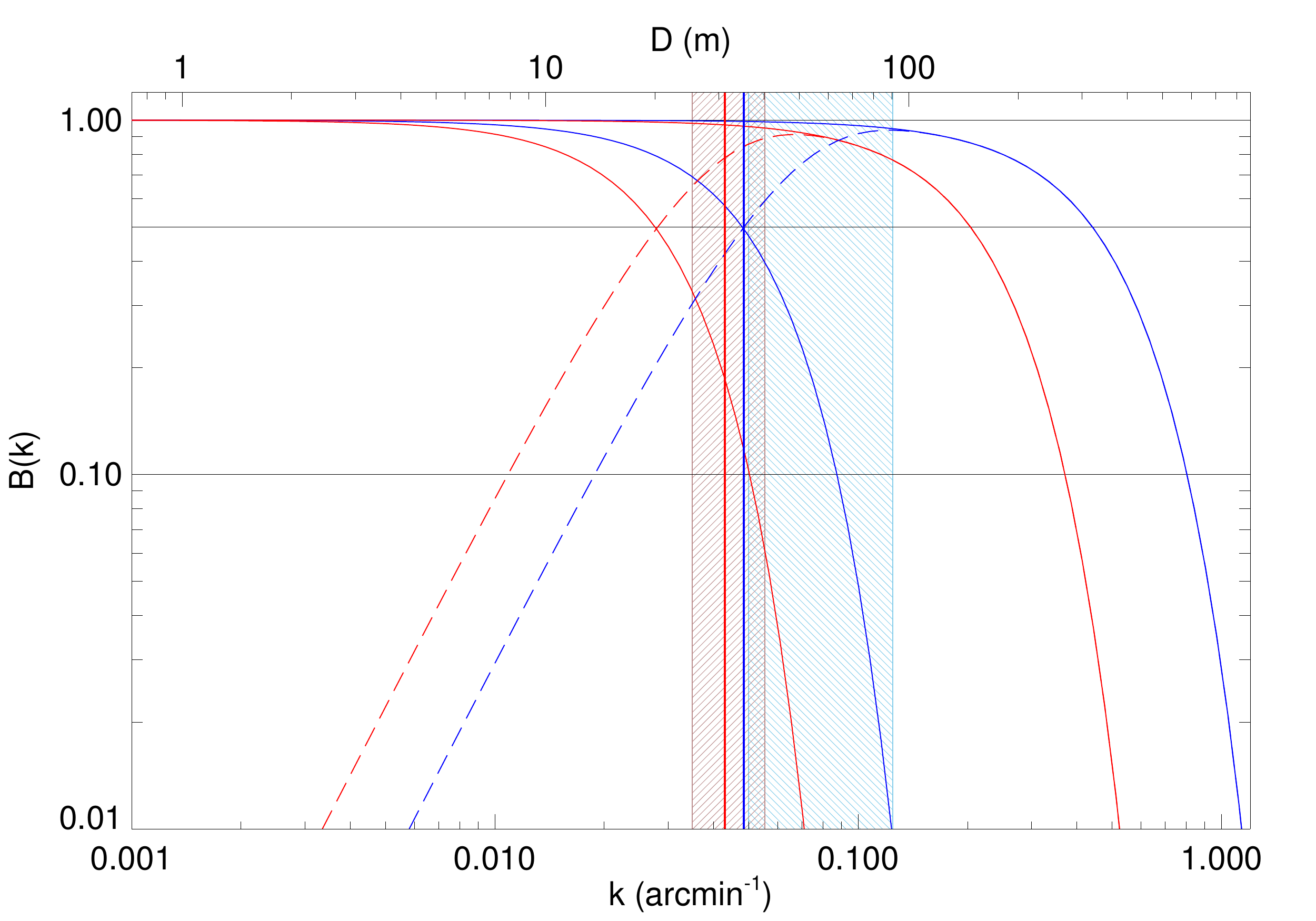}
\caption{
Beams $B(k)$ in the Fourier domain for the interferometers
(synthesized/restored beams: solid curves on right extending to higher
$k$) and for the single dishes used to supply the short-spacing \hi\
data (solid curves on left).  Curves, each represented as a Gaussian,
show how visibilities in the \uv\ domain are diminished by the beam.
Vertical line indicates shortest unprojected baseline of the
interferometer.  These considerations are relevant to the \uv\ ranges
adopted for cross calibration of data sets (shaded).  They are also
relevant for combining data; the complementary weighting functions of
\uv\ data (dashed curves) cross at 0.5.  For \dhigls\ see
Appendix~\ref{overrange} and Section~\ref{combinegbtdrao}, respectively.
Upper: CGPS (red) and \dhigls\ (blue) both use the \draost\ but
incorporate different single dish data, from DRAO 26 m and GBT 100 m,
respectively.
Lower: SGPS (red) combining ATCA and Parkes 64 m and VGPS (blue)
combining VLA and GBT 100 m.  In both of these surveys the feathering
of the single dish data follows the beam profile and the combination
range spans all $k$.
}
\label{overlapfig}
\end{figure}

As discussed in Section~\ref{gbt_observations}, an essential step in
wide-field imaging with interferometers is addition of short-spacing
data from a single dish.  In Figure~\ref{overlapfig} we summarize
information relevant to the \uv\ ranges available and adopted in cross
calibrating and combining the data sets in \dhigls\ and in previous
\hi\ surveys: the CGPS \citep{landecker2000,taylor2003,higgs2005},
VGPS \citep{stil2006}, and SGPS \citep{mccl2005}.

Each combination of survey facilities is characterized by two beams,
one for the synthesized (restored) beam, extending to higher $k$, and
one for the single dish.  The curves of $B(k)$ in
Figure~\ref{overlapfig} show how the visibilities in the \uv\ domain
are diminished by the beam.  The cutoff from the single dish beam and
the presence of noise clearly limit the \uv\ range for overlapping
data at the high-$k$ end.
Near the low-$k$ end, vertical lines represent the minimum unprojected
baseline in the interferometer.  This does not take into account
foreshortening or mosaicking.  A counter-consideration relating to
foreshortening of the shortest baseline is shadowing for compact
configurations and low elevation observations.
The shaded regions in Figure~\ref{overlapfig} show the baseline ranges
used in cross calibrating the data.

In combining the two data sets, in the case of the \draost\ the
deconvolved single dish \uv\ data are used below the lower bound of
this range and the interferometric data above the upper bound.  At
intermediate $k$ in between, they are combined after weighting in the
Fourier domain using the complementary functions shown as dashed
curves in the upper panel of Figure~\ref{overlapfig}.
In the case of the SGPS and VGPS 
one of the weighting functions used for the combination of data is
defined by the single dish beam (therefore not appearing as dashed in
the lower panel of Figure~\ref{overlapfig}),
in effect reversing the deconvolution of that data set; this and the
complementary weighting function
define a combination range spanning a range in $k$ much broader than
the range used for cross calibration.

\subsection{Reflections on Choices for Different \hi\ Surveys}
\label{overlapchoices}

CGPS (Figure~\ref{overlapfig}, upper, red).  Short-spacing data were
added for each separate synthesis before mosaicking.  For the single
dish data, the 26 m is quite undersized for the task.  The lower bound
adopted is lower than the minimum baseline (8.6 m versus 12.9 m; see
\citealp{higgs2005}).  Even so, the single dish data needed in the
overlap range are well down in the 26 m beam.

SGPS (lower, red).  The Parkes 64 m beam is also cutting off data
significantly over the entire overlap range.  This was mitigated
somewhat by mosaicking of the ATCA data \citep{mccl2005}, which
reduces the effective minimum baseline by $D/2$, where $D$ is the
diameter of the antennas making up the interferometer.  Still the
feathering implies an extrapolation of the interferometric data to low
$k$.  The reduction by the synthesized beam is barely relevant.

VGPS (lower, blue).  The minimum baseline of the VLA is quite large
(36 m), so that the overlap range is affected by the beam of the GBT
despite its 100 m aperture.  The lower and upper bounds adopted for
cross calibration (not given in \citealp{stil2006}) were 36 m and 90 m
(Jeroen Stil, private communication).

\dhigls (upper, blue).  The mosaicking by {\tt SUPERTILE} was not
designed to achieve the $D/2$ reduction in minimum baseline.  Still,
relative to the facilities combined in other surveys, there is a
potentially much broader \uv\ overlap range because the minimum
baseline is small, the GBT is large, and the synthesized beam is
small.  Relative noise is also a consideration.  Details are given in
Section~\ref{gbt_observations}.

\section{\ghigls\ and \draost\ Cross Calibration}
\label{crosscalib}

Here we describe the evaluation of the scale factor, $f_{\rm cc}
\equiv I_{\rm GBT}/I_{\rm DRAO}$, required for the accurate
combination of the mosaics made from the \draost\ data with the lower
resolution single dish \ghigls\ data.
Note that this factor is the inverse of that used by the {\tt MIRIAD}
\citep{sault1995} task {\tt IMMERGE}, which corrects short-spacing
data to the scale of the interferometric data.

\subsection{Preparation of the Data}
\label{prepdata}

The cross calibration comparison was done at the lower resolution of
the \ghigls\ data and in the \uv\ domain.  The evaluation of $f_{\rm
  cc}$ was carried out optimally using the full 2D information.

We used the data for the \DF\ region to refine our approach, as
illustrated below.
The \ghigls\ SPIDER data are in the Galactic GLS projection whereas
the \DFi\ mosaic is in the ICRS NCP projection.  To minimize
reprocessing of the \ghigls\ data and to ensure the same spatial
coverage for the cross calibration, the \DFi\ cube was reprojected to
the same Galactic GLS projection as the \ghigls\ data,
adopting the same 18\arcsec\ pixel grid spacing as in the original
\DFi\ cube.
We also processed the weight map in the same way, to track the region
in which the two data sets could be compared, and similarly the
synthesized beam map. The velocity channels were interpolated to
correspond to those in the SPIDER cube.

For the cross calibration the signal needs to stand out from the
noise.
With regard to the noise, we compared the data in the region where
the noise level in the \DFi\ mosaic is no more than a factor of two
larger than the noise minimum (see the white dashed contour in
Figure~\ref{coverageDF}).
With regard to the signal, we compared the data near the peak in the
LVC \hi\ spectrum.  Specifically, we used a map of \wh, the integral
of the \hi\ spectra, including only $N_{\rm ch}$ channels near the
peak of the \hi\ standard deviation spectrum (see
Figure~\ref{sigmavelDF}).  We estimated the relative signal-to-noise
ratio (S/N) as a function of $N_{\rm ch}$ using $W_{\rm sd} / \sqrt
N_{\rm ch}$, where $W_{\rm sd}$ is the standard deviation about the
mean (alternatively it can be estimated via analysis of the power
spectrum).  We adopted the range 1.6~\kms\ $ < \vlsr < 6.5$~\kms\ to
create the optimal \wh\ map, which results in a S/N about twice that
of a single channel.

As in Appendix~\ref{powerspecnonrect}, starting with the original
\ghigls\ data in the image plane we first removed the median value of
the data (in this case \wh) within the white dashed contour, then
apodized with a 35\arcmin\ taper centered on the contour to reduce the
data to zero, and beyond that in the encompassing square image padded
with zeros.  We then computed the 2D Fourier transform
$\tilde{f}(k_x,k_y)$ of this modified image.  The tapering and padding
with zeroes, applied equally to two fields to be compared, will not
affect the derived scale factor.  As discussed in
Appendix~\ref{powerspecnonrect}, there is no significant effect on the
power law exponent either.  We note that this scheme is similar to
that used by \citet{bert2016} for power spectra of non-rectangular
regions in the determination of a cross calibration factor for
\Herschel/SPIRE and \Planck/HFI data.

To produce the corresponding product for the \draost\ data, starting
with \wh\ from the finely-gridded reprojected \DFi\ data described
above,
we deconvolved with the synthesized beam, 
convolved with the 2D \ghigls\ beam, and
regridded to the coarser \ghigls\ image grid. 
We then selected the region, removed the median, apodized, and
zero-padded in the same way as for the \ghigls\ data,
and finally computed the 2D Fourier transform.

\begin{figure}
\centering
\includegraphics[angle=90,width=1.0\linewidth]{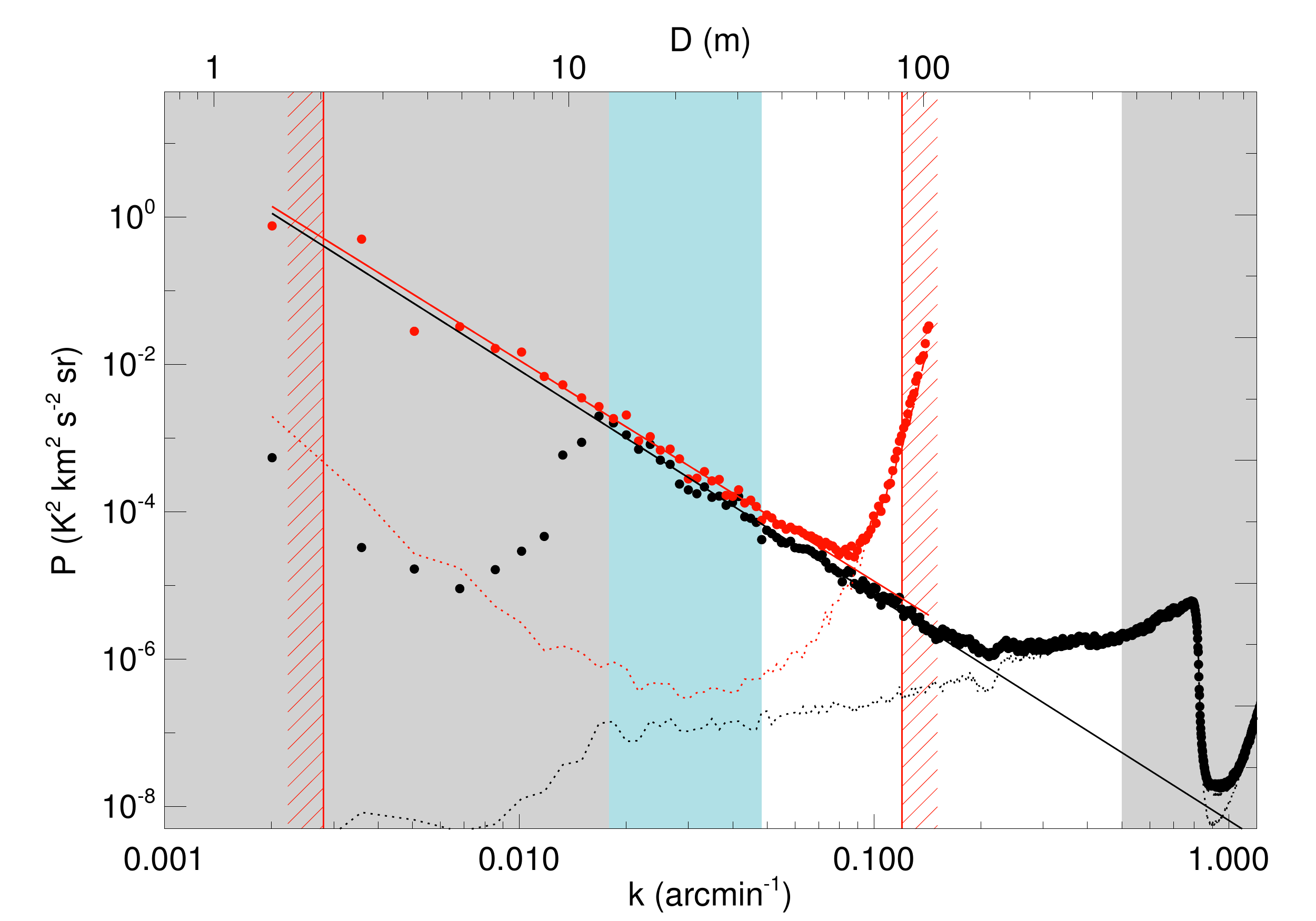}
\caption{
Power spectra for the DF region for spectral data integrated over
channels near the peak in the standard deviation spectrum (see text).
Deconvolved \draost\ data shown in black and \ghigls\ GBT data shown
in red.  \draost\ power spectrum is fit to Equation~(\ref{plawd}) as
for single channel case in Figure~\ref{noiseDFdrao_ps}.  Fit to the
\ghigls\ power spectrum follows same model, but with a different noise
template and a different range in $k$ as indicated by vertical hatched
regions in red.
Blue band identifies overlapping range of $k$ used for cross
calibration, well outside influence of noise in the \ghigls\ data and
any uncertainties in the \ghigls\ beam.
}
\label{df_ps_idl-eps-converted-to.pdf}
\end{figure}

Solely for visualizing the relative calibration of the data we also
produced another closely related product from the \draost\ image data,
omitting the steps of convolving with the 2D \ghigls\ beam and
regridding to the coarser \ghigls\ image grid.  As in
Appendix~\ref{appdfmps} we computed the 1D power spectrum from these
2D data and from those for \ghigls.
Figure~\ref{df_ps_idl-eps-converted-to.pdf} shows the power spectra of the two data
products and the power law fits calculated for the different but
overlapping $k$ ranges appropriate to the two data sets.  The power
spectrum exponent for the \ghigls\ data, $-3.00 \pm 0.05$, is very
close to that for the \draost\ data, $-3.06 \pm 0.03$.  This is an
important check that has not been possible for the other \hi\ surveys
mentioned.
Furthermore, it is clear that a vertical shift is required to align
the \draost\ data with those from the GBT data.  The amount of the
shift is $f_{\rm cc}^2$.

\subsection{The Overlap Range}
\label{overrange}

Next we consider the ``overlap range'' in $k$, or in practice an
annulus in the \uv\ domain, in which the data are to be compared.

On the low side this range is constrained by the minimum baseline of
the interferometer.  As above for the \draost\ we chose \bmincross~m
($\kmincross\, {\rm arcmin}^{-1}$).  This is a conservatively higher
choice relative to the minimum baseline, compared to other \hi\
surveys (Figure~\ref{overlapfig}).  Nevertheless, it is still not far
into the range of $k$ that is affected by the beam of the
(short-spacing) single dish data, as was necessary for
the VGPS (larger minimum baseline), CGPS (smaller single dishes), or
SGPS (combination of the two).

The choice for the high-$k$ side depends on properties of the GBT
data.  It needs to take into account the relative noise of the two
data sets, which can be assessed using power spectrum model fits
according to Equation~(\ref{plawd}).  As is apparent from the results
shown in Figure~\ref{df_ps_idl-eps-converted-to.pdf}, because of the effect of the
\ghigls\ beam, as we move to higher values of $k$ in the power
spectrum, the GBT noise becomes larger relative to not only the
\draost\ noise but also the \hi\ signal.  Another consideration is
that data in the entire range of $k$ being considered are affected by
the \ghigls\ beam, and any uncertainties in the beam would produce
more pronounced effects at higher values of $k$ (the synthesized beam
is not a consideration).  To be conservative, we adopted \bmaxcross~m
($\kmaxcross\, {\rm arcmin}^{-1}$).

Despite these conservative choices, the overlap range is relatively
broad and much less impacted by the beam of the single dish data as
compared to the other surveys (Figure~\ref{overlapfig}); this provides
the opportunity for a more critical assessment of the cross
calibration determination.

\subsection{Quantitative Comparisons of the Data}
\label{ccoptions}

\begin{figure}
\centering
\includegraphics[angle=90,width=1.0\linewidth]{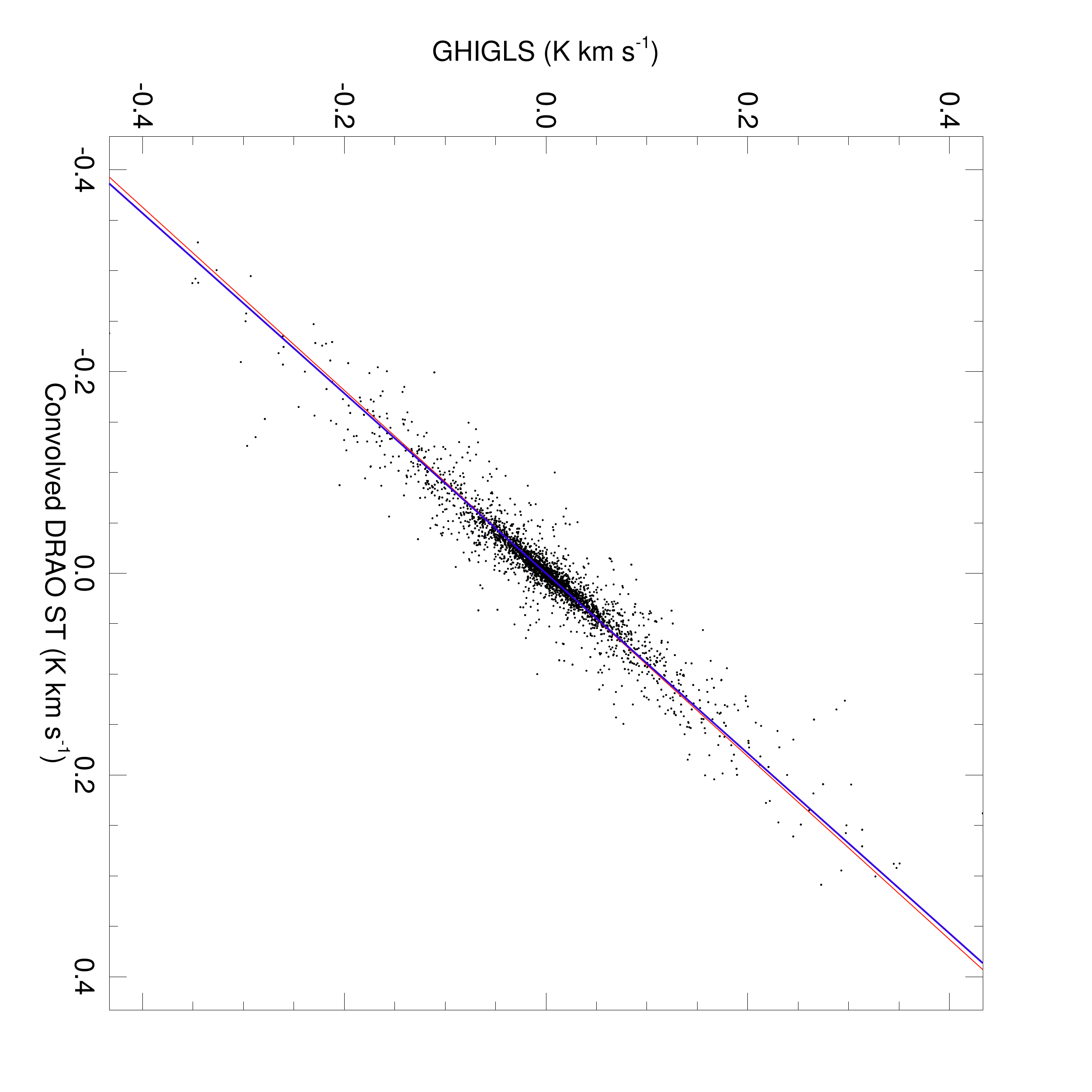}
\caption{
Correlation of complex Fourier-transformed data (real and imaginary
parts are kept distinct and plotted separately) in annulus in \uv\
domain corresponding to overlapping $k$ range, illustrated for DF
integrated spectral data used in Figure~\ref{df_ps_idl-eps-converted-to.pdf}.  Blue
line shows the least absolute deviation fit, with slope $ I_{\rm
GBT}/I_{\rm DRAOST}=\ccal \pm \estccal$ (for systematic uncertainties,
see text).  Red line, from ordinary least squares fit, is very similar
(slope $1.10\pm0.01$).
}
\label{compareDFvis}
\end{figure}

There are a number of different methods for determining the cross
calibration factor.
The one we adopted, as in {\tt IMMERGE}, was to correlate the complex
2D Fourier-transformed data (keeping real and imaginary parts
distinct) within the overlap-range annulus in the \uv\ domain.  To
mitigate against any residual cross, we masked two rectangular regions
each with a width of three pixels centered along each axis (i.e.,
along $u=0$ and $v=0$).  The intercept from this correlation is
consistent with zero within its error (Figure~\ref{compareDFvis}) and
subsequently in fitting the data we enforced this by adding data at
$(0,0)$ explicitly.

As judged from the noise power spectra, the \ghigls\ data have greater
uncertainty than those from the \draost. This is confirmed by the
standard deviation of the values of the modulus in rings of constant
$k$.  Therefore, we performed the usual ``y on x" regression implied
by the format of Figure~\ref{compareDFvis}.  As in {\tt IMMERGE}, for
the linear fit we minimized the absolute deviation (L1-norm).  For the
\DF\ field we found $f_{\rm cc} = \ccal \pm \estccal$.  We also used
ordinary least squares fitting (i.e., minimizing the L2-norm), finding
$1.10 \pm 0.01$.  Furthermore, the bisector slope is also the same
within 1~$\sigma$.  These checks reinforce that the result is robust.

In the same way we found independent values of $f_{\rm cc}$ for four
other regions, \EN, \UM, \DR, and \PO\ (see Table~\ref{fcc_table}).
The \draost\ syntheses included in these mosaics had different values
of $F_{\rm DRAO}/F_{\rm NVSS}$ applied, indicating the efficacy of
that part of the calibration.

\begin{table}
\caption{Values of cross calibration scale factor $f_{\rm cc} =
I_{\rm GBT}/I_{\rm DRAO}$ }
\centering
\begin{tabular}{lc}
\hline
\hline
Region & $f_{\rm cc}$ \\
\hline
\DF & $ \ccal \pm \estccal$ \\
\EN & $1.10 \pm 0.02$ \\
\UM & $1.09 \pm 0.01$ \\
\DR & $1.12 \pm 0.02$ \\
\PO & $1.17 \pm 0.02$ \\
\hline
\end{tabular}
\label{fcc_table}
\end{table}

\subsection{Systematic Errors}
\label{fcheck}

We have investigated the systematic errors that might arise.  One is
from the choice of overlap range.  For example, the values of $f_{\rm
cc}$ using data from the independent annuli between 13 m and 26.4 m
and between 26.4 m and 35 m are $1.09 \pm 0.01$ and $1.16 \pm 0.01$,
respectively.

Deconvolving the \ghigls\ data and not convolving the \draost\ data
with the \ghigls\ beam changes the effective weighting going into the
slope.  We found $f_{\rm cc} = 1.13 \pm 0.01$.

We investigated correlating values of the Fourier amplitude (modulus
of the visibility) in 2D.  This again changes the weighting. We also
note that at higher $k$ the amplitudes are smaller but there are more
independent data samples.  We found $f_{\rm cc} = 1.14\pm0.01$.

We also correlated power spectrum moduli in 2D, finding $f_{\rm cc}$
from the square root of the slope.  This is akin to fitting a common
power law to the two sets of data.  Again the weighting changes and we
found $f_{\rm cc} = 1.12 \pm 0.01$.  We note that correlating power
spectrum moduli in 1D (a weighted fit is indicated) is less
satisfactory, because of the compression of the data in $k$ and
azimuthal coverage before the correlation (though it should make no
difference for noise-free data).

From these investigations, we estimate that the systematic error in
$f_{\rm cc}$ is \esyccal.

We also looked at the cross calibration factor synthesis by synthesis.
The \draost\ data have to be corrected for the effect of the primary
beam, which limits the spatial coverage to about 90\arcmin\ and makes
the data noisier at the low end of the $k$ range.  An annulus between
25 and 50~m ($0.034 < k < 0.069$ arcmin$^{-1}$) was selected.  We
found a mean cross calibration factor $1.13\pm0.01$, consistent with
the above.
As another check, we used \draost\ data that were not scaled to NVSS
and found that the values of $I_{\rm GBT}/I_{\rm DRAO}$ were well
correlated with the inverse of $F_{\rm DRAO}/F_{\rm NVSS}$
(Section~\ref{register}).
This indicates that if cross calibration synthesis by synthesis were
sufficiently precise, there would be no point in first doing the synthesis
by synthesis scaling to NVSS.

We adopted a common cross calibration factor of $\ccal \pm \estccal \pm
\esyccal$ for all seven DHIGLS regions.   A correction of the same order was found by
\citet{pido2015} for combining GBT and VLA data.  This is mentioned
not for direct comparison to our $f_{\rm cc}$, but to point out that a
non-unity cross calibration factor is not unprecedented.

\section{Estimated Optical Depth Corrections}
\label{hiopac}

It is generally the case for the \dhigls\ spectra that the \hi\
optical depth $\tau$ is small.  This can be illustrated as follows.
For optically thick \hi\ emission, the brightness temperature $\Trb$
saturates at the spin temperature $\Trs$.  For optically thin
emission, on the other hand, $T_{\mathrm b} = \tau\, T_{\mathrm s}$.
A plausible value of the CNM spin temperature is $\Trs = 80$~K, the
collisional temperature found from intermediate-latitude H$_2$
observations for values of \nh\ near $10^{20}$~\cmm\
\citep{gill06,wakk06}.  This is not too different from the
distribution $67 \pm 14$~K found by \citet{rachford2009} for
translucent lines of sight at lower latitude.  
A single $\Trs$ along any line of sight is in any case only an
approximation.  For CNM gas in thermal equilibrium \citet{wolfire2003}
show (in their Figure~10) that the gas temperature (and so $\Trs$) is
roughly inversely related to $\vnh$.  For comparison to $\Trs$, even
for the channel map in \DF\ with the brightest emission (at 3.45~\kms)
the average $T_{\mathrm b}$ is 12~K; the dispersion is 9~K with a low
exponential tail to about 40~K.  The total emission is enhanced by
contributions from the WNM gas as well, for which the absorption is
much smaller.  We conclude that optical depth effects on the \hi\
emission spectrum are typically small.

Column density can be calculated with an approximate correction for
optical depth assuming a single spin temperature $\Trs$ using
\begin{equation}
\label{taucorrect}
\nhm   = C\, \Trs\ \int - \ln \left( 1 - \frac{\Trb}{\Trs} \right) \mathrm{d}v
\end{equation}
rather than simply $C\,$\wh.  We used $\Trs = 80$~K.  
Corrections are small, consistent with the results of \citet{lee15}
for the Perseus molecular cloud region; they found significant
corrections only for column densities much greater than $40
\times10^{19}$~\cmm, larger than typically found in our intermediate
latitude regions.

\subsection{Resolution Dependence}
\label{resdep}

The correction in Equation~(\ref{taucorrect}) clearly depends on the
contrast of the observed brightness $\Trb$ relative to $\Trs$.
Because peaks in the intrinsic \hi\ emission on the sky are diminished
if the observing beam is larger than the spatial structure, the
estimated optical depth correction is reduced at lower resolution.

We estimated the impact on \ghigls\ \nh\ maps by comparing results
from data at two differing resolutions, assuming $\Trs = 80$~K.
For the first map, \nh\ was created from the \dhigls\ cube using
Equation~(\ref{taucorrect}) and then convolved to the \ghigls\
resolution and regridded to 3\farcm5 pixels.  We call this
\nh(\dhigls).  For the second map we convolved the \dhigls\ cube to
the \ghigls\ resolution, regridded, and then computed \nh.  We call
this \nh(\ghigls).  The resulting difference, expressed as $\Delta =$
\nh(\dhigls)$/C \, - \, $ \nh(\ghigls)$/C$, is positive.

Within the high signal to noise region bounded by the white dashed
contour in \DFG, and for the LVC component, we found $\log(\Delta) =
0.02\pm0.2$ for $\Delta$ in units of K~\kms, meaning that at the lower
resolution of actual \ghigls\ \nh/$C$ is underestimated by of order 0
to 2~K~\kms\ for this region.  From Table~3 in \citet{mart15}, the
uncertainty for the LVC component of (\ghigls\ field) SPIDER,
$\sigma_{\nhm} = 0.3 \times 10^{19}$~cm$^{-2}$, or $\sigma_{\nhm}/C
\sim 2$~K~\kms, comparable to the maximum additional $\Delta$
correction.  Thus the uncertainty of \nh\ cited in \citet{mart15}
might be somewhat underestimated in this systematic way by not
accounting for the resolution dependence of optical depth effects.
Further discussion of the effect of the optical depth correction and
the dependence on $\Trs$ can be found in the analysis of the \ghigls\
data in \citet{mart15}.

Whether there is a significant effect on the estimated optical depth
correction for the DHIGLS data would depend on whether the bright
spatial structures are still unresolved at the resolution of the
\draost.  There is the further uncertainty in the correction because
the value of $\Trs$ is not known precisely and is unlikely to be
constant.

\section{\URSA\ Revisited}
\label{discdetailsursa}

For the LVC gas in the \dhigls\ \UMG\ region we found a power law
exponent $-2.89 \pm 0.04$ for the \nh\ map (Section~\ref{nhanalysis})
and $-3.03 \pm 0.09$ for the centroid velocity map
(Section~\ref{centroid}).

In their analysis of LVC gas in early \draost\ data of two adjacent
intermediate-latitude pointings in Ursa Major (\URSA; these correspond
to the FP2 and NN pointings among our current 16 pointings for the
\UMi\ mosaic), \citet{mamd2003} found an exponent $-3.6 \pm 0.2$ for
both the \nh\ map and centroid velocity map.
Particular features in their data processing and analysis of the \nh\
map might have contributed to the difference with respect to our
value.
(i) To mitigate against the effects of noise, the \draost\ data were
filtered.  Filtering on small angular scales (high $k$) might remove
not only noise but also some \hi\ signal power at these $k$ values,
artificially steepening the power spectrum.
(ii) No explicit allowance was made for the effect of the synthesized
beam and this was stated to contribute to the reported uncertainty of
the exponent.
(iii) Their \draost\ data were combined with zero-spacing data from
the 26~m single dish at DRAO, which does not allow for the same cross
calibration as done here for the GBT data (see
Appendix~\ref{crosscalib} and Figure~\ref{overlapfig}).  If the
\draost\ data calibration were low relative to the 26~m data, the
power spectrum of \nh\ would be artificially steepened.

It is additionally puzzling that the steeper power spectrum in this
earlier work appeared to be corroborated by analysis of entirely
independent Leiden/Dwingeloo \hi\ data in a large enclosing field in
the \ncpl\ that extended the power spectrum to larger angular scales
(smaller $k$); see Figures 14 and 15 in \citet{mamd2003}.
That power spectrum produced from the Leiden/Dwingeloo data ranged
from $k = 0.001$\,arcmin$^{-1}$ to about \kminvalst.  According to
Figure~\ref{overlapfig} power should be reduced in this range, by the
square of the 35\farcm7 beam of the 25~m Dwingeloo telescope,
steepening the apparent power spectrum.  However, this was not
accounted for in their analysis.
We explored this further using a 7\deg\ by 10\deg\ region selected
from the \ghigls\ \ncpl\ field and a related large 25\deg\ by 25\deg\
field from LAB data.  We made \nh\ maps selecting LVC in the velocity
range observed by \citet{mamd2003} ($-6.6 < v <
19.4$~\kms). Accounting for the respective beams, we find power
spectra of \nh\ with
$\expon = -2.68\pm0.14$ and 
$-2.68\pm0.24$ for \ghigls\ and LAB, respectively, consistent with
what we found for the \dhigls\ \UMG\ region.
The results for the analysis of the corresponding centroid velocity
maps are similar:
$\expon = -2.54\pm0.08$ and 
$-2.42\pm0.10$ for \ghigls\ and LAB, respectively.  If the correction
for the beam is ignored in the power spectrum model, it is clear that
the fit will result in a steeper power law unless the upper limit of
the fitted $k$-range is reduced appropriately.
Note also that for the combined LAB survey, the effective beam of
40\arcmin\ and 0\pdeg5 beam sampling result in a further reduction in
power at high $k$ \citep[see Appendix A in][]{higgs2005}, exacerbating
the steepening.

Concerns related to (i) and (ii) above would apply to the power
spectrum analysis of the centroid velocity map and the velocity
channel analysis in \citet{mamd2003}.
For the latter, the S/N first increases as additional channels are
added to the emission being integrated and the relative noise is
reduced.  Not accounting for this decrease in the relative noise level
(for example by not including a model of the noise in the power
spectrum model) could result in an artificial steepening of the power
law as the noise is reduced, depending on the range in $k$ over which
the noise dominates.


\section{Data and Analysis for Additional \dhigls\ Regions}
\label{additionalfields}

As summarized in Table~\ref{beamsize}, the \dhigls\ data also include
three other mosaics with fewer pointings and two single pointing
syntheses.  These have been processed as described for the \DFi\ and
\ENi\ mosaics.  An overview for each is provided in the following
subsections.  Note the different angular scales and the different
ranges for the colorbars in the figures below.

\begin{figure}
\centering
\includegraphics[angle=0,width=0.9\linewidth]{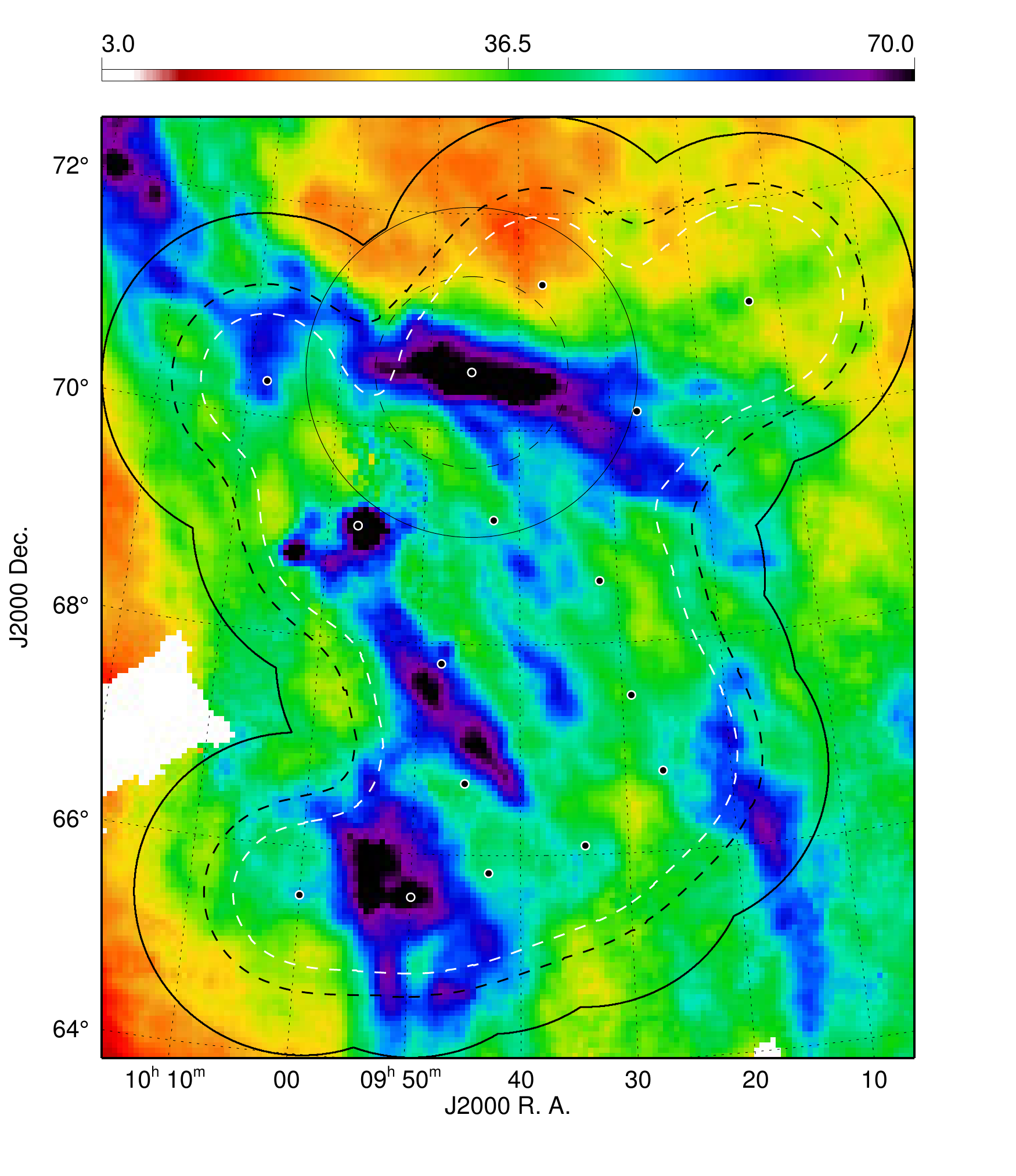}
\caption{
Same as Figure~\ref{coverageDF}, but for 16 pointings making up the
mosaic of the \UM\ region, with the highlighted NN pointing at
$(9^{\rm h}45^{\rm m}21^{\rm s},\, 70^\circ32\arcmin11\arcsec)$.
}
\label{coverageUM}
\end{figure}

\begin{figure}
\centering
\includegraphics[clip=true,trim=10 20 10 10 ,angle=0,width=0.8\linewidth]{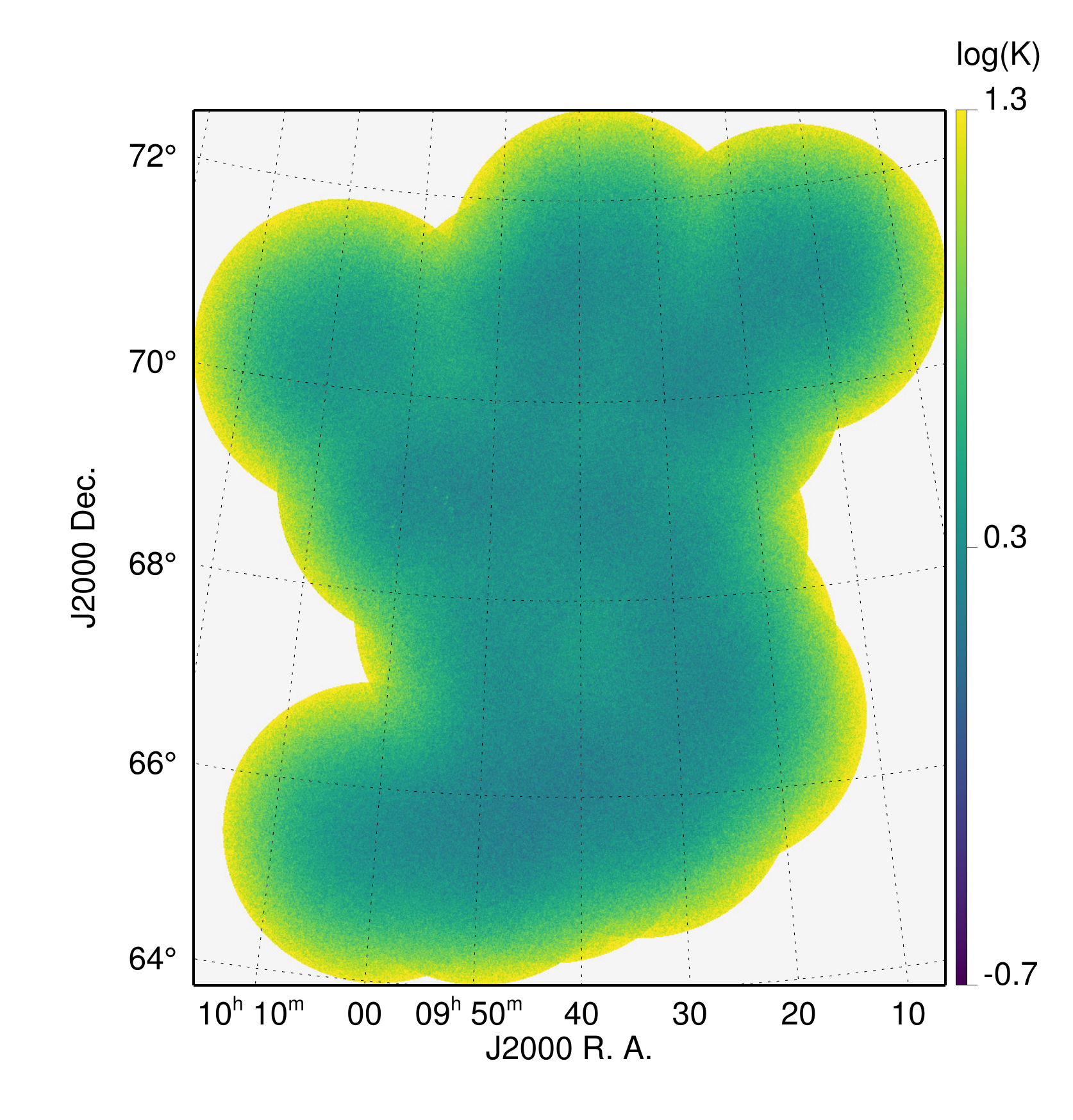}
\includegraphics[clip=true,trim=10 20 10 10 ,angle=0,width=0.8\linewidth]{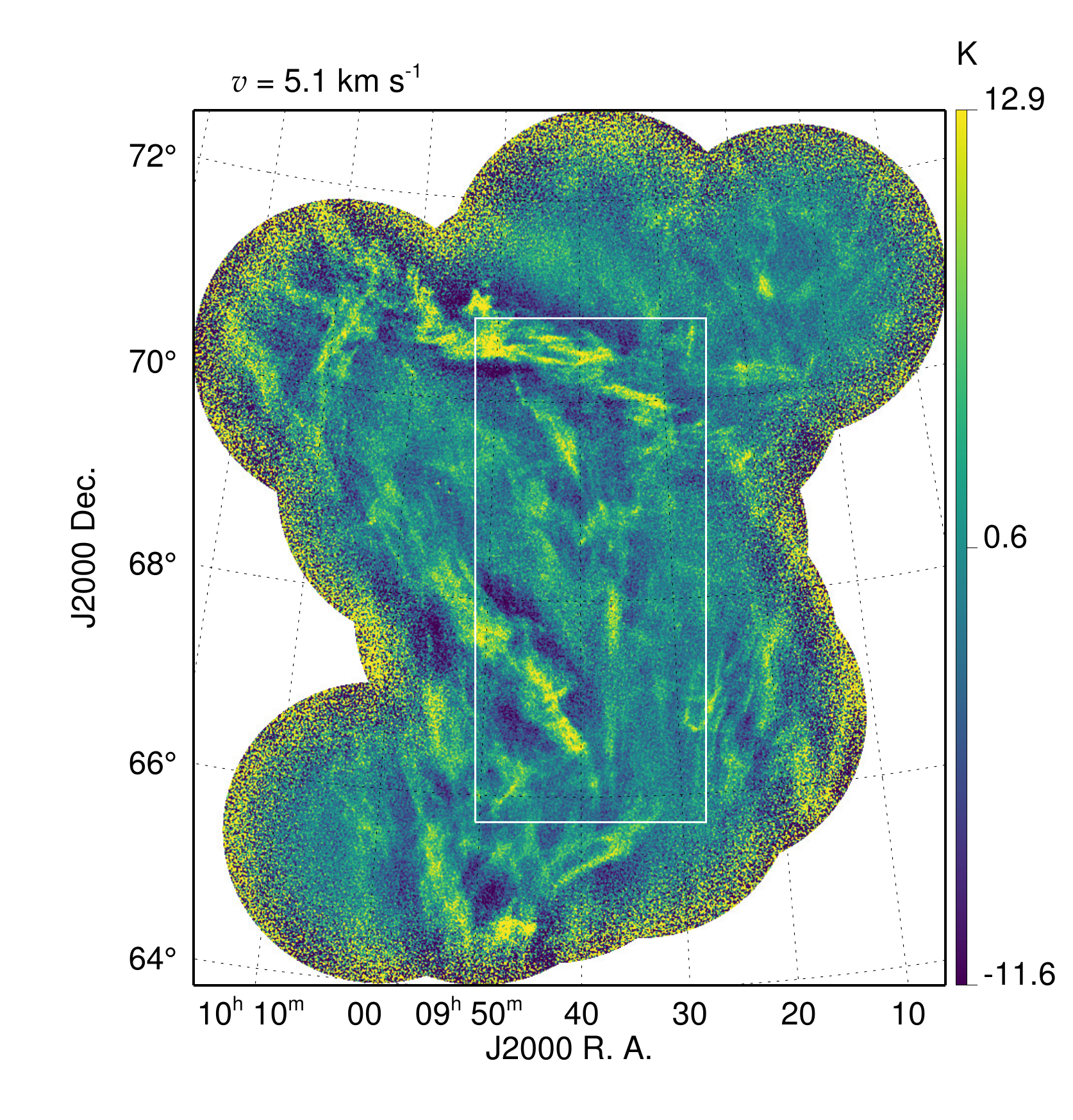}
\caption{
Upper: Noise map for \UMi\ mosaic as in Figure~\ref{noiseDFdrao}.
Lower: Representative channel map ($v = 5.1$~\kms) for \UMi\ mosaic as
in Figure~\ref{noiseDFdraohi}.  White rectangle indicates region used
for power spectrum analysis in Figure~\ref{noiseOTdrao_ps}.
}
\label{noiseUMdrao} 
\end{figure}

\begin{figure}
\centering
\includegraphics[angle=0,width=0.8\linewidth]{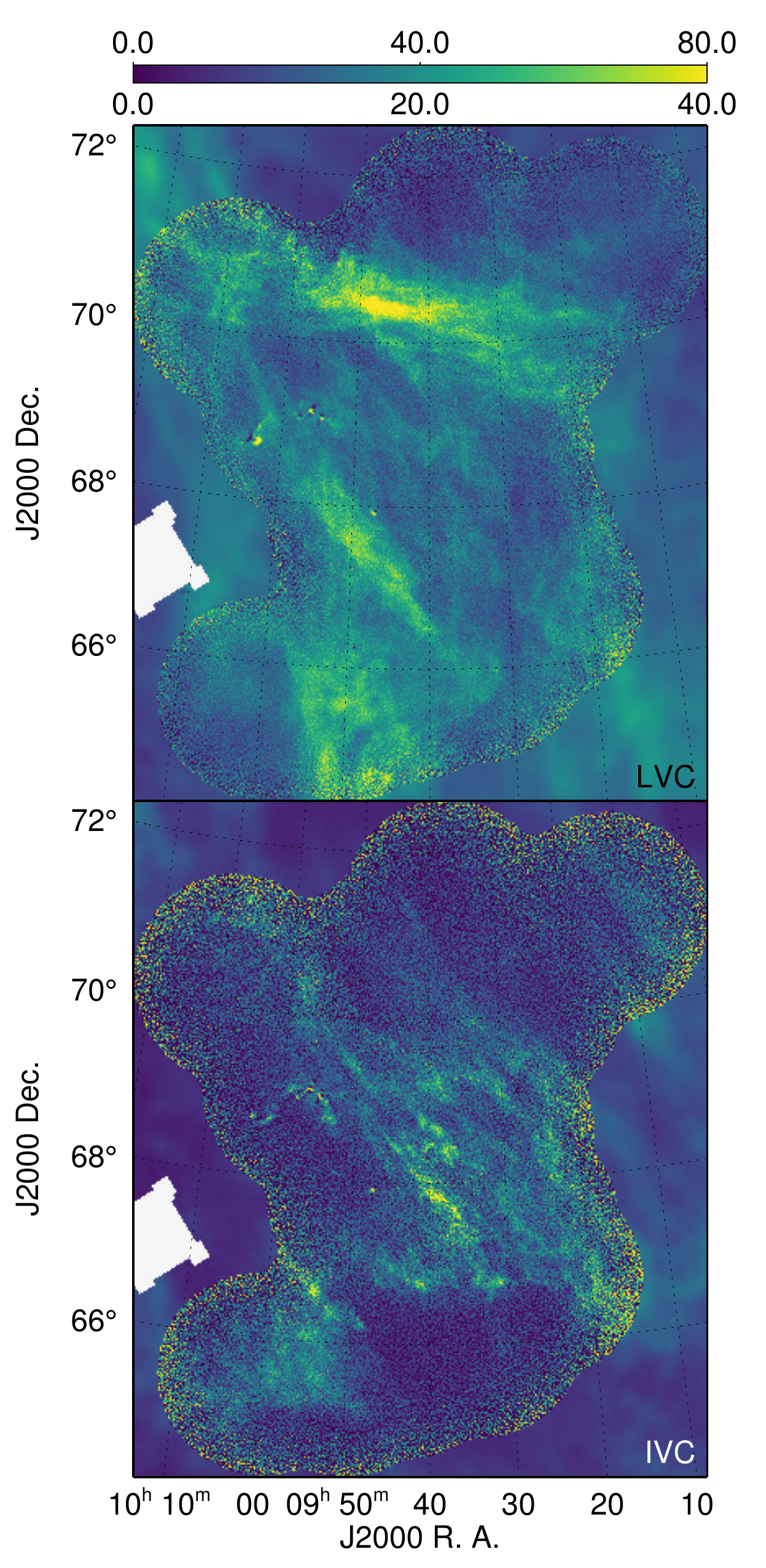}
\caption{
Maps of \nh\ for VCs in \UMG\ in units \colunits\ for the velocity
ranges in Table~\ref{compvel_table}.  Upper: LVC.  Lower: IVC.  Areas
not covered by the \ghigls\ \ncpl\ combined field are white.
}
\label{UMAcomponents}
\end{figure}

\begin{figure}
\centering
\includegraphics[angle=0,width=0.8\linewidth]{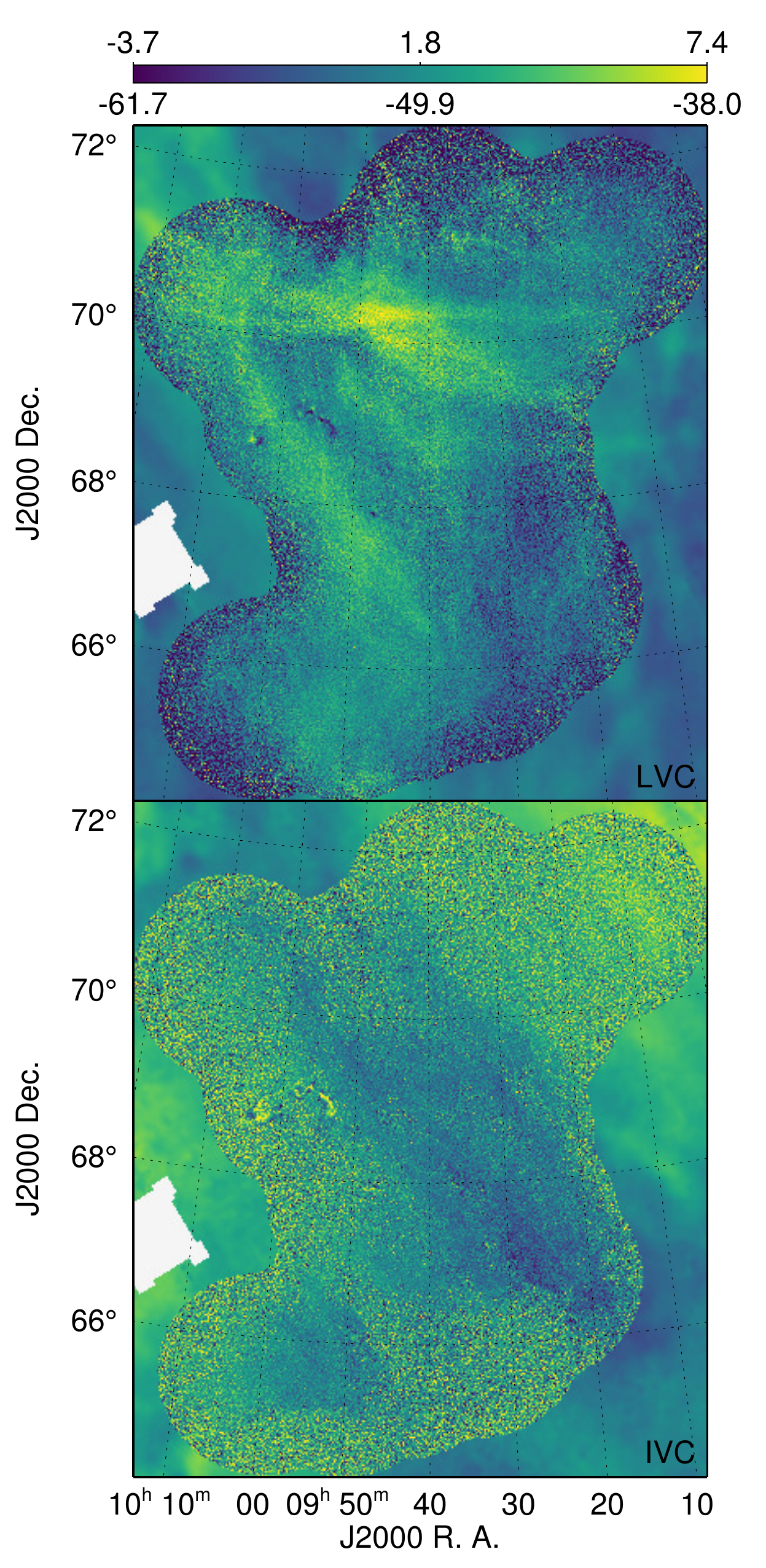}
\caption{
Centroid velocity maps for \UMG\ LVC (upper) and IVC (lower) VCs. Units are
\kms.
}
\label{componentsUMv_ps}
\end{figure}

\subsection{The \UM\ Region}
\label{umsummary}

The separation in the irregular grid of pointing centers
(Figure~\ref{coverageUM}) for the \UMi\ mosaic in the \ghigls\ UMA
field is somewhat smaller than the FWHM of the \draost\ primary beam.
The upper panel of Figure~\ref{noiseUMdrao} shows the resulting noise
map of \UMi.  The coverage is relatively sparse compared to \ENi\ and
\DFi, and so the noise level is higher, though at 2~K is still lower
than at the centre of a single synthesis.  This is slightly better
than achieved in the CGPS, but the signal in this region is of course
much lower than in the Galactic plane.
An example signal channel from the \UMi\ mosaic is shown in the lower
panel of Figure~\ref{noiseUMdrao}.

The \nh\ maps produced from the \UMG\ data cube are shown in
Figure~\ref{UMAcomponents}, assuming $\Trs = 80$~K.  The corresponding
centroid velocity maps (Section~\ref{centroid}) are shown in
Figure~\ref{componentsUMv_ps}.  The power spectra are discussed in
Section~\ref{nhanalysis} using data within the white dashed contour in
Figure~\ref{coverageUM}.

Color maps made from the \dhigls\ cube are presented in the lower
panels of Figures~\ref{DFthree} and \ref{DFthreeIVC}.  Narrow emission
and absorption lines in both LVC and IVC gas are discussed in
Section~\ref{showoff}.

\clearpage

\begin{figure}
\centering
\includegraphics[clip=true,trim=10 20 10 10 ,angle=0,width=0.8\linewidth]{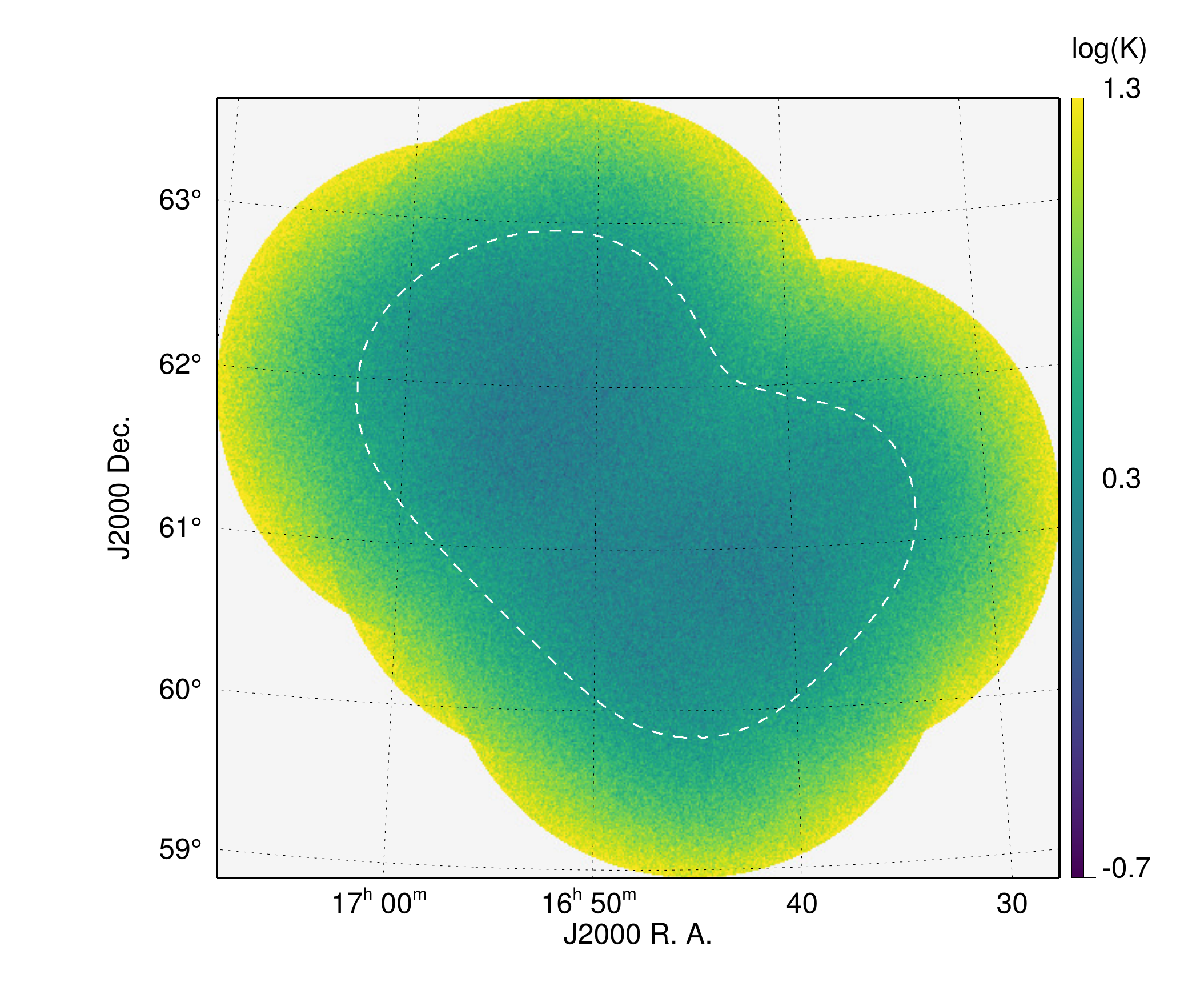}
\includegraphics[clip=true,trim=10 20 10 10 ,angle=0,width=0.8\linewidth]{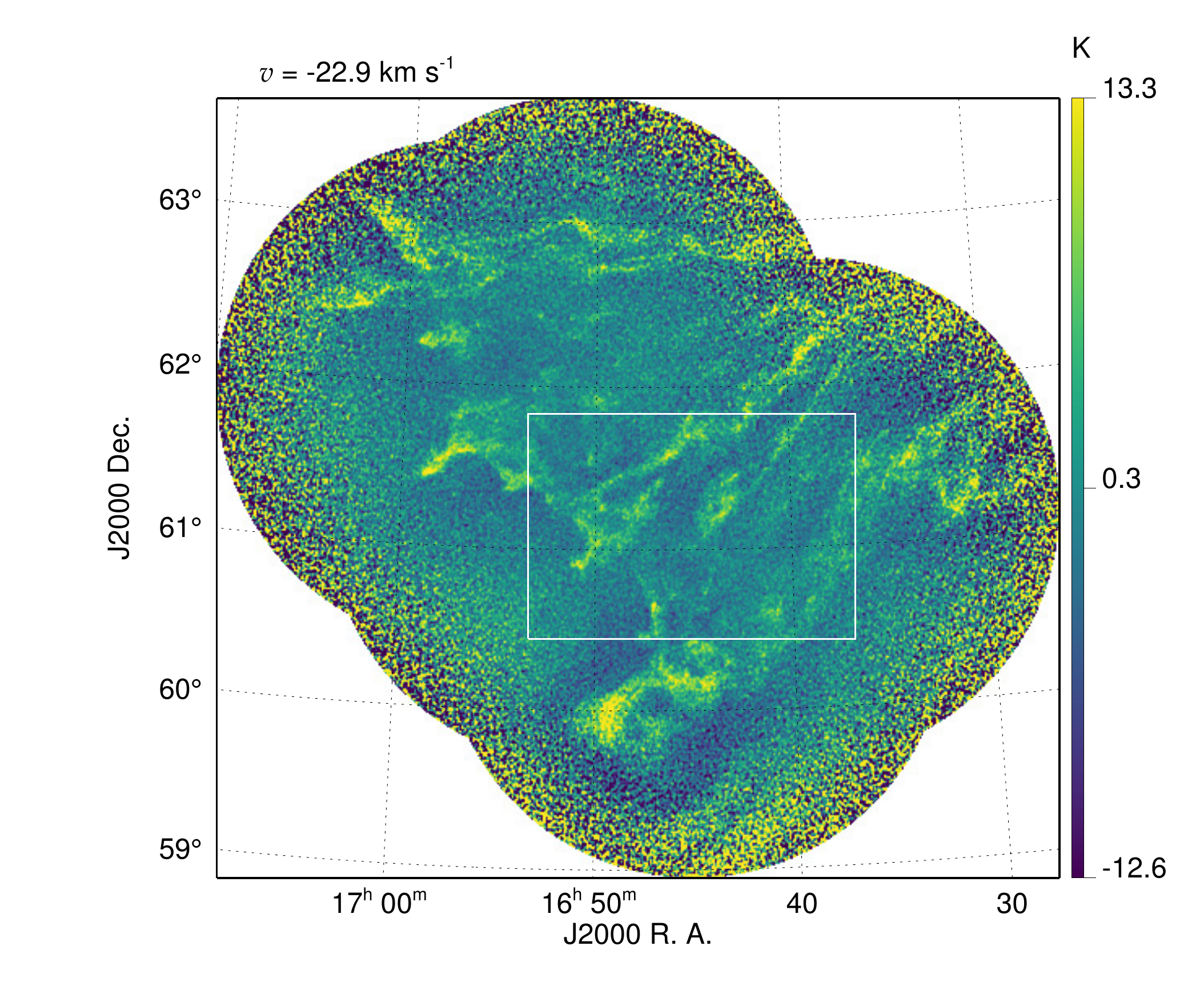}
\caption{
Noise map and single channel map ($v = -22.9$~\kms) for \DRi\ mosaic
as in Figures~\ref{noiseDFdrao} and \ref{noiseDFdraohi}.
White rectangle region used for analysis of the single-channel power
spectrum in Figure~\ref{noiseOTdrao_ps} and white dashed contour used
for analysis of the VCs in Figure~\ref{componentsDR_ps}.
}
\label{noiseDRdrao} 
\end{figure}

\begin{figure}
\centering
\includegraphics[angle=0,width=0.8\linewidth]{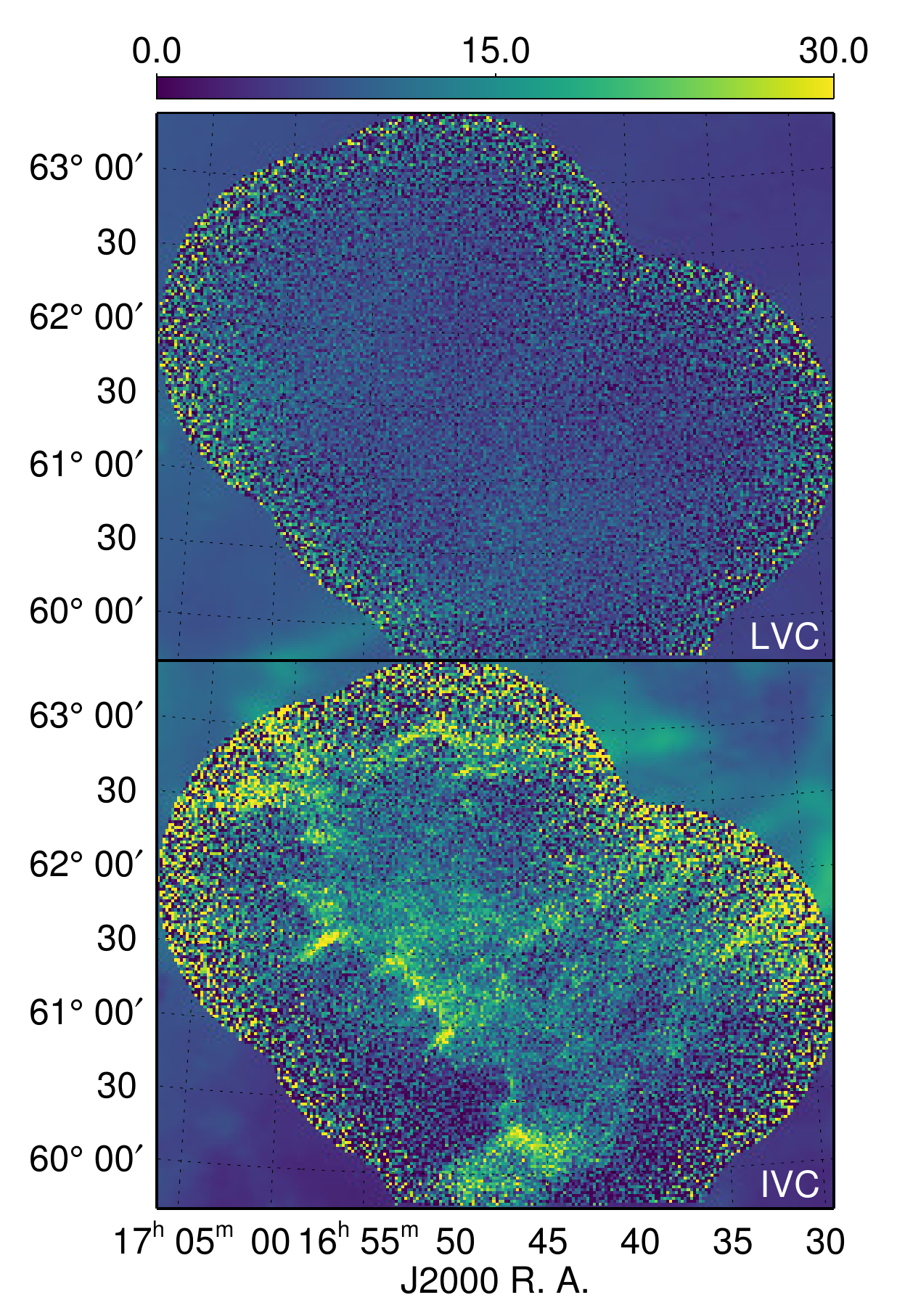}
\caption{
Maps of \nh\ for VCs in \DRG\ in units \colunits\ for the velocity
ranges in Table~\ref{compvel_table}. 
Upper: LVC.  Lower: IVC.
}
\label{DRACOcomponents}
\end{figure}

\begin{figure}
\centering
\includegraphics[clip=true,trim=40 0 0 0, angle=0,width=1.1\linewidth]{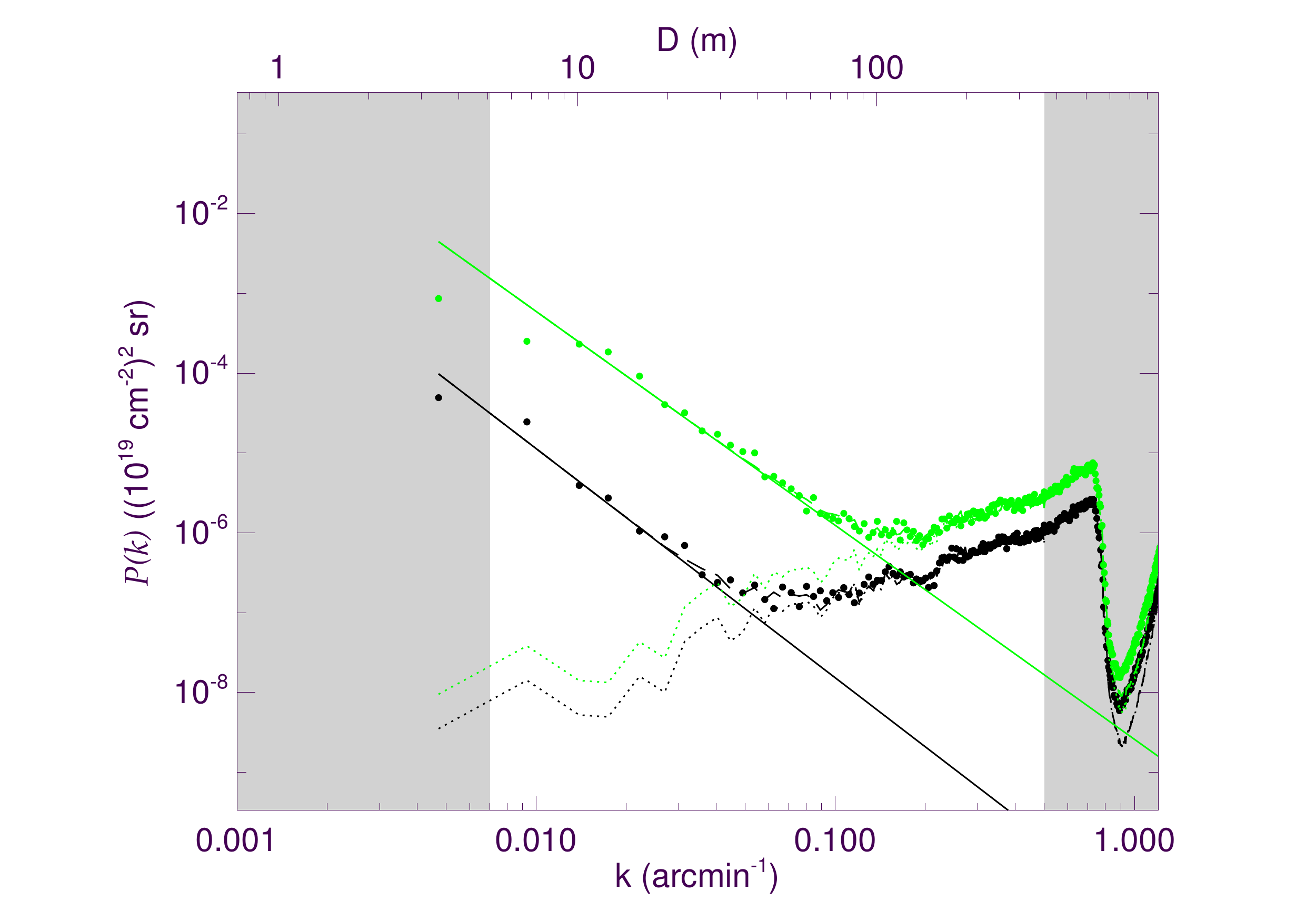}
\caption{
As in Figure~\ref{componentsDF_ps}, for power spectra of two \DRG\ VCs:
LVC (black) and IVC (green).
}
\label{componentsDR_ps}
\end{figure}

\subsection{The \DR\ Region}
\label{drsummary}

The five pointing centers for the \DRi\ mosaic in the \ghigls\ DRACO
field can be seen in Figure~\ref{field_locations} and the noise map
for \DRi\ is given in Figure~\ref{noiseDRdrao}.  The minimum noise
level, 1.5~K, is only slightly better than at the center of a single
synthesis, reflecting the relatively sparse coverage.
Also given in Figure~\ref{noiseDRdrao} is a channel map from \DRi\,
representative of IVC gas with significant \hi\ signal.

The strong IVC emission that defines the Draco nebula can also be seen
in the \nh\ map in the lower part of Figure~\ref{DRACOcomponents}.
For data within the white dashed contour, the power law exponent is
$-2.68\pm0.07$ (Figure~\ref{componentsDR_ps}).  For comparison, the
exponent evaluated for the smaller rectangular region is $-2.57 \pm
0.18$ and that for the single channel of \DRi\ analyzed in
Section~\ref{enmnoise} is $-2.27 \pm 0.14$
(Figure~\ref{noiseOTdrao_ps}).
On the other hand the LVC is very weak and so the exponent is less
well determined: $-2.87\pm0.24$.

A color map made from the \dhigls\ cube in the IVC range is presented 
in Figure~\ref{DRthree}.  Narrow emission and absorption lines in the 
IVC gas are discussed in Section~\ref{showoff}.

\clearpage

\begin{figure}
\centering
\includegraphics[clip=true,trim=10 20 10 10 ,angle=0,width=0.8\linewidth]{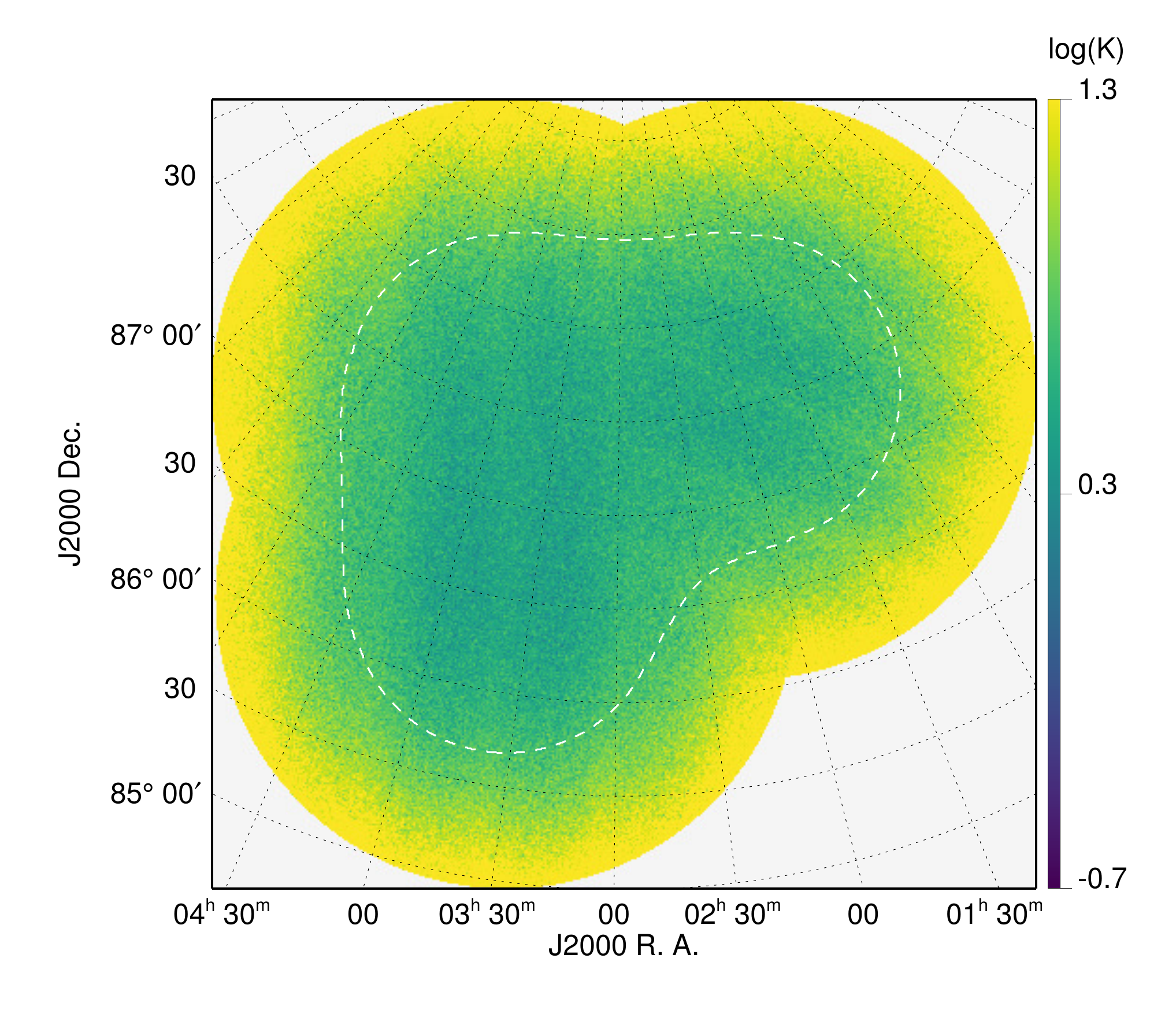}
\includegraphics[clip=true,trim=10 20 10 10 ,angle=0,width=0.8\linewidth]{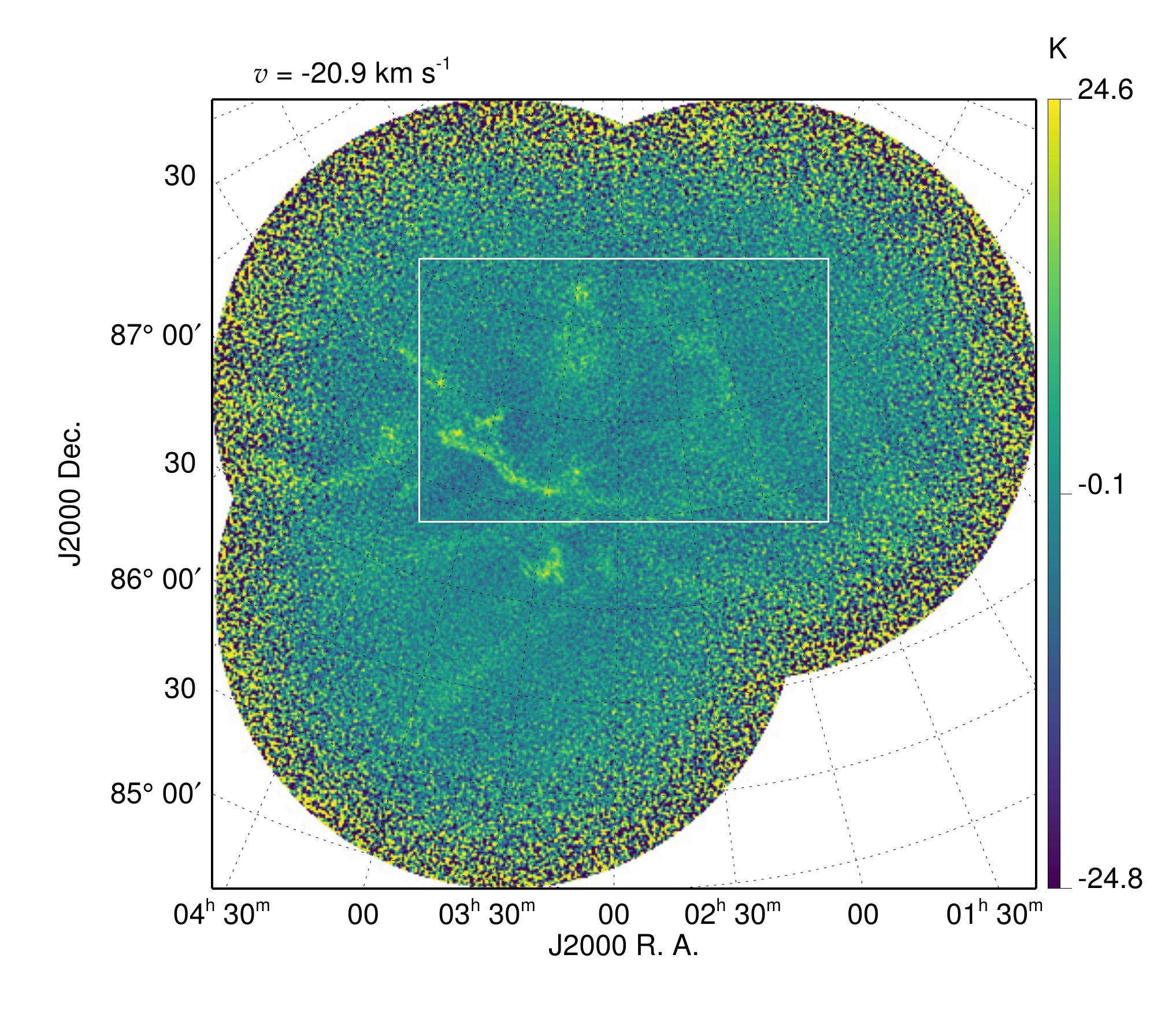}
\caption{
Similar to Figure~\ref{noiseDRdrao}, for \POi\ mosaic.  White
rectangle region used for analysis of the single-channel power
spectrum in Figure~\ref{noiseOTdrao_ps} and white dashed contour used
for analysis of the VCs in Figure~\ref{componentsPO_ps}.
}
\label{noisePOdrao} 
\end{figure}

\begin{figure}
\centering
\includegraphics[angle=0,width=0.8\linewidth]{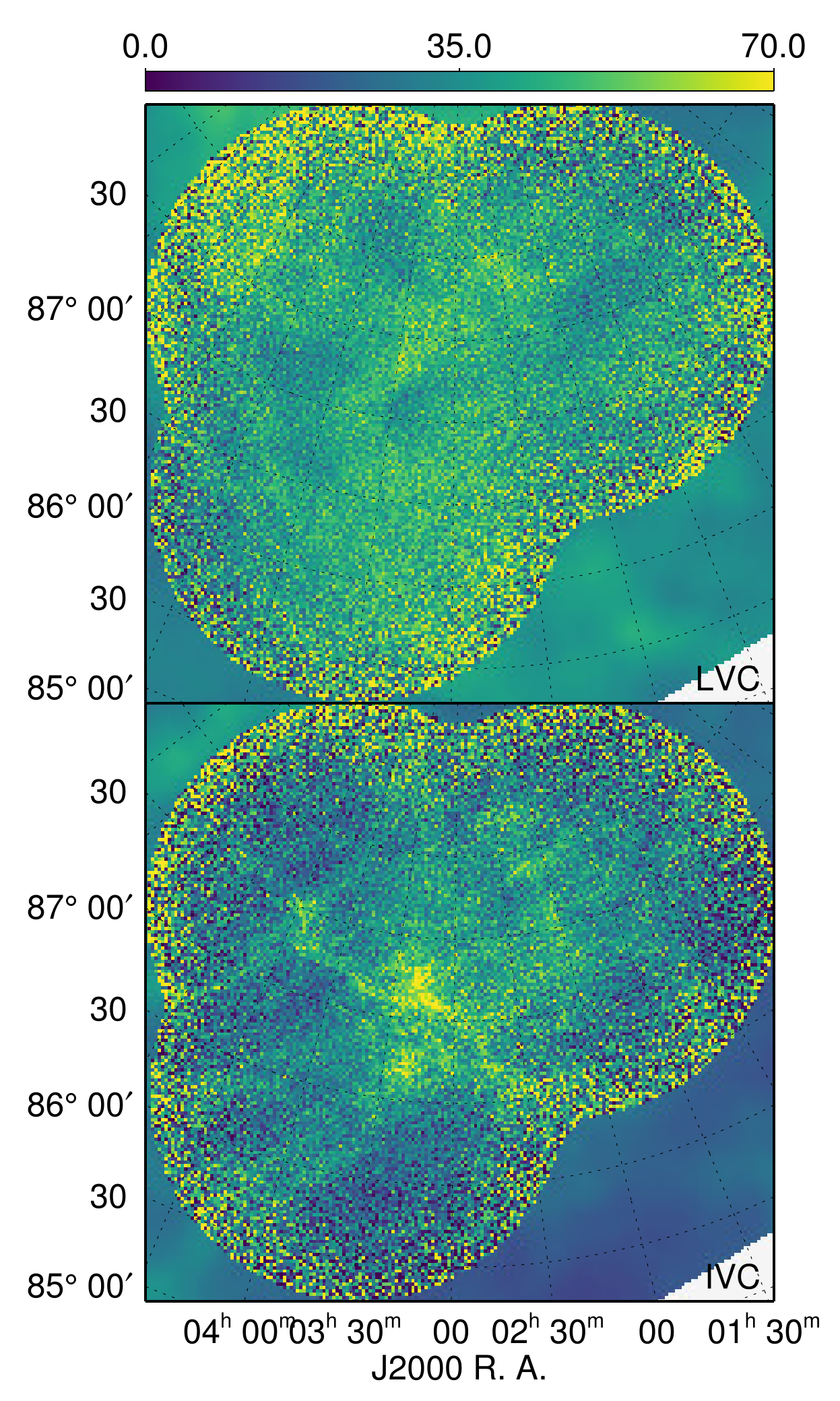}
\caption{
Maps of \nh\ for VCs in \POG\ in units \colunits\ for the velocity
ranges in Table~\ref{compvel_table}.  Upper: LVC.  Lower: IVC.
}
\label{POLcomponents}
\end{figure}

\begin{figure}
\centering
\includegraphics[clip=true,trim=40 0 0 0, angle=0,width=1.1\linewidth]{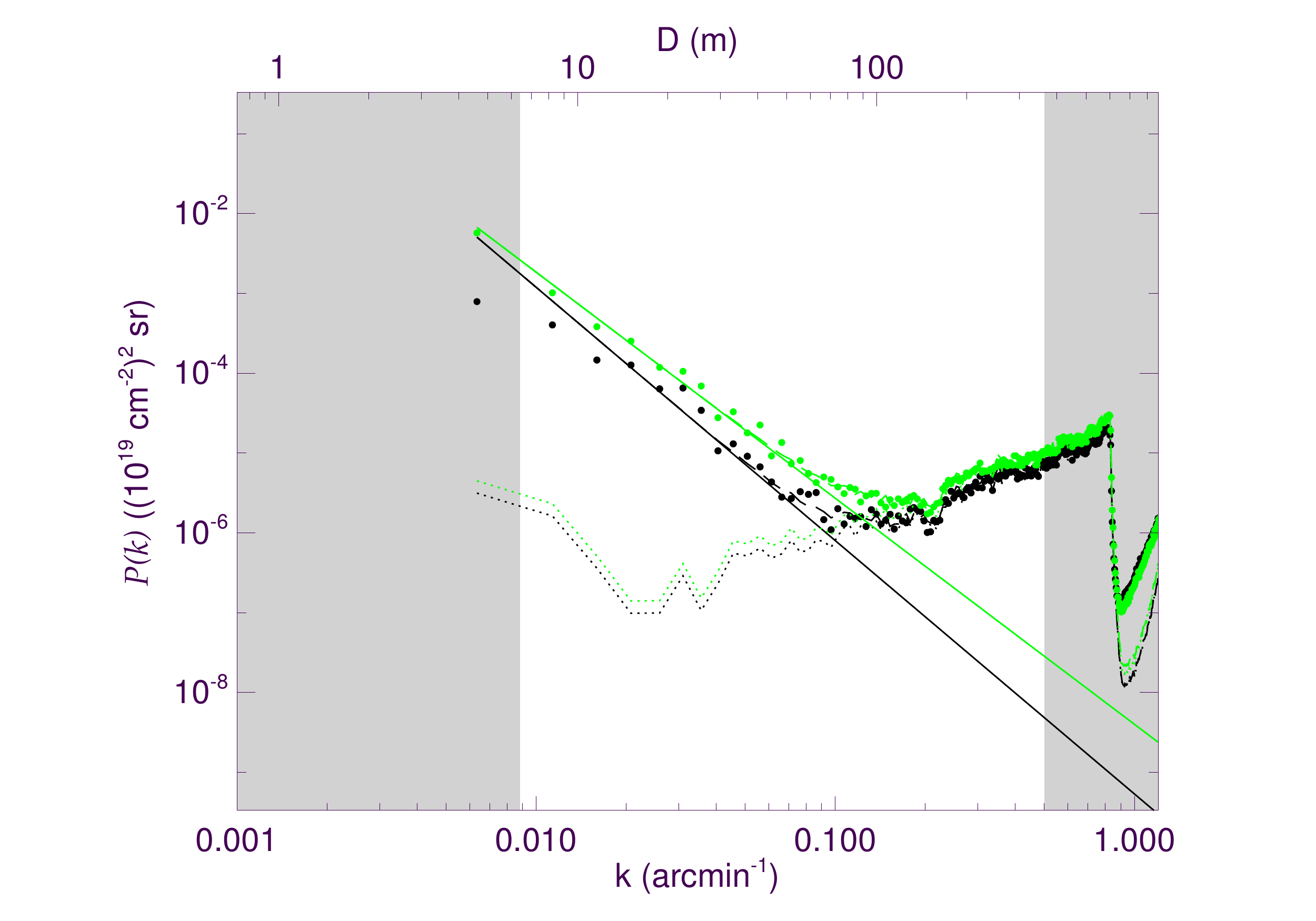}
\caption{
As in Figure~\ref{componentsDF_ps}, for power spectra of two \POG\ VCs:
LVC (black) and IVC (green).
}
\label{componentsPO_ps}
\end{figure}

\subsection{The \PO\ Region}
\label{posummary}

The three pointing centers for the \POi\ mosaic in the \ghigls\ POL
field can be seen in Figure~\ref{field_locations} and the noise map of
\POi\ is given in Figure~\ref{noisePOdrao}.  The noise level is
similar to that at the center of a single synthesis
(Figure~\ref{noiseDF20}).  The lower panel of Figure~\ref{noisePOdrao}
shows a representative channel map of \POi\ in the IVC range.

The \nh\ maps produced from the \POG\ data cube follow in
Figure~\ref{POLcomponents}.  The power law exponent for the IVC data
within the white dashed contour is $-2.82\pm0.08$
(Figure~\ref{componentsPO_ps}).  For comparison, the exponent
evaluated for the smaller rectangular region is $-2.80 \pm 0.15$ and
that for the single channel of \POi\ analyzed in
Section~\ref{enmnoise} is $-2.4 \pm 0.2$
(Figure~\ref{noiseOTdrao_ps}).
The LVC power spectrum has a somewhat small amplitude and a similar
exponent: $-3.17\pm0.16$.

\clearpage

\begin{figure}
\centering
\includegraphics[angle=0,width=0.8\linewidth]{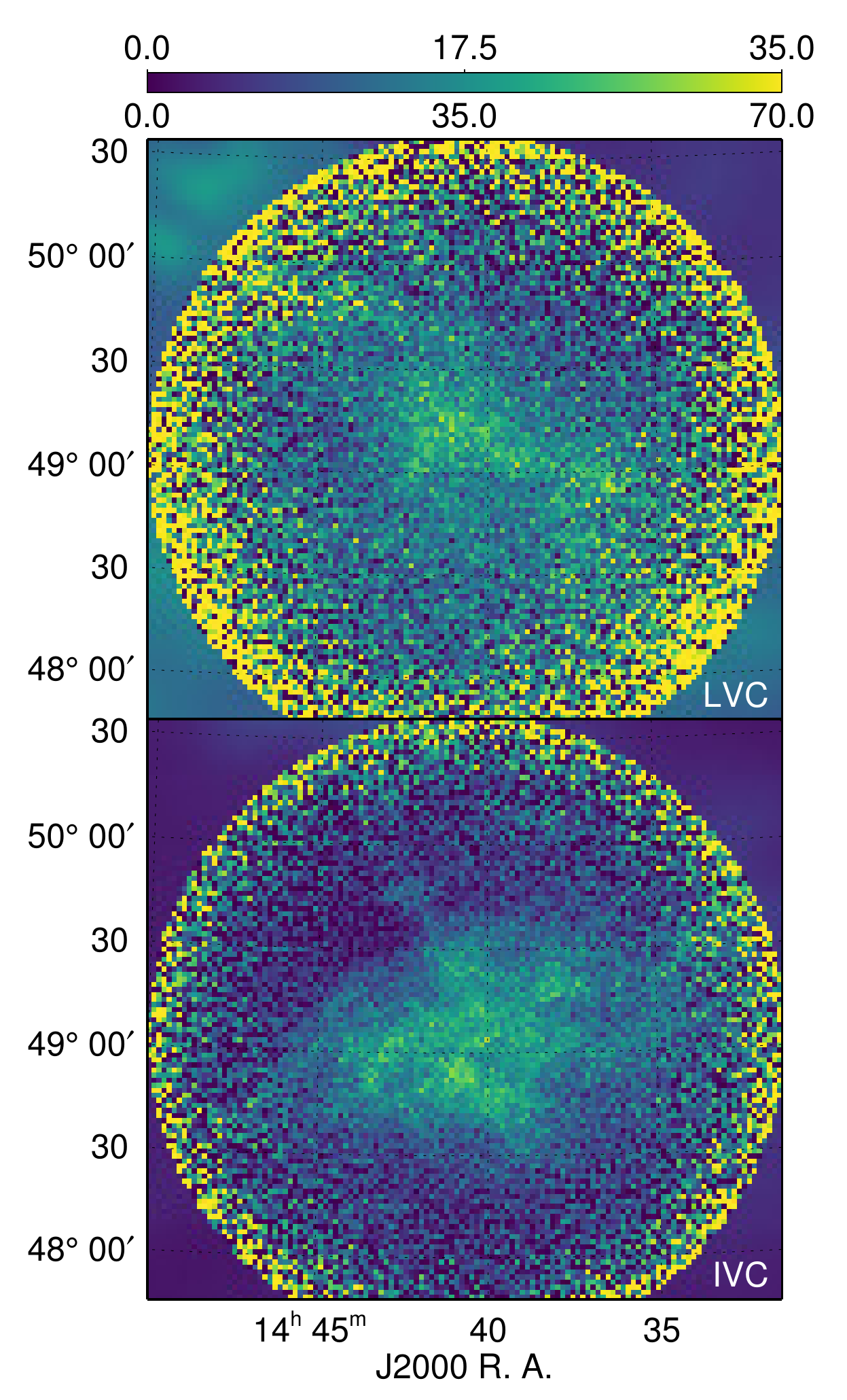}
\caption{
\nh\ component maps from single-synthesis region \MGG\ within \ghigls\
field G86, for the velocity ranges in Table~\ref{compvel_table}.
Upper: LVC.  Lower: IVC.  Units are \colunits.
}
\label{G86components}
\end{figure}

\begin{figure}
\centering
\includegraphics[angle=0,width=0.8\linewidth]{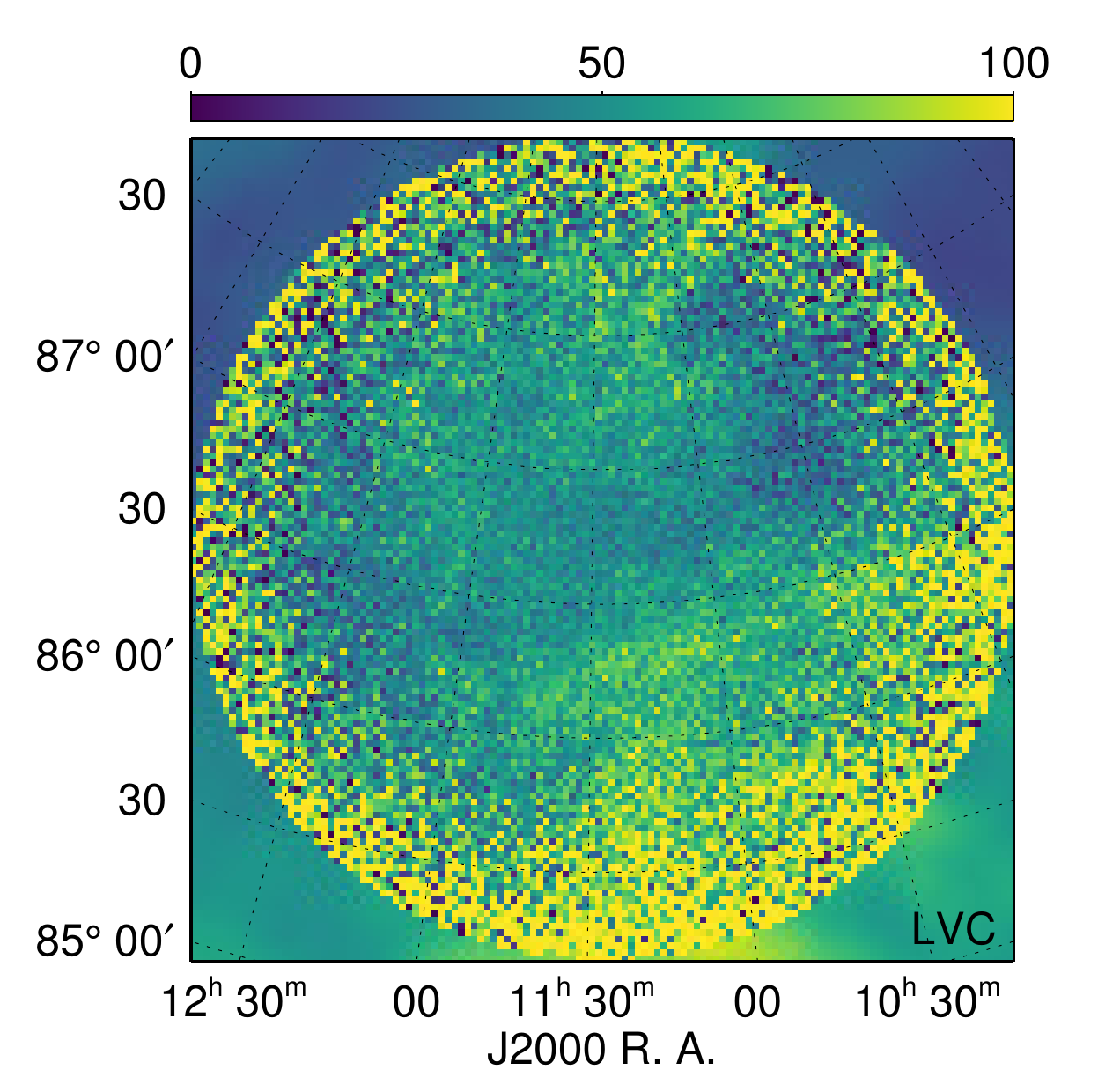}
\caption{
\nh\ map of single-synthesis region \MCG\ within \ghigls\ field POL,
for the LVC velocity range in Table~\ref{compvel_table}.  Units are
\colunits.
}
\label{POL2components}
\end{figure}

\subsection{Single Syntheses MG in G86 and MC in POL}
\label{mgsummary}

The noise maps for the two single syntheses, MG in the \ghigls\ G86
field and MC in the \ghigls\ POL field, are like that shown for DF20
in Figure~\ref{noiseDF20}, with slightly different values of $\sin
\delta$.

In Figure~\ref{G86components}, it can be seen that compared to the LVC
in \MGG, the IVC has a higher column density and more spatial structure.
The \nh\ map for LVC in \MCG\ follows in Figure~\ref{POL2components}
(IVC is much less significant).

\FloatBarrier

\subsection{Other Data from the \draost}
\label{otherst}

In this paper we are focusing on the mosaics from \hi\ spectral-line data.
However, we note that, as was the case for the CGPS, survey products
derived from observations using the \draost\ also include
complementary continuum mosaics at 408 MHz and 1420 MHz and
polarization mosaics in four bands near 1420 MHz.
Such data enable many studies \citep{kothes2010}.  Point source
catalogs can be constructed with flux density, spectral index, and
polarization information.  The polarization properties of radio
galaxies have been analyzed using the deepest continuum data then
available, in \EN\ \citep{taylor2007,grant2010}.  Rotation measures of
point sources can be used to probe the magnetic field
\citep{brown2003}, in this case at higher latitude including the
Galactic halo \citep{mao2012}.  Diffuse emission at 408 MHz and at
1420 MHz in polarization can be mapped \citep{landecker2010}, now at
intermediate latitude.  The potential correlation with \hi\ structures
can also be explored, but all of these studies are beyond the scope of
this paper.

\end{appendix}



\end{document}